\RequirePackage[loading]{tracefnt}
\documentclass[final]{jfm}
\usepackage{graphicx,amssymb}
\usepackage{epstopdf,epsfig}
\usepackage{mathtools}
\usepackage{stackengine}
\usepackage{caption}
\usepackage{subcaption}
\usepackage[section]{placeins}
\usepackage{hyperref}
\hypersetup{colorlinks=true,linkcolor=black,citecolor=black}
\usepackage[sort&compress]{natbib}
\usepackage{cleveref}
\usepackage{booktabs}
\usepackage{multirow}

\graphicspath{{./figs/}}

\crefformat{equation}{(#2#1#3)}
\crefmultiformat{equation}{(#2#1#3)}%
{ and~(#2#1#3)}{, (#2#1#3)}{ and~(#2#1#3)}

\newcommand{\abs}[1]{\lvert#1\rvert}

\newcommand{\Ren}{\mathrm{Re}}
\renewcommand{\i}{i}
\newcommand{\ee}{\varepsilon}
\newcommand{\dd}{\bar{\delta}}
\renewcommand{\t}[1]{\tilde{#1}}
\newcommand{\h}[1]{\hat{#1}}
\newcommand{\hh}[1]{\hat{\hat{#1}}}
\newcommand{\vinf}{v_{\infty }}
\newcommand{\pinf}{p_{\infty }}
\newcommand{\cb}{\bar{\chi }}
\renewcommand{\H}{\mathcal{H}}
\renewcommand{\Re}{\operatorname{\mathrm{Re}}}
\renewcommand{\Im}{\operatorname{\mathrm{Im}}}

\renewcommand{\thesection}{\arabic{section}}
\renewcommand{\theequation}{\thesection\arabic{equation}}
\numberwithin{equation}{section}

\shorttitle{Acoustics and stability of viscous shear layers over lined walls}
\shortauthor{D.~Khamis and E.~J.~Brambley}

\title{Viscous effects on the acoustics and stability of a shear layer over an impedance wall\footnote{A preliminary version of some of this work was presented at the 21st AIAA/CEAS Aeroacoustics Conference, Dallas, 2015, paper 2015-2229}}

\author{Doran Khamis\aff{1}
  \and Edward James Brambley\aff{1}
  \corresp{\email{e.j.brambley@damtp.cam.ac.uk}}}

\affiliation{\aff{1}Department of Applied Mathematics and Theoretical Physics, University of Cambridge,
Cambridge, UK}

\volume{810}
\pagerange{489--534}
\pubyear{2017}
\doi{jfm.2016.737}
\date{21 April 2016; revised 1 August 2016; accepted 1 November 2016.}

\begin{document}



\maketitle

\begin{abstract}
The effect of viscosity and thermal conduction on the acoustics in a shear layer above an impedance wall is investigated numerically and asymptotically by solving the linearised compressible Navier--Stokes equations (LNSE).  It is found that viscothermal effects can be as important as shear, and therefore including shear while neglecting viscothermal effects by solving the linearised Euler equations (LEE) is questionable.  In particular, the damping rate of upstream propagating waves is found to be under-predicted by the LEE, and dramatically so in certain instances.  The effects of viscosity on stability are also found to be important.  Short wavelength disturbances are stabilised by viscosity, greatly altering the characteristic wavelength and maximum growth rate of instability.  For the parameters considered here (chosen to be typical of aeroacoustic situations), the Reynolds number below which the flow stabilizes ranges from $10^5$ to $10^7$.

By assuming a thin but nonzero-thickness boundary layer, asymptotic analysis leads to a system of boundary layer governing equations for the acoustics. This system may be solved numerically to produce an effective impedance boundary condition, applicable at the wall of a uniform inviscid flow, that accounts for both the shear and viscosity within the boundary layer.  An alternative asymptotic analysis in the high frequency limit yields a different set of boundary layer equations, which are solved to yield analytic solutions. The acoustic mode shapes and axial wavenumbers from both asymptotic analyses compare well with numerical solutions of the full LNSE.  A closed-form effective impedance boundary condition is derived from the high-frequency asymptotics, suitable for application in frequency-domain numerical simulations.  Finally, surface waves are considered, and it is shown that a viscous flow over an impedance lining supports a greater number of surface wave modes than an inviscid flow.
\end{abstract}

\begin{keywords}
aeroacoustics, boundary layer stability, compressible flows.
\end{keywords}

\section{Introduction}

Wave propagation in a steady flow over an acoustically lined wall has been widely studied due to its applications to noise damping in acoustically lined aeroengines.  Early work considered a uniform inviscid slipping mean flow with fluctuating inviscid acoustic perturbations, as this allows analytic solution in terms of trigonometric functions in Cartesian ducts, and Bessel functions in cylindrical or annular ducts.  The lined wall was usually modelled by matching normal particle displacement in the fluid to the normal displacement of the wall, now termed the \citet{myers1980}, or Ingard--Myers~\citep{ingard1959} boundary condition, although some authors chose to match normal velocity instead~\citep{rice1969,doak&vaidya1970,ko1971}.  The effect of mean flow shear on the acoustic perturbations has also been studied~\citep[e.g.][]{ko1972,pridmore-brown1958,tack&lambert1965,mungur&gladwell1969}, usually by Fourier transforming the linearised Euler equations leading to the \citet{pridmore-brown1958} equation, which must in general be solved numerically. For a non-slipping inviscid mean flow, it was found independently by \citet{eversman&beckemeyer1972} and \citet{tester1973a} that, in the limit of a vanishingly thin inviscid shear layer, continuity of normal displacement is recovered at the lined wall.  This proved that the Myers boundary condition was the correct boundary condition for an infinitely thin inviscid slipping flow over a lined wall, and put an end to the confusion surrounding whether displacement or velocity should be matched at the boundary.  However, \citet{gabard2013} showed that, for parameters representative of aeroengines, the Myers boundary condition in some cases over-predicted sound attenuation by over 10dB when compared with the linearised Euler equations, showing that the limit of a vanishingly thin shear layer can be a poor assumption in practice \citep[as was also suggested by][]{eversman1973}.

Fundamental problems were more recently found with the mathematical formulation of slipping flow over a lined wall~\citep{brambley2009}. When the Myers condition is applied to a uniform mean flow the system is illposed, meaning rigorous mathematical stability analyses cannot be performed. The illposedness is regularised by considering a non-slipping inviscid mean flow with a finite region of shear (rather than a vortex sheet), and more recent works have sought to modify the Myers boundary condition to account for this thin shear layer~\citep{myers&chuang1984,joubert2010,rienstra&darau2011,brambley2011a,khamis&brambley2016a}. The boundary conditions resulting from these studies are wellposed, and in particular that of \citet{brambley2011a} both matches well with solutions to the full linearised Euler equations~\citep{gabard2013} and allows for the spatial and temporal stability of inviscid shear flow over a lined wall to be investigated~\citep{brambley2013} via rigorous Briggs--Bers~\citep{briggs1964,bers1983} analysis \citep[see the appendix of][]{brambley2009}.  All of these studies neglected viscosity.

It has been suggested that to accurately correlate theoretical predictions with experimental observations, viscous effects need to be taken into account~\citep{buraketal2008,buraketal2009,renou&auregan2010,renou&auregan2011}.  When \citet{boyeretal2011} numerically solved the inviscid Euler equations to attempt to identify theoretically an experimentally observed hydrodynamic instability in flow over a lining~\citep{marxetal2010}, they found they could not predict the growth rate of the unstable mode, whereas the real part of the wavenumber was reasonably well predicted; \citet{boyeretal2011} did not consider viscous effects, presumably as the Reynolds number of the experiments \citet{marxetal2010} ($\Ren\sim 2.5\times10^{5}$ by the definition in this paper) was considered high enough for viscosity to be negligible, and instead three dimensional and non-parallel flow effects were blamed for the discrepancy.  However, \citet{marx&auregan2013} did include viscosity, and found that the unstable surface wave mode found in the experimental study~\citep{marxetal2010} is very sensitive to viscosity.

A number of studies have considered the effect of viscous dissipation on sound propagation in shear flow over an impedance lining.  \Citet{nayfeh1973} considered the case where the acoustic boundary layer is thin compared with the mean flow boundary layer, expanding to first order in the acoustic boundary layer thickness.  \Citet{aureganetal2001} considered an arbitrary ratio of mean to acoustic boundary layer thickness under the assumption that both were small, but also assumed a low Mach number flow, expanding to first order in the Mach number.  They found an effective boundary condition that shifted between continuity of normal displacement and continuity of mass flux across the boundary layer, depending on the ratio of the mean flow and acoustic boundary layer thicknesses.  \Citet{brambley2011b} extended this work to relax the assumption of low Mach number, with the only remaining assumption being that the boundary layer was thin.  High frequency asymptotics of the viscous boundary layer model led to a recovery of the Myers boundary condition to leading order (conservation of normal displacement), while low frequency asymptotics led to conservation of mass flux at leading order, both in agreement with \citet{aureganetal2001} (since the acoustic boundary layer thickness scales as $1/\sqrt{\omega}$ for frequency $\omega $).  The model of \citet{brambley2011b} was effectively a viscous Myers condition, since it considered viscothermal shear flow in the limit of a vanishingly thin boundary layer thickness.  Importantly, the viscous Myers condition does not by itself regularise the illposedness of the inviscid Myers boundary condition.  Brambley found no closed-form solution for the acoustics in the viscous Myers model, however.  The studies by \citet{dokumaci2014} and \citet{mikhail&el-tantawy1994} find analytical solutions for viscous acoustics by making several limiting simplifications. \Citet{mikhail&el-tantawy1994} considered a hard-walled duct with no mean flow, and assumed the viscous fluid to be non-heat conducting. \Citet{dokumaci2014} assumed a viscous uniform mean flow and invoked the Myers condition at the wall.  Owing to the relatively low Reynolds number $\Ren \approx 10^{3}$ they considered, \citet{mikhail&el-tantawy1994} found viscous effects to be felt far outside the acoustic boundary layer, so that their solutions in the core of the duct could not be considered inviscid; in aeroacoustic situations where $\Ren \gtrsim 10^{5}$ are typical, the acoustic mode shapes in the core of the duct have generally converged to the inviscid case~\citep{khamis&brambley2015}.

At a lined wall, waves can exist that are primarily oscillations of the wall and do not propagate into the fluid, remaining localised near the boundary. These waves consequently do not exist for hard walls.  \Citet{rienstra2003} coined the term ``surface waves'' to describe these waves. Modelling the boundary layer and impedance lining using the Myers boundary condition, \citet{rienstra2003} showed that a possible four surface wave modes could exist; using the modified Myers condition \citet{brambley2013} found that a nonzero thickness shear layer could support more surface wave modes than a vortex sheet at the wall, increasing the possible number to six.  The analysis of surface waves in the literature to date has been entirely inviscid.  It is known that viscosity has the greatest effect very close to a boundary where surface waves exist, and it is therefore likely that viscosity is important for the accurate prediction of both the number and position of surface wave modes.

Acoustic liners are commonly manufactured using a perforated facing sheet, which is therefore inhomogeneous on the small scale of the distance between perforations.  The mean flow above such liners could also be expected to be inhomogeneous at the same small scale.  A common simplification in the literature is to average over this small scale and thereby model the liner as a homogeneous boundary, which has been shown to give reasonable accuracy in practice~\citep[e.g.][]{boyeretal2011}.  However, for this assumption to be valid, the acoustic wavelengths and boundary layers considered must lie within limits defined by the hole diameters and spacings of the perforated facing sheet~\citep{dai&auregan2016}.  Recent numerical and experimental work on liners in grazing flow has found that small-scale inhomogeneities of the liner may lead to liner self-noise and increased drag when compared with a boundary layer over a flat plate~\citep{tametal2014,zhang&bodony2016}, and shown that nonlinear effects may be important in accurately modelling the liner response~\citep{zhang&bodony2012}.  In the current work we forego the complications of inhomogeneities and nonlinearity in order to concentrate on the effects of viscosity.

In this paper, first the effects of viscosity and thermal conduction on the propagation of sound in a lined duct with shear flow are analysed by solving the linearised Navier--Stokes equations (\S\ref{sec:governing}) numerically. Details of the numerical solution are given in~\S\ref{sec:numericalmethod}.  Comparisons are made with inviscid computations in inviscid shear flow (\S\ref{sec:numericalresults}). To quantify the results, separate comparisons are made between the inviscid shear flow computations and analytical solutions for the acoustics in inviscid uniform flow.  Since it is widely accepted that the effects of shear are important to the acoustics in a duct~\citep{gabard2013}, we use this second comparison as a baseline against which to judge the relative importance of viscosity. Following this, the viscothermal sheared boundary layer above a lined wall is investigated asymptotically for a thin-but-finitely-thick boundary layer of thickness $\delta$, for both an $\mathcal{O}(1)$ frequency and in a high frequency limit, in~\S\ref{sec:asymptotics}.  In the latter case, analytical forms for the acoustics in the boundary layer are found, and a closed-form effective impedance boundary condition is derived which includes the effects of both shear and viscosity.  The accuracy of these asymptotics are compared in \S\ref{sec:asympresults}, together with their predictions for surface waves on the lined surface.

\section{Governing equations and nondimensionalisation} \label{sec:governing}

A fluid can be described by six variables: three orthogonal components of the velocity $\mathbf{u} = (u, v, w)$, and three state variables $(P, \rho, T)$, the pressure, density, and temperature respectively.  The dynamics of a viscous, compressible perfect gas are governed by the Navier--Stokes equations~\citep{landau&lifshitz,stewart1942}
\begin{subequations}
\begin{gather}
\frac{\partial \rho }{\partial t} + \nabla \cdot (\rho \mathbf{u}) = 0, \\
\rho \frac{D\mathbf{u}}{Dt} = -\nabla p + \nabla \cdot \sigma, \\
\rho \frac{DT}{Dt} = \frac{Dp}{Dt} + \nabla \cdot (\kappa \nabla T) + \sigma _{ij} \frac{\partial u_{\i}}{\partial x_{j}}, \\
T = \frac{\gamma }{\gamma -1}\frac{p}{\rho }, \label{constitutivelaw}
\end{gather}
\label{fullgoverning}%
\end{subequations}
where the material derivative $D/Dt = \partial/\partial t + \mathbf{u}.\nabla $, $\kappa $ is the thermal conductivity, $\gamma $ is the ratio of specific heats, and 
\begin{equation}
\sigma _{ij} = 2\mu \left(\frac{\partial u_{\i}}{\partial x_{j}} + \frac{\partial u_{j}}{\partial x_{\i}}\right) + \mu ^{S} \nabla \cdot \mathbf{u} \delta _{ij}. \label{stresstensor}
\end{equation}
is the viscous stress tensor, for the shear viscosity $\mu $ and the second~\citep{tritton1988} viscosity $\mu ^{S} = \mu ^{B} - 2\mu /3$, where $\mu ^{B}$ is the bulk viscosity.  All quantities in \cref{fullgoverning} and \cref{stresstensor} have been nondimensionalised by the following scheme. We imagine a cylindrical duct with a constant axial base flow at its centreline. With a star denoting a dimensional variable and a subscript $0$ denoting a centreline value, we have density $\rho ^{*}= \rho_{0}^{*}\rho$; temperature $T^{*} = c_{0}^{*2}/c_{p}^{*}T$, where $c_{p}^{*}$ is the specific heat at constant pressure; and velocity $\mathbf{u}^{*} = c_{0}^{*}\mathbf{u}$, where $c_{0}^{*}$ is the centreline speed of sound.  Lengths are scaled by the duct radius $l^{*}$, pressure and viscous stresses by $c_{0}^{*2}\rho _{0}^{*}$, and time by $l^{*}\!/c_{0}^{*}$.  Coefficients of viscosity (shear and bulk) are scaled by $c_{0}^{*}l^{*}\!\rho _{0}^{*}$, and thermal conductivity by $c_{0}^{*}l^{*}\!\rho_{0}^{*}c_{p}^{*}$. In such a scheme, the dimensionless centreline base flow quantities take the values $\rho _{0} = 1$, $T_{0} = 1/(\gamma -1)$, $P_{0} = 1/\gamma $, and $U_{0} = M$, where $M$ is the centreline Mach number.  The cylindrical duct has a dimensionless radius of unity. The base flow is approximated as being everywhere parallel, hence the base pressure is constant across the duct cross-section, $P \equiv P_{0}$.  Moreover, we assume a non-swirling mean flow, with $\mathbf{U}\cdot \mathbf{e}_{\theta} = 0$.

We define the Reynolds number with respect to the centreline sound speed, $\Ren = c_{0}^{*}l^{*}\!\rho_{0}^{*}/\mu _{0}^{*}$.  The Prandtl number is similarly defined by centreline variables, $\Pr = \mu _{0}^{*}c_{p}^{*}/\kappa _{0}^{*}$.  These definitions allow the viscosity and thermal conductivity, assumed to have some dependence on radial position and temperature through the function $\H(r,T)$, to be expressed in terms of $\Ren$ and $\Pr$ as 
\begin{align}
\mu &= \frac{\H}{\Ren},&
\mu ^{B} &= \frac{\H}{\Ren}\frac{\mu _{0}^{B*}}{\mu _{0}^{*}},&
\kappa &= \frac{\H}{\Ren \Pr}. \label{visctherm}
\end{align}
Previous studies of viscothermal acoustic propagation have used constant values~\citep{aureganetal2001,nayfeh1973} or a linear temperature dependence~\citep{brambley2011b} for the molecular viscosities and thermal conductivity. Radial dependence of the viscosity is important when modelling a turbulent eddy viscosity~\citep{marx&auregan2013}.  For now we leave $\H$ as a general function of position and temperature.

Each flow variable is assumed to have a time-averaged base flow part and a time harmonic acoustic part such that $Q_{\mathrm{total}}(\mathbf{r},t) = Q(\mathbf{r}) + \epsilon _{a}q(\mathbf{r})\exp{\{\i\omega t\}}$, where $\epsilon _{a}\ll 1$ is the magnitude of acoustic oscillations, and $\omega $ is the nondimensional frequency (the Helmholtz number).

    \subsection{Steady base flow} \label{sec:baseflow}
    
\begin{figure}
\captionsetup[subfigure]{aboveskip=0pt}
    \centering
        \includegraphics*[width=0.8\textwidth]{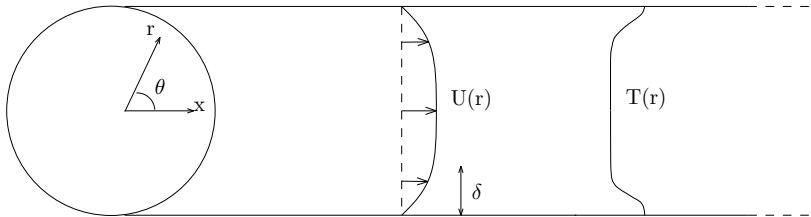}
    \caption{Schematic of parallel flow in a cylindrical duct with radially-varying temperature.} \label{fig:ductgeom}
\end{figure}

Rather than solve the steady Navier--Stokes equations, or indeed solve the unsteady Navier--Stokes equations (possibly together with a sub-grid-scale turbulence model) and then time average, we here approximate the steady base flow as a simple analytic profile, which we do not therefore require to satisfy~\cref{fullgoverning}.  This is in order to compare with inviscid results with similar assumptions and with empirical profiles derived from experiments.  We consider flow along a cylinder with coordinate system $\mathbf{r} = (x, r, \theta)$, as shown in \cref{fig:ductgeom}, and take the steady base flow velocity and temperature to be independent of $x$ and $\theta$, and the base flow velocity to be in the axial direction.  This is a reasonable approximation for viscous flow provided we are sufficiently far from a leading edge that the boundary layer varies little over the $x$ range to be considered~\citep{brambley2011b}.  At the duct wall, we also require no slip, $U(1)=0$, and thermal equilibrium, $T_r(1)=0$, where a subscript denotes differentiation.  Subject to these constraints, no restriction is placed on the mean flow velocity profile $U(r)$ and temperature profile $T(r)$.  The base flow density is given by $1/\!\rho(r) = (\gamma - 1)T(r)$ from the constitutive law \cref{constitutivelaw} and the fact that the base flow pressure is constant.  Note that if $T(r)$ were a constant independent of $r$ then the base flow sound speed $c_0(r)$ would also be constant, although this is not assumed in what follows.

All results presented here are for a hyperbolic velocity and temperature profile,
\begin{subequations}
\begin{align}
U(r) &= M\tanh\!\left(\frac{1-r}{\delta}\right)\! + M\!\left(\!1-\tanh\!\left(\frac{1}{\delta}\right)\!\!\right)\!\!\left(\frac{1 + \tanh(1/\delta)}{\delta}r + (1+r)\!\!\right)\!(1-r) \label{Utanh} \\
T(r) &= T_{0} + \tau\!\left(\!\cosh{\left(\frac{1-r}{\delta }\right)}\!\right)^{\!-1}, \label{Tcosh}
\end{align}
\label{baseflow}%
\end{subequations}
where $\delta$ is a measure of boundary layer thickness, with $U(1-\delta )\approx 0.76M$ and $U(1-3\delta )\approx 0.995M$.  Motivated by the compressible Blasius boundary layer temperature profile, we take $\tau = 0.104$ to three significant figures in what follows.  We vary $M$ and $\delta$ later.

    \subsection{The linearised Navier--Stokes equation}

To derive the linearised Navier--Stokes equations (LNSE) we Fourier transform the acoustic quantities in the axial coordinate, and define a Fourier series in the azimuthal coordinate.  Then we consider a single mode, writing $q(\mathbf{r}) = \t{q}(r)\exp{\{-ikx -im\theta \}}$ where $k$ is the axial wavenumber and $m$, an integer, is the azimuthal mode number. Linearising \cref{fullgoverning} about the base flow, retaining only terms at $\mathcal{O}(\epsilon _{a})$, leads to the LNSE
\begin{subequations}
\begin{align}
0 =&\, \i(\omega - U k)\gamma \t{p} - \i (\omega - U k)(\gamma - 1)\rho \t{T} - \i k\t{u} + T\left(\frac{\t{v}}{T}\right)_{r} + \frac{1}{r}\t{v} - \frac{\i m}{r}\t{w} \label{lns_mass} \\
0 =&\, \i \rho(\omega - U k)\t{u} + \rho U_{r}\t{v} - \i k\t{p} - \frac{1}{\Ren}\Bigg\{ (\H\t{u}_{r} + U_{r}\t{\H})_{r} + \frac{1}{r}(\H\t{u}_{r} + U_{r}\t{\H}) - \frac{m^{2}}{r^{2}}\H\t{u} \notag \\
&\,- (2+\beta)k^{2}\H\t{u} - \i k(1+\beta)(\H\t{v})_{r} + \i k \beta \H_{r}\t{v} - \frac{\i k}{r}(1+\beta)\H\t{v} - \frac{k m}{r}(1+\beta)\H\t{w}\!\Bigg\} \label{lns_umom}\\
0 =&\, \i \rho(\omega - U k)\t{v} + \t{p}_{r} - \frac{1}{\Ren}\Bigg\{(2+\beta)(\H\t{v}_{r})_{r}- \frac{2}{r}\H_{r}\t{v} - \Big(k^{2} + \frac{m^{2}}{r^{2}}\Big)\H\t{v} + (2+\beta)\Big(\frac{\H\t{v}}{r}\Big)_{r}\notag \\
&\,  - \i k (1+\beta)(\H\t{u})_{r} + \i k(\H_{r}\t{u} - U_{r}\t{\H}) - \i m(1+\beta)\Big(\frac{\H\t{w}}{r}\Big)_{r}\! + \frac{\i m}{r}\H_{r}\t{w} + \frac{2\i m}{r^{2}}\H\t{w}\!\Bigg\}\label{lns_vmom}\\
0 =&\, \i \rho(\omega - U k)\t{w} - \frac{\i m}{r}\t{p} - \frac{1}{\Ren}\Bigg\{\!-\frac{k m}{r}(1+\beta)\H\t{u} - \frac{\i m}{r^{2}}(3+\beta)\H\t{v} - \frac{\i m}{r}(1+\beta)(\H\t{v})_{r}  \notag \\
&\,+ \frac{\i m}{r}\beta \H_{r}\t{v} - \left(\!k^{2} + \frac{m^{2}}{r^{2}}\right)\!\H\t{w} + (\H\t{w}_{r})_{r} - \frac{m^{2}}{r^{2}}(1+\beta)\H\t{w} + \H\!\left(\frac{\t{w}}{r}\right)_{\!r}\! - \frac{1}{r}\H_{r}\t{w}\Bigg\} \label{lns_wmom} \\
0 =&\, \i \rho(\omega - U k)\t{T} + \rho T_{r}\t{v} - \i(\omega - U k)\t{p} - \frac{1}{\Ren}\Bigg\{U_{r}^{2}\t{\H} + 2\H U_{r}\t{u}_{r} - 2\i k\H U_{r}\t{v}\notag \\
&\,+\frac{1}{\Pr}\!\left(\!(\H\t{T}_{r} + T_{r}\t{\H})_{r} + \frac{1}{r}(\H\t{T}_{r} + T_{r}\t{\H}) - \left(\!k^{2} + \frac{m^{2}}{r^{2}}\right)\!\H\t{T}\right)\!\!\Bigg\}, \label{lns_energy}
\end{align}\label{lns}%
\end{subequations}
where a subscript denotes differentiation.  We have introduced the shorthand $\beta = \mu _{0}^{B*}/\mu _{0}^{*} - 2/3$, where $\mu _{0}^{B*}/\mu _{0}^{*} = 0.6$ will be used in all computations~\citep[consistent with the range of possible values stated in][]{pinkerton1947,greenspan1959,cramer2012}.  The density perturbation has been eliminated from the system \cref{lns} using the constitutive law, \cref{constitutivelaw}:
\begin{equation}
\t{\rho } = \frac{\gamma }{\gamma - 1}\frac{\t{p}}{T} - \frac{\rho }{T}\t{T}. \label{elim_density}
\end{equation}
The system \cref{lns} is closed by assigning a functional form to $\H(r,T)$ and Taylor expanding; to leading order in the acoustic perturbations we find
\begin{equation}
\t{\H} = \left.\frac{\partial \H(r,T)}{\partial T}\right\vert_{(r,T)}\t{T}. \label{viscousH}
\end{equation}
In this work we forego the radial dependence of the viscosity and choose a linear temperature dependence as a `leading order' approximation of the true dependence~\citep{kadoyaetal1985}.\footnote{The reference gives the viscosity of dry air as $\mu = B(A_{1}T + A_{0.5}T^{0.5} + \sum_{\i=0}^{-4}A_{\i}T^{\i}) + \Delta \mu_{\rho}$, where $B$ and the $A_{j}$ are empirical fitting parameters and the $\Delta \mu_{\rho}$ is some small ``excess'' viscosity that depends on the density.} Thus, we define
\begin{equation}
\H(r,T) = \frac{T}{T_{0}}. \label{viscousdependence}
\end{equation}
The shear and bulk viscosities are assigned the same temperature dependence, as their ratio $\mu ^{B*}/\mu ^{*}$ is relatively insensitive to temperature variations in both air and water~\citep[][chap.~10]{pierce1981}.

The LNSE, as given in \cref{lns} with \cref{viscousdependence}, form a system of five linear ordinary differential equations in $r$ for the five acoustic quantities $(\t{p}, \t{u},\t{v}, \t{w}, \t{T})$. The system is first order in $\t{p}$ and second order in $\t{u}$, $\t{v}$, $\t{w}$ and $\t{T}$ so we may apply nine boundary conditions.  At $r=0$, each variable must satisfies a regularity condition.  At the duct wall $r=1$, no slip specifies $\t{u}(1) = \t{w}(1)=0$ while the isothermal assumption of the duct wall specifies $\t{T}(1)=0$.  As the ninth boundary condition we choose to normalise our solution by specifying $\t{p}(1) = 1$ in order to force a nonzero solution, which therefore leaves $\t{v}(1)$ unconstrained.  Note that a unique solution with these boundary conditions may be expected for any values of $\omega$, $k$ and $m$.

At the wall, the acoustic pressure drives a nonzero radial velocity given by the impedance $Z$ (nondimensionalised by $Z^* = \rho_0^*c_0^*Z$).  Because of the no slip condition, this unambiguously implies the additional boundary condition
\begin{equation}
\frac{\t{p}(1)}{\t{v}(1)} = Z, \label{dispersion}
\end{equation}
which is a dispersion relation relating allowable values for $\omega$, $k$ and $m$.  Each allowable value of $(\omega, k, m)$ is referred to as a duct mode.  The impedance $Z$ may depend upon $\omega$, $k$ or $m$ through an appropriate, causal liner model~\citep{rienstra2006}.  For example, a mass--spring--damper model of the boundary with a mass $d$, spring constant $b$ and damping coefficient $R$ gives the impedance
\begin{equation}
Z(\omega ) = R + \i \omega d - \i b/\omega. \label{msdZ}
\end{equation}
However, for now we make no assumption of the specific form of $Z(\omega, k)$.

\section{Numerical method} \label{sec:numericalmethod}

Here, we describe the method used here to solve the LNSE~\cref{lns} numerically.  The domain of the problem is $r\in[0,1]$, although near the wall at $r=1$ we will consider a thin boundary layer. We therefore choose to discretise the domain non-uniformly, with more grid points clustered near the boundary $r=1$.  This non-uniform grid is then mapped to a uniformly-spaced computational grid, $\psi \in[0,1]$, using the map
\begin{equation}
r = \frac{\tanh{S\psi }}{\tanh{S}}, \label{domainmap}
\end{equation}
where $S>0$ is a stretching parameter.  Larger values of $S$ allow more points to be clustered near $r=1$. The LNSE~\cref{lns} are then rewritten using $\psi$ as the independent variable using
\begin{equation}
\frac{\mathrm{d}}{\mathrm{d}r} = \frac{\partial \psi }{\partial r}\frac{\mathrm{d}}{\mathrm{d}\psi }. \label{derivmap}
\end{equation}
It was found that this mapping allowed the numerical derivatives to be calculated more stably than using directly a nonuniform computational discretisation for $r$.

The computational domain is then discretised into $N$ equally-spaced points, forming a matrix equation $A\mathbf{x} = \mathbf{b}$ where $A$ is a $5N\times5N$ sparse matrix and $\mathbf{x}$ is the solution vector. The vector $\mathbf{b}$ contains zeros in $5N-1$ entries (the homogeneous equations and eight boundary conditions), with the the pressure normalisation being enforced by the remaining (nonzero) entry.  In forming $A$, radial derivatives are approximated using a 6th order 7-point centred finite difference stencil~\citep{brambley2015,brambley2016}, since exponential growth and decay is expected in the $r$-direction as well as oscillations. The system is solved using a sparse matrix solver, with in general $N=8000$ being sufficient for convergence to typical errors of $\lesssim 10^{-8}$ (see \cref{sec:appendixNUM}), and $S$ chosen large enough to place at least $400$ points inside the boundary layer irrespective of its thickness.  The convergence of the numerical method is further evidenced by the good agreement with the asymptotics seen in~\cref{sec:asympresults}.

The same solver may be used to produce the inviscid results by numerically setting $1/\Ren = 0$.  In the inviscid case, only two boundary conditions may be applied, which we take to be the $\t{p}(1) = 1$ normalisation and the $\t{v}(0)$ regularity condition, with regularity of the other variables at $r=0$ following automatically in this case.

For a given azimuthal mode $m$ and impedance model for $Z$, a Newton--Raphson iteration is used to find complex values of $k$ (or $\omega $) given $\omega $ (or $k$) such that the dispersion relation~\cref{dispersion} is satisfied.  More information is given in \cref{sec:appendixNUM}.

\section{Comparisons of viscous and shear effects} \label{sec:numericalresults}   

We make two types of comparisons in this study: comparing sheared viscous solutions of the LNSE (labelled sv) with sheared inviscid solutions (labelled si), both of which are found using the numerical method described above; and comparing sheared inviscid~(si) solutions with uniform flow inviscid solutions (labelled ui).  The uniform inviscid solutions are found analytically by setting $(U(r), \rho(r), T(r), 1/\Ren) = (M, 1, 1/(\gamma-1), 0)$, giving solutions in terms of Bessel functions,
\begin{align}
\t{p}_{\mathrm{ui}}(r) &= \frac{J_{m}(\alpha r)}{J_{m}(\alpha)},&
\t{v}_{\mathrm{ui}}(r) &= \frac{\i\alpha J'_{m}(\alpha r)}{(\omega - Mk)J_{m}(\alpha)},&
\alpha ^{2} &= (\omega - Mk)^{2} - k^{2},
 \label{uniform}
\end{align}
where where $J'_{m}(\alpha r)$ denotes the first derivative of $J_{m}$ with respect to its argument. In equations \cref{uniform} we make no assumption about the impedance boundary at $r=1$.

\begin{figure}
\captionsetup[subfigure]{aboveskip=0pt}
    \centering
    \begin{subfigure}[t]{0.9\textwidth}
        \centering        
        \includegraphics*[width=1.\textwidth]{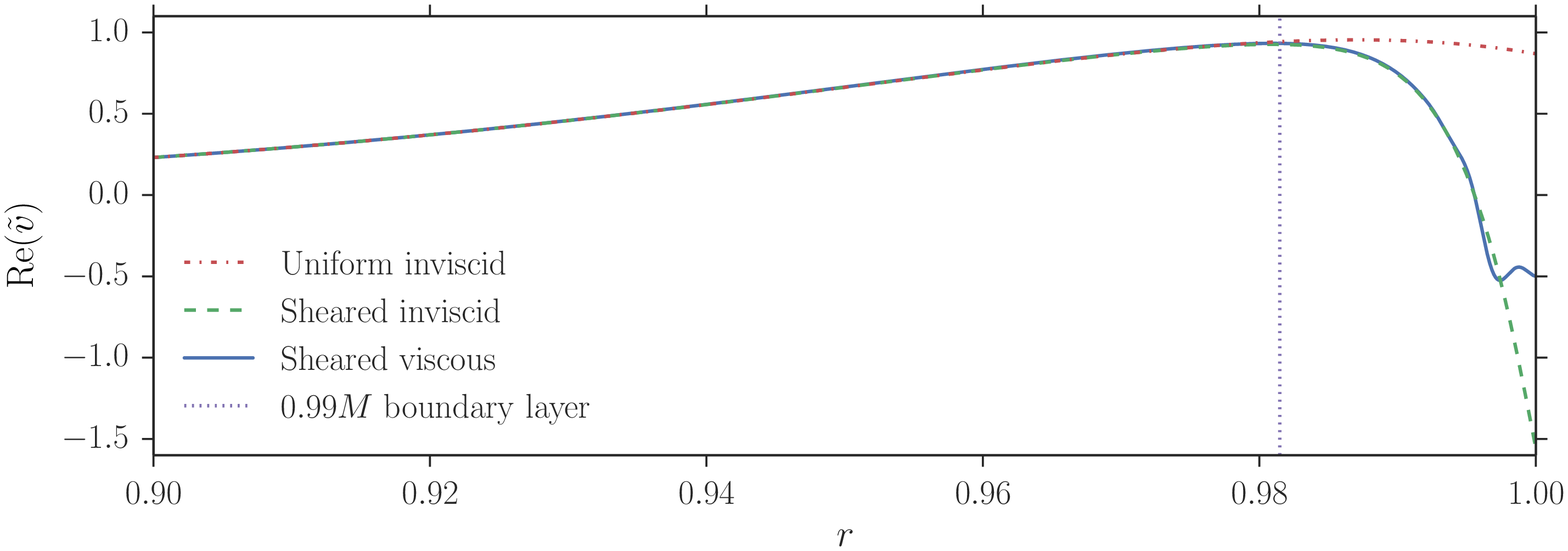}
    \caption{}
    \label{fig:examplemodeshapev}
    \end{subfigure}%
    
    \begin{subfigure}[t]{0.9\textwidth}
        \centering
        \includegraphics*[width=1.\textwidth]{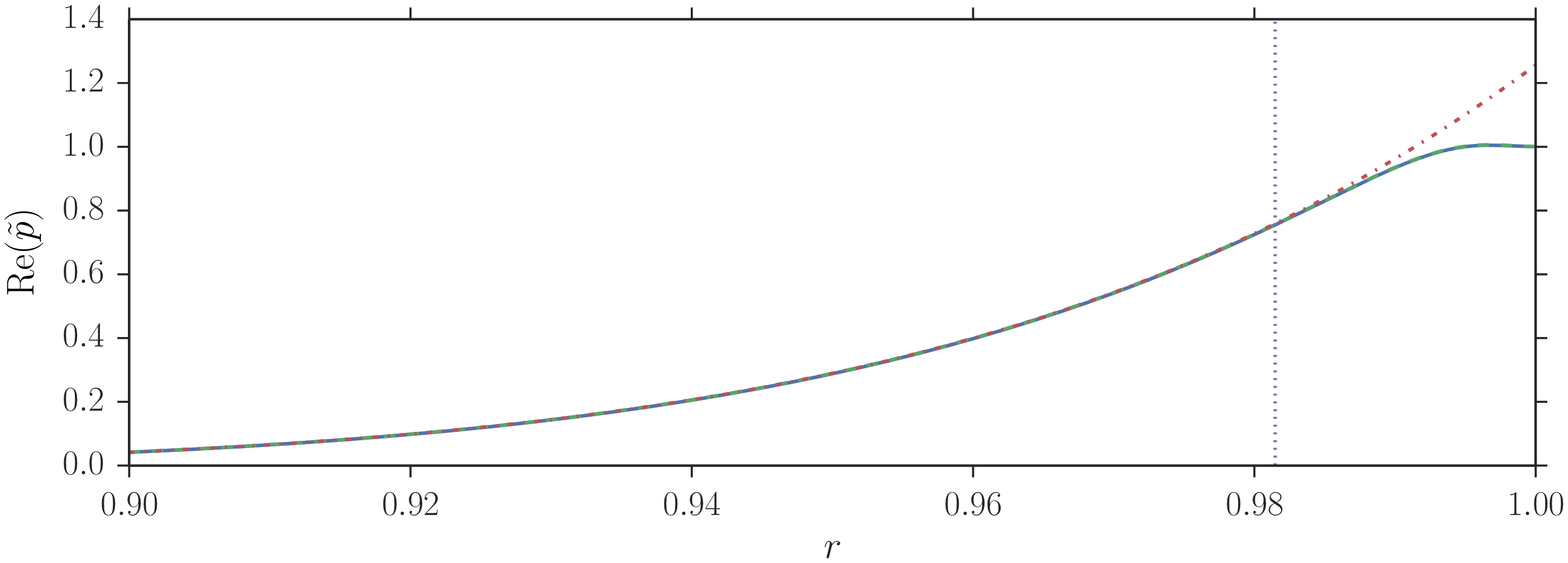}
    \caption{}
    \label{fig:examplemodeshapep}
    \end{subfigure}%
    \caption{The (a) $\Re(\t{v})$ and (b) $\Re(\t{p})$ mode shapes of the three solutions: $\t{q}_\mathrm{ui}$ (uniform inviscid, dashdot), $\t{q}_\mathrm{si}$ (sheared inviscid, dashed), and $\t{q}_\mathrm{sv}$ (sheared viscous, solid) where in (a) $\t{q} = \t{v}$ and in (b) $\t{q} = \t{p}$. The dotted vertical line lies at $r=1-3\delta $, where $U \simeq 0.99M$. Parameters are $\omega = 5$, $k = 26 - 14i$, $m=12$, $M=0.5$, $\delta = 7\times 10^{-3}$, $\Ren = 1\times 10^{6}$, with the hyperbolic base flow \cref{baseflow}. The mean squared error between the solutions for $r<1-5\delta$ is $10^{-6}$.} \label{fig:examplemodeshape}
\end{figure}

As an initial illustrative example, \cref{fig:examplemodeshape} compares the mode shapes of the three solutions, the uniform inviscid solutions rescaled (by varying $\t{p}(1)$) to match the numerical solutions in the core of the duct $r<1-3\delta $ where shear and viscothermal effects are negligible~\citep[as anticipated from][]{khamis&brambley2015}. Shear and viscothermal effects are seen only to be significant within the boundary layer region $r>1-3\delta $, where they produce $\mathcal{O}(1)$ effects at the wall $r=1$.  The plots in \cref{fig:examplemodeshape} would correspond to a mode (i.e.\ a solution of the dispersion relation~\cref{dispersion}) if $Z = -1.27+0.97\i$ for the sheared, viscous numerics, if $Z = -0.64+0.02\i$ for the sheared, inviscid numerics, and if $Z = 0.12+0.35\i$ for the inviscid uniform flow solution.  This suggests that these three solutions will have significantly different interactions with an impedance wall, as we see next.

\subsection{Impedance errors}

Setting $Z_\mathrm{sv} = \t{p}_\mathrm{sv}(1)/\t{v}_\mathrm{sv}(1)$ and $Z_\mathrm{si} = \t{p}_\mathrm{si}(1)/\t{v}_\mathrm{si}(1)$, we define the impedance error due to assuming an inviscid fluid (henceforth referred to as viscous impedance error) as 
\begin{equation}
\min\{\abs{Z_\mathrm{sv}-Z_\mathrm{si}}, \abs{1/Z_\mathrm{sv} - 1/Z_\mathrm{si}}\},
\end{equation}
which was chosen to handle correctly near-zero and near-infinite impedances.  With this definition, this impedance error is also the admittance error for the admittance $Y = 1/Z$.  Note that, since we are not solving the dispersion relation \cref{dispersion}, the impedance $Z$ at the wall is not prescribed here; we are merely comparing the impedance produced by the viscous and inviscid equations.  Similarly, the impedance error associated with neglecting base flow shear in an inviscid system (henceforth referred to as shear impedance error) is given by 
\begin{equation}
\min\{\abs{Z_\mathrm{si}-Z_\mathrm{ui}}, \abs{1/Z_\mathrm{si} - 1/Z_\mathrm{ui}}\},
\end{equation}
where $Z_\mathrm{ui} = \t{p}_\mathrm{ui}(1)/\t{v}_\mathrm{ui}(1)$. To avoid division by zero in the calculation of $Z_\mathrm{ui}$, a small imaginary part is added to the frequency in the following computations, equal to $1\%$ of $\Re(\omega)$.

There are many parameters that affect the acoustics in the boundary layer, and the importance of both shear and viscosity will depend on the particular values used.  In order to draw meaningful conclusions, we calculate the impedance error for axial wavenumbers across a section of the complex plane $\abs{\Re(k)}\le100$, $\abs{\Im(k)}\le100$, and choose two main parameters to investigate: the frequency and Mach number.

\begin{figure}
\captionsetup[subfigure]{aboveskip=1pt}
    \centering
    \begin{subfigure}[t]{0.33\textwidth}
        \centering
        \caption{}
        \includegraphics*[width=1.\textwidth]{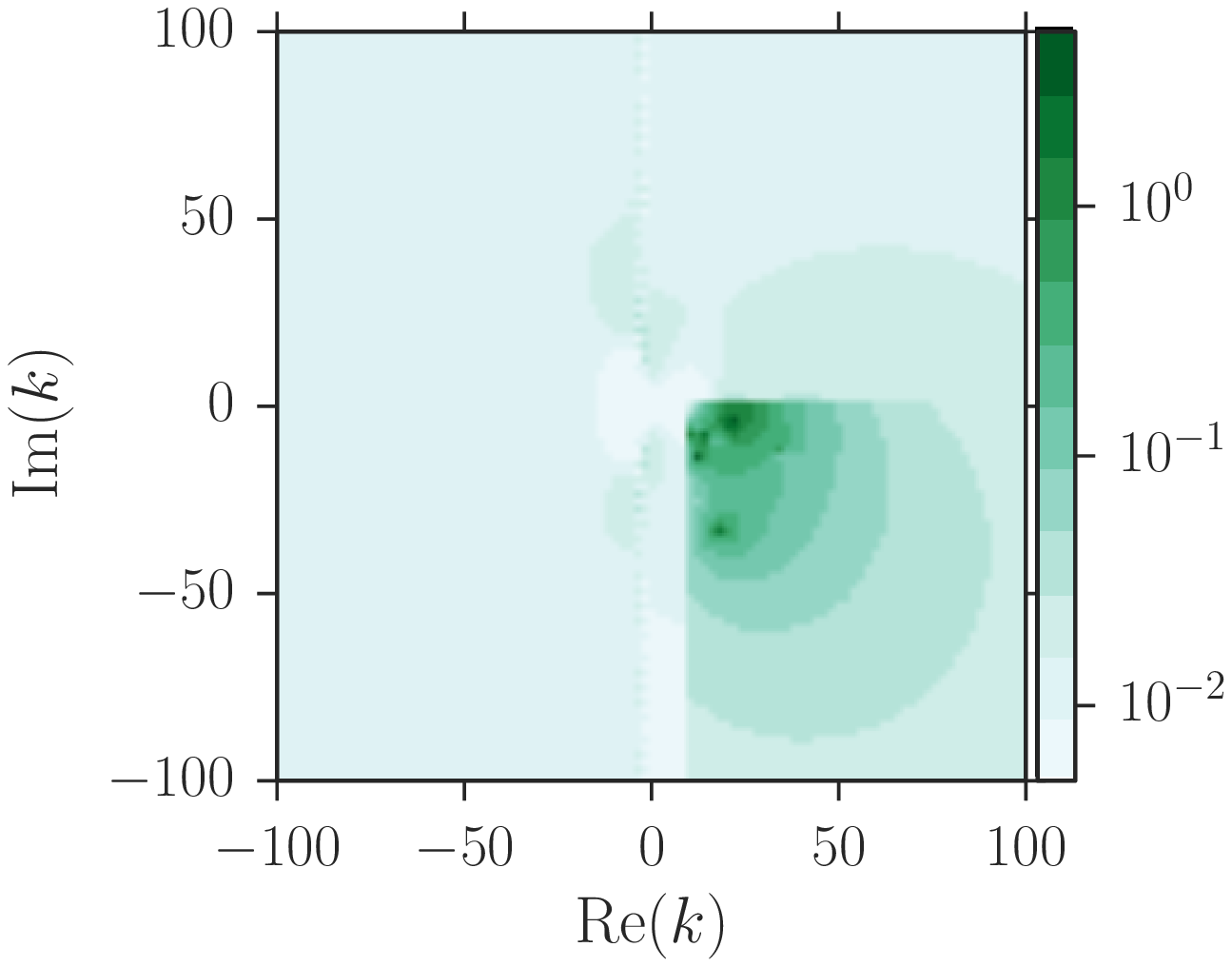}
    \label{fig:lowfreqv1}
    \end{subfigure}%
    ~ 
    \begin{subfigure}[t]{0.33\textwidth}
        \centering
        \caption{}
        \includegraphics*[width=1.\textwidth]{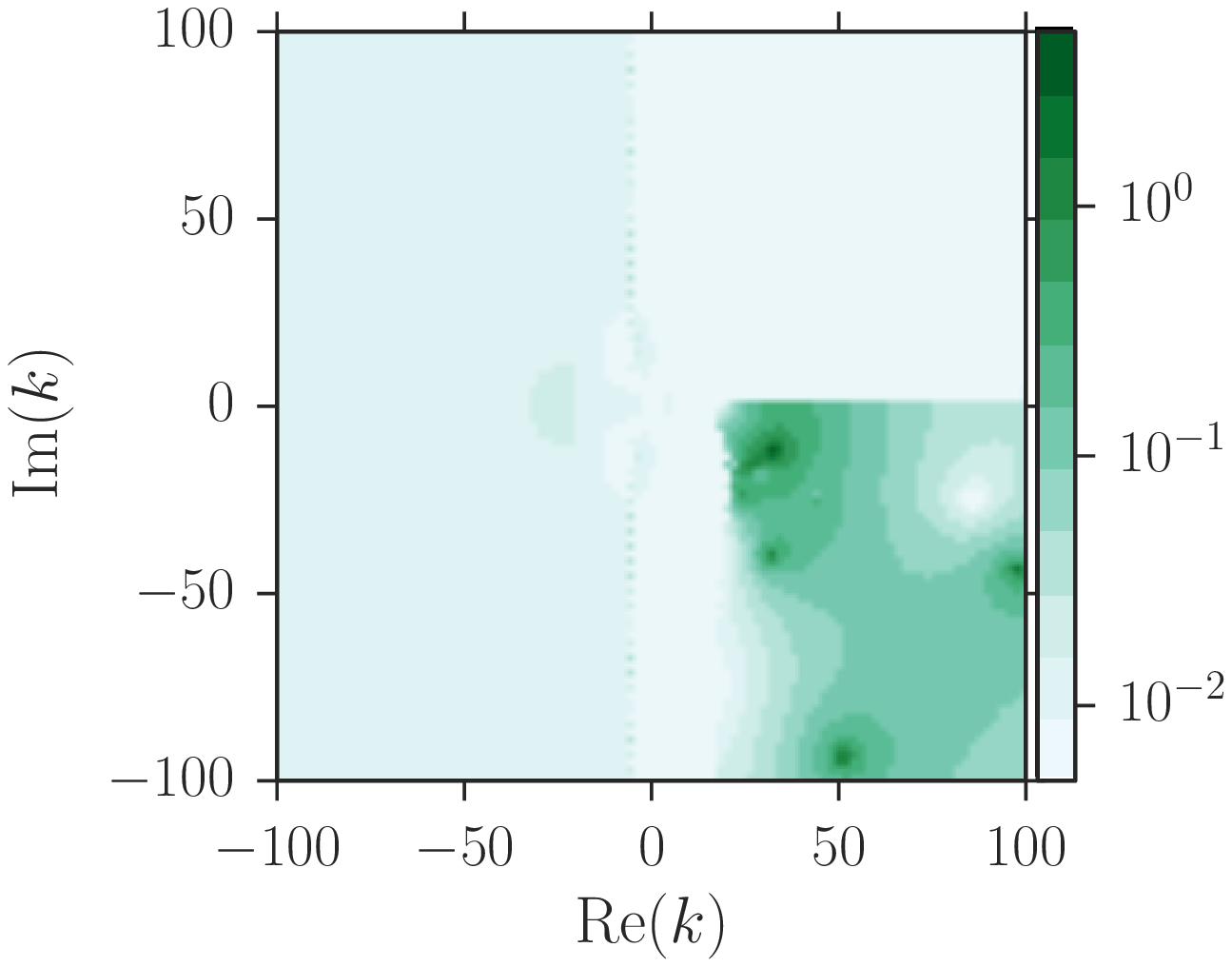}
    \label{fig:midfreqv1}
    \end{subfigure}%
    ~ 
    \begin{subfigure}[t]{0.33\textwidth}
        \centering
        \caption{}
        \includegraphics*[width=1.\textwidth]{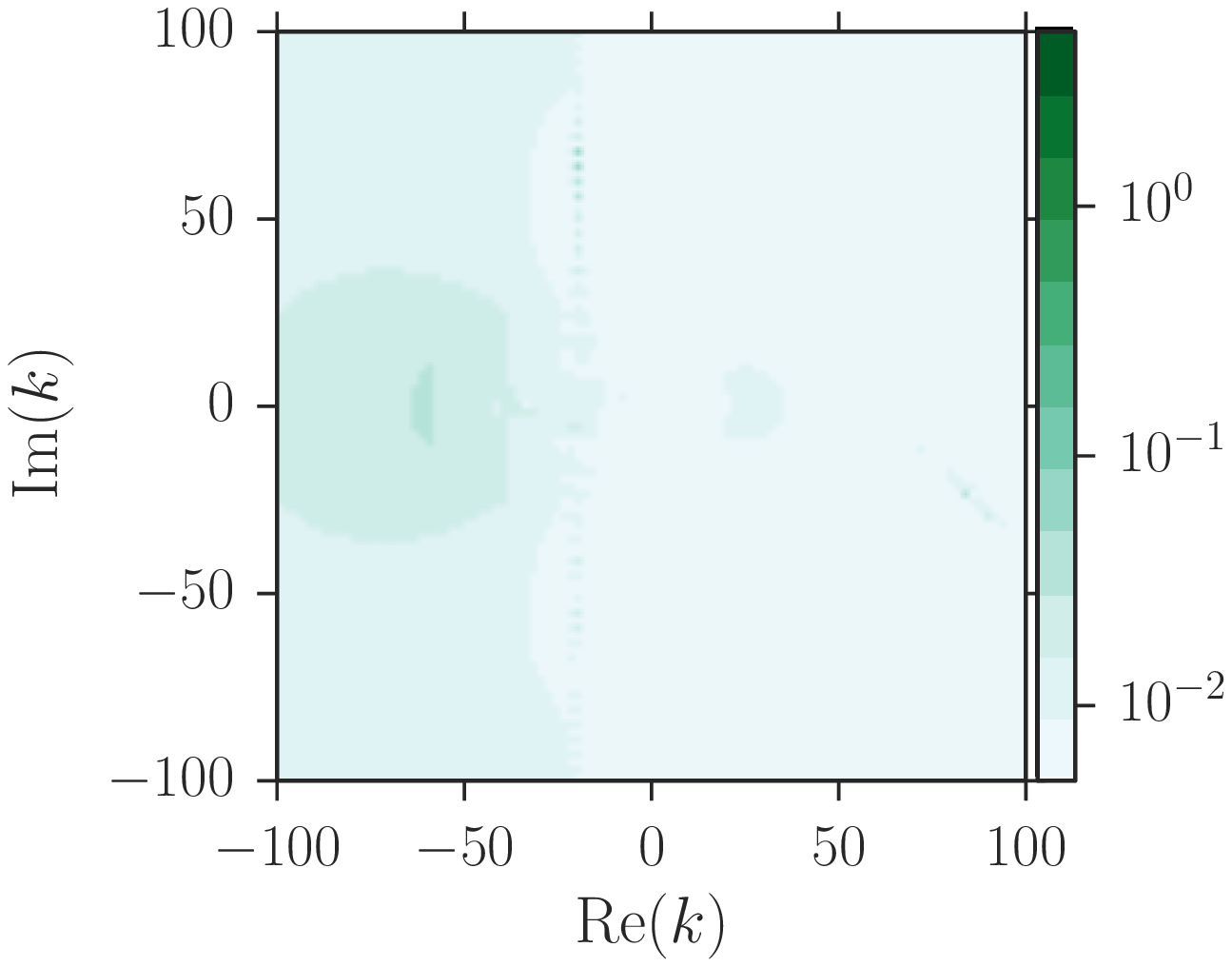}
    \label{fig:highfreqv1}
    \end{subfigure}%

    \begin{subfigure}[t]{0.33\textwidth}
        \centering
        \caption{}
        \includegraphics*[width=1.\textwidth]{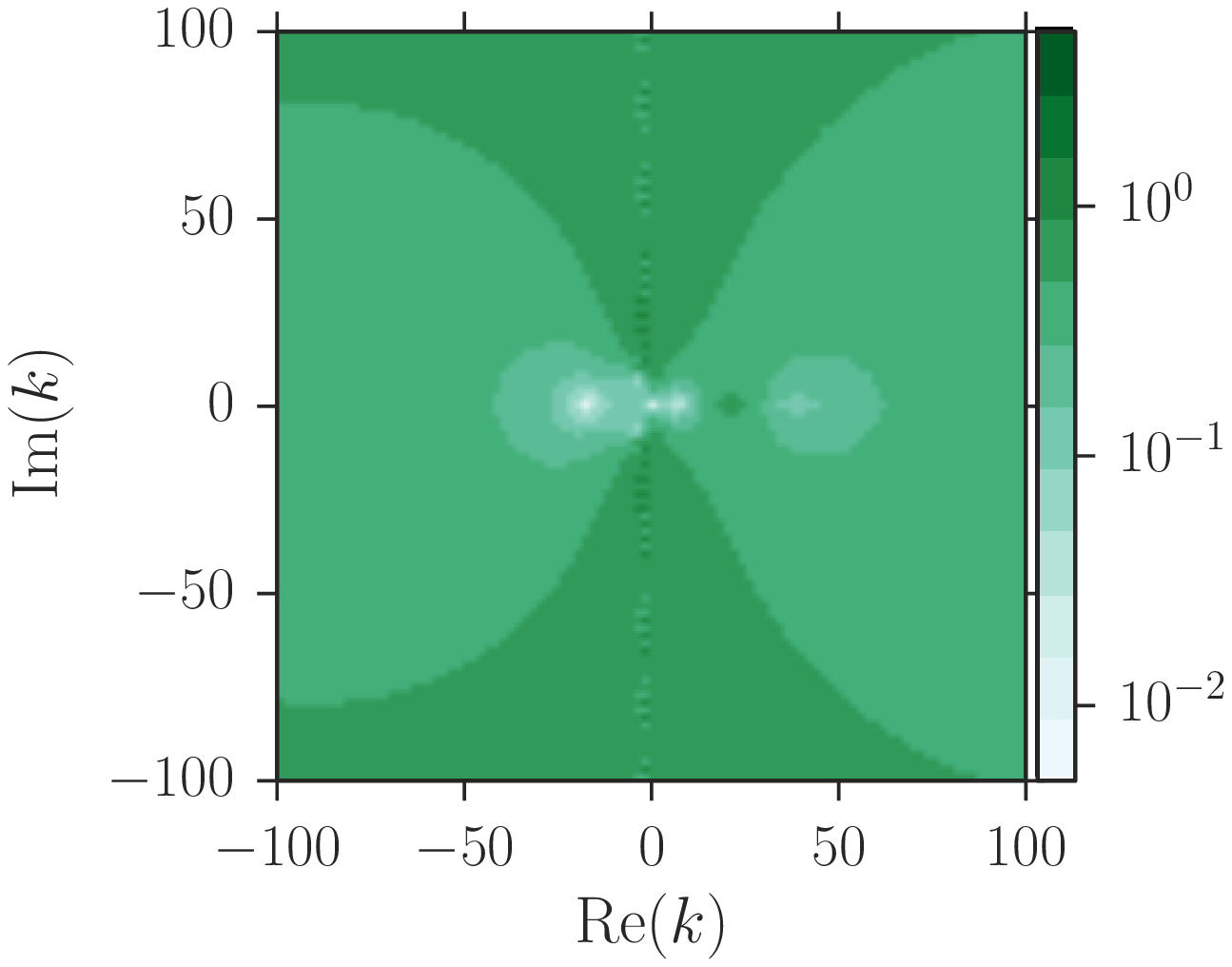}
    \label{fig:lowfreqs1}
    \end{subfigure}%
    ~ 
    \begin{subfigure}[t]{0.33\textwidth}
        \centering
        \caption{}
        \includegraphics*[width=1.\textwidth]{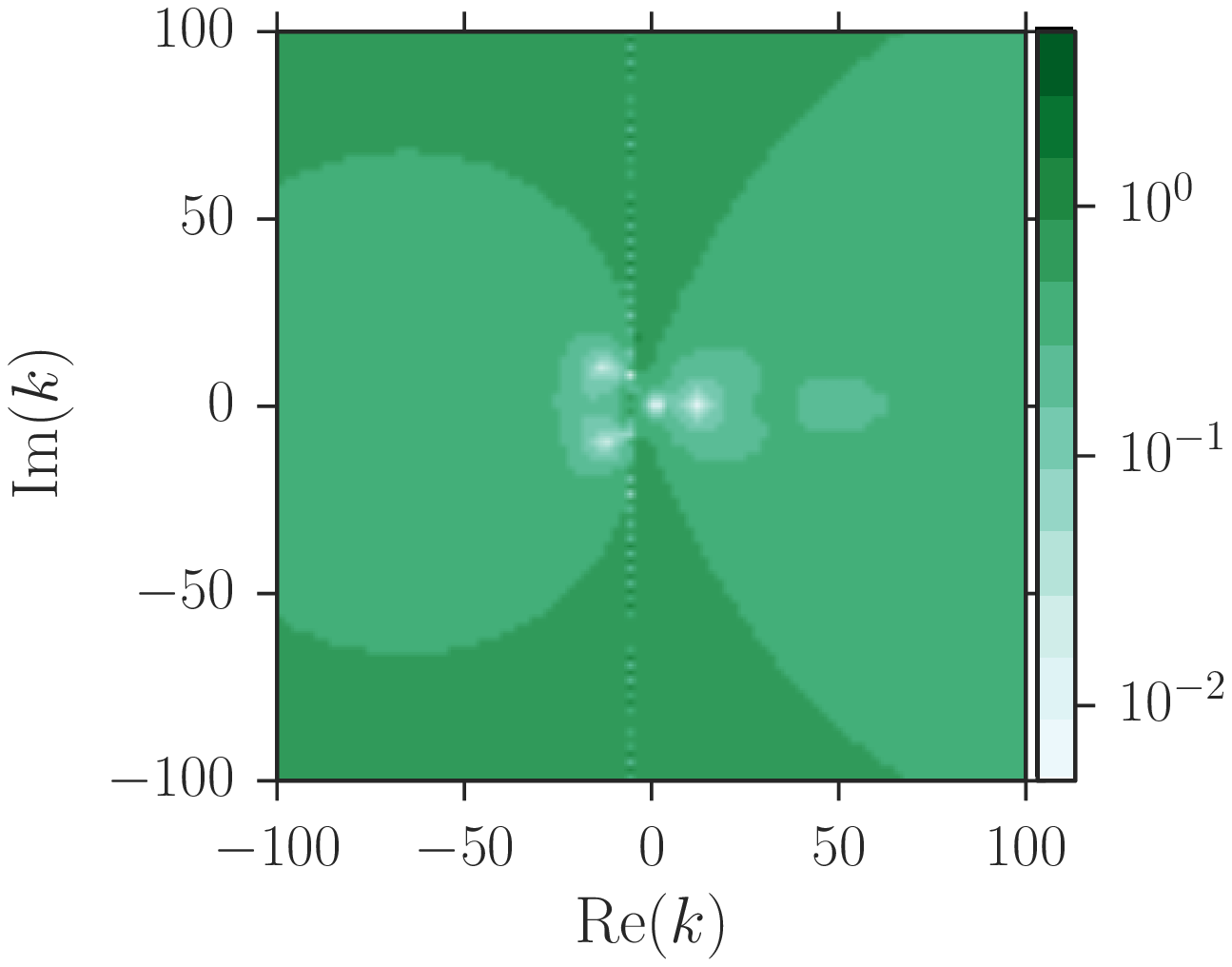}
    \label{fig:midfreqs1}
    \end{subfigure}%
    ~ 
    \begin{subfigure}[t]{0.33\textwidth}
        \centering
        \caption{}
        \includegraphics*[width=1.\textwidth]{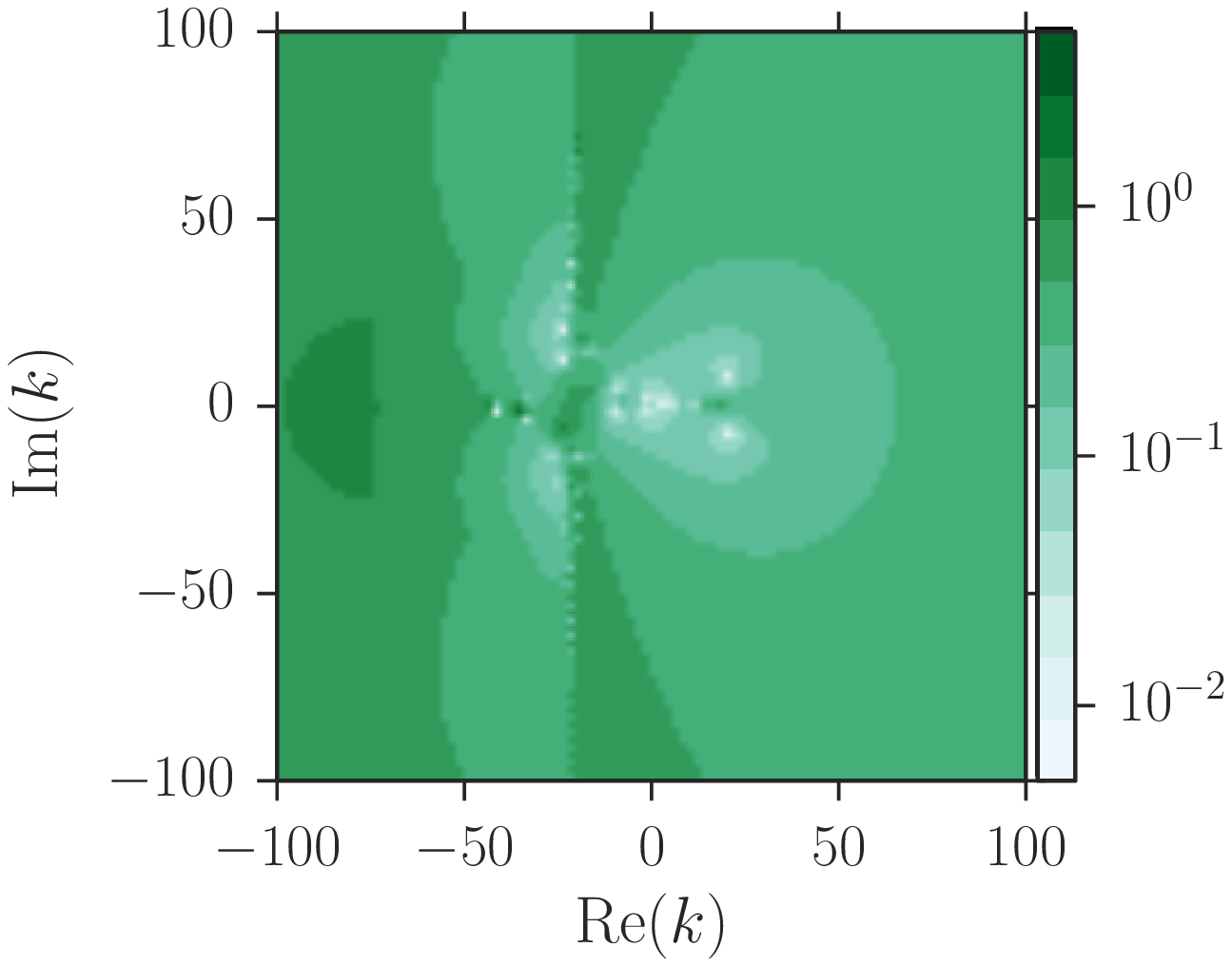}
    \label{fig:highfreqs1}
    \end{subfigure}%
    \caption{(a,b,c) viscous impedance error, comparing the sheared inviscid numerics with the sheared viscous numerics. (d,e,f) shear impedance error, comparing the uniform inviscid numerics with the sheared inviscid numerics. (a,d) $\omega = 4 + 0.04\i$; (b,e) $\omega = 8 + 0.08\i$; (c,f) $\omega = 31 + 0.31\i$. Parameters are $m = 6$, $M = 0.5$, $\Ren = 5\times 10^{5}$, $\delta = 3\times 10^{-2}$. Base profiles as in \cref{baseflow}.} \label{fig:viscshearadmittanceerrs1}
\end{figure}

The impedance errors are plotted in the wavenumber plane in \cref{fig:viscshearadmittanceerrs1} for $M=0.5$, $\delta = 3\times 10^{-2}$ and $\Ren = 5\times 10^{5}$.  For these parameters, for large sectors of the $k$-plane and for all three frequencies, the shear error is an order of magnitude larger than the viscous error.  However, at low frequencies, \cref{fig:lowfreqv1,fig:midfreqv1}, the viscous error in the region $\Re(k)> \omega/M$, $\Im(k) < 0$ is comparable to the shear error in the same region, \cref{fig:lowfreqs1,fig:midfreqs1}.  This region, bounded by the viscous branch cut $k = \omega/M - \i q$ and inviscid branch cut $k = \omega/M + q$ was labelled the ``anomalous region'' by \citet{brambley2011b}, as the behaviour in this region is unlike the rest of the $k$-plane and its origin is unknown.

\begin{figure}
\captionsetup[subfigure]{aboveskip=1pt}
    \centering
    \begin{subfigure}[t]{0.33\textwidth}
        \centering
        \caption{}
        \includegraphics*[width=1.\textwidth]{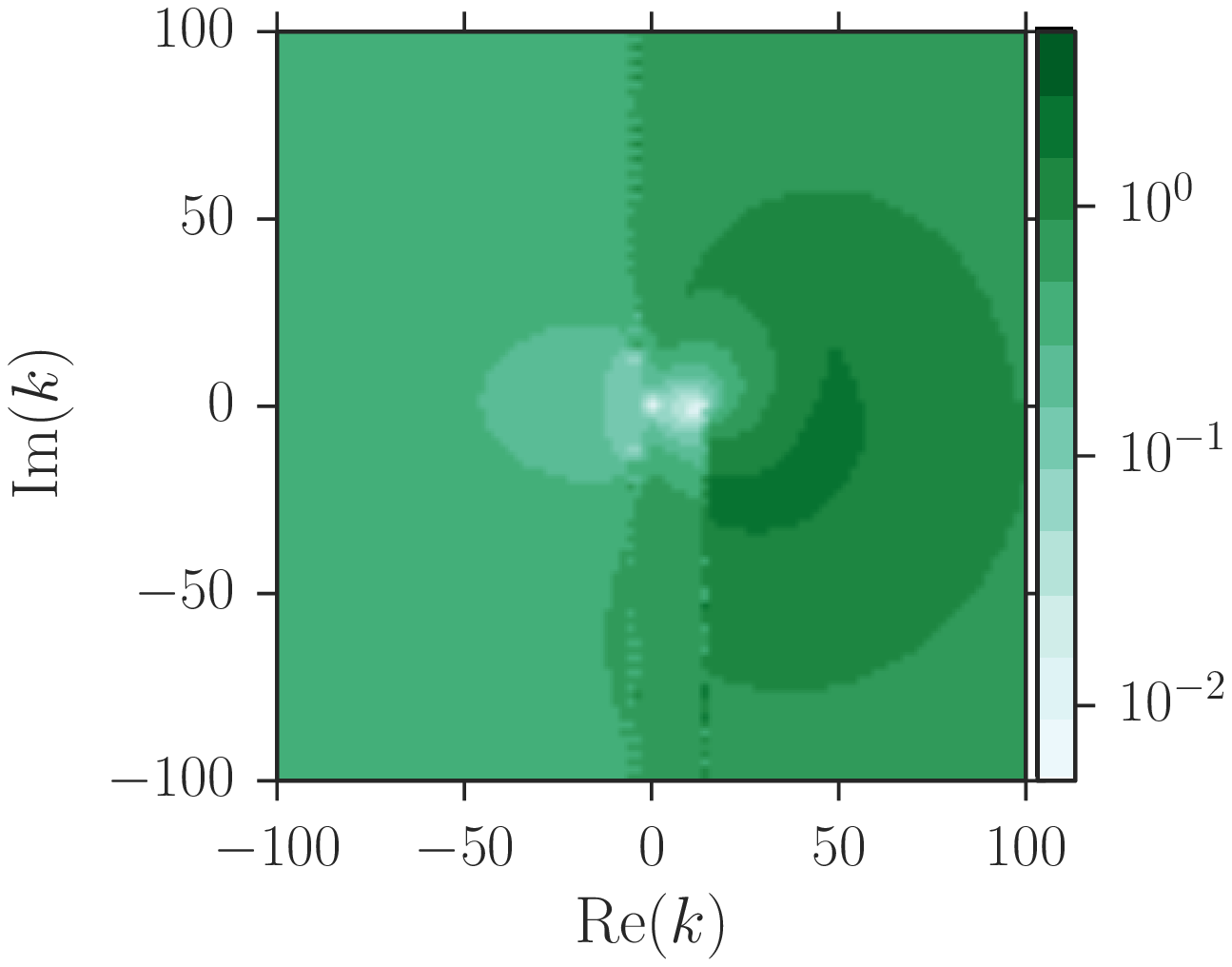}
    \label{fig:lowfreqv2}
    \end{subfigure}%
    ~ 
    \begin{subfigure}[t]{0.33\textwidth}
        \centering
        \caption{}
        \includegraphics*[width=1.\textwidth]{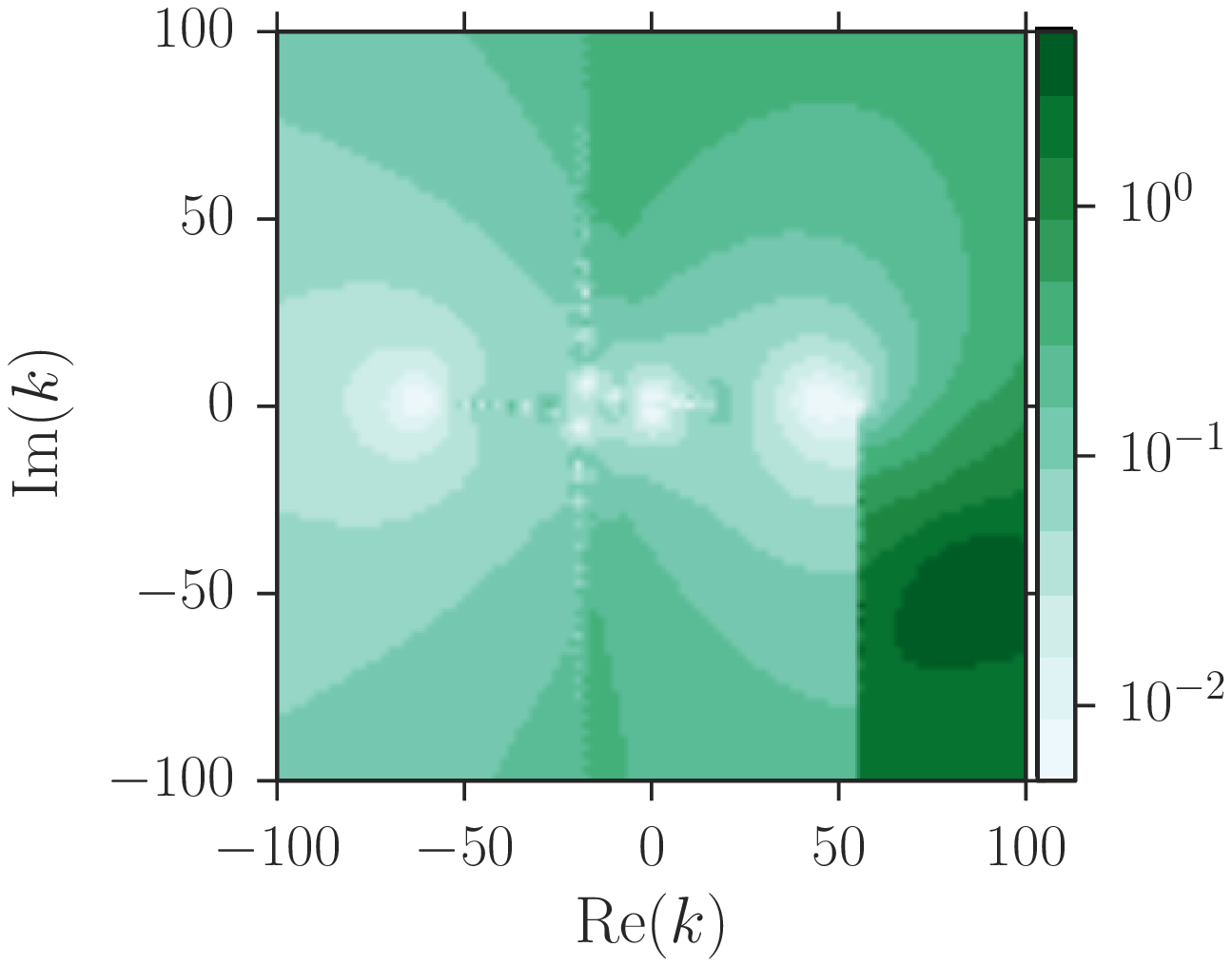}
    \label{fig:midfreqv2}
    \end{subfigure}%
    ~ 
    \begin{subfigure}[t]{0.33\textwidth}
        \centering
        \caption{}
        \includegraphics*[width=1.\textwidth]{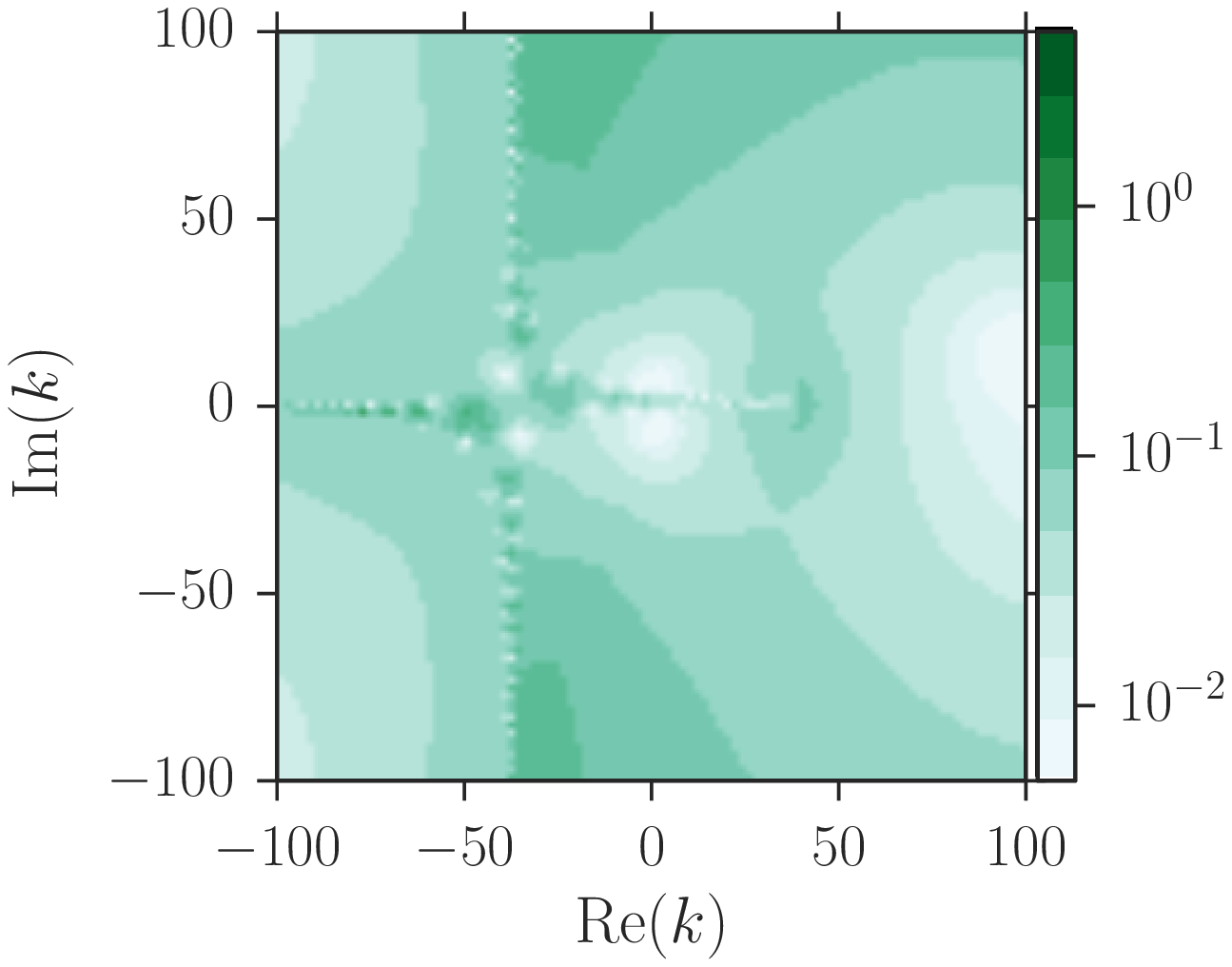}
    \label{fig:highfreqv2}
    \end{subfigure}%

    \begin{subfigure}[t]{0.33\textwidth}
        \centering
        \caption{}
        \includegraphics*[width=1.\textwidth]{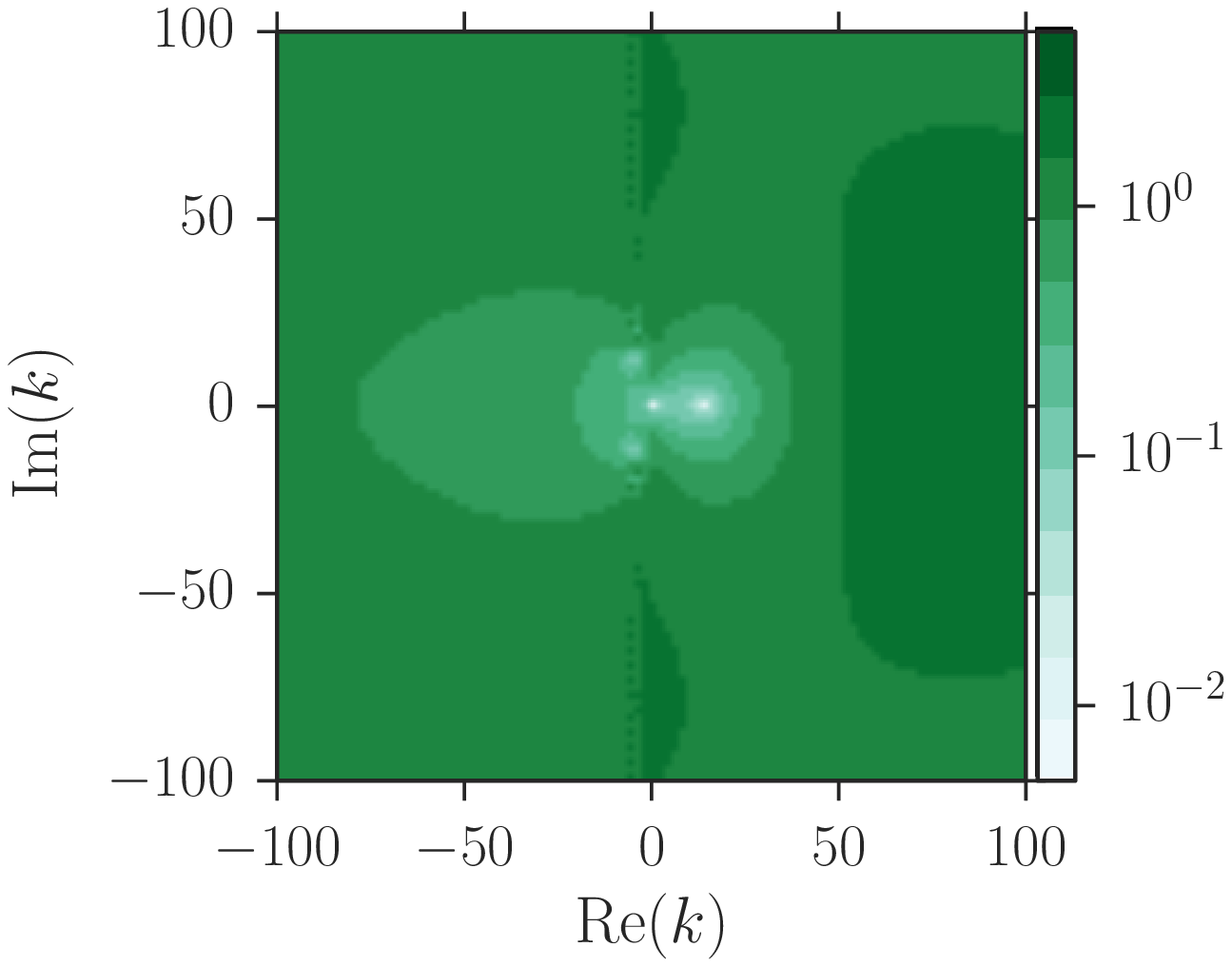}
    \label{fig:lowfreqs2}
    \end{subfigure}%
    ~ 
    \begin{subfigure}[t]{0.33\textwidth}
        \centering
        \caption{}
        \includegraphics*[width=1.\textwidth]{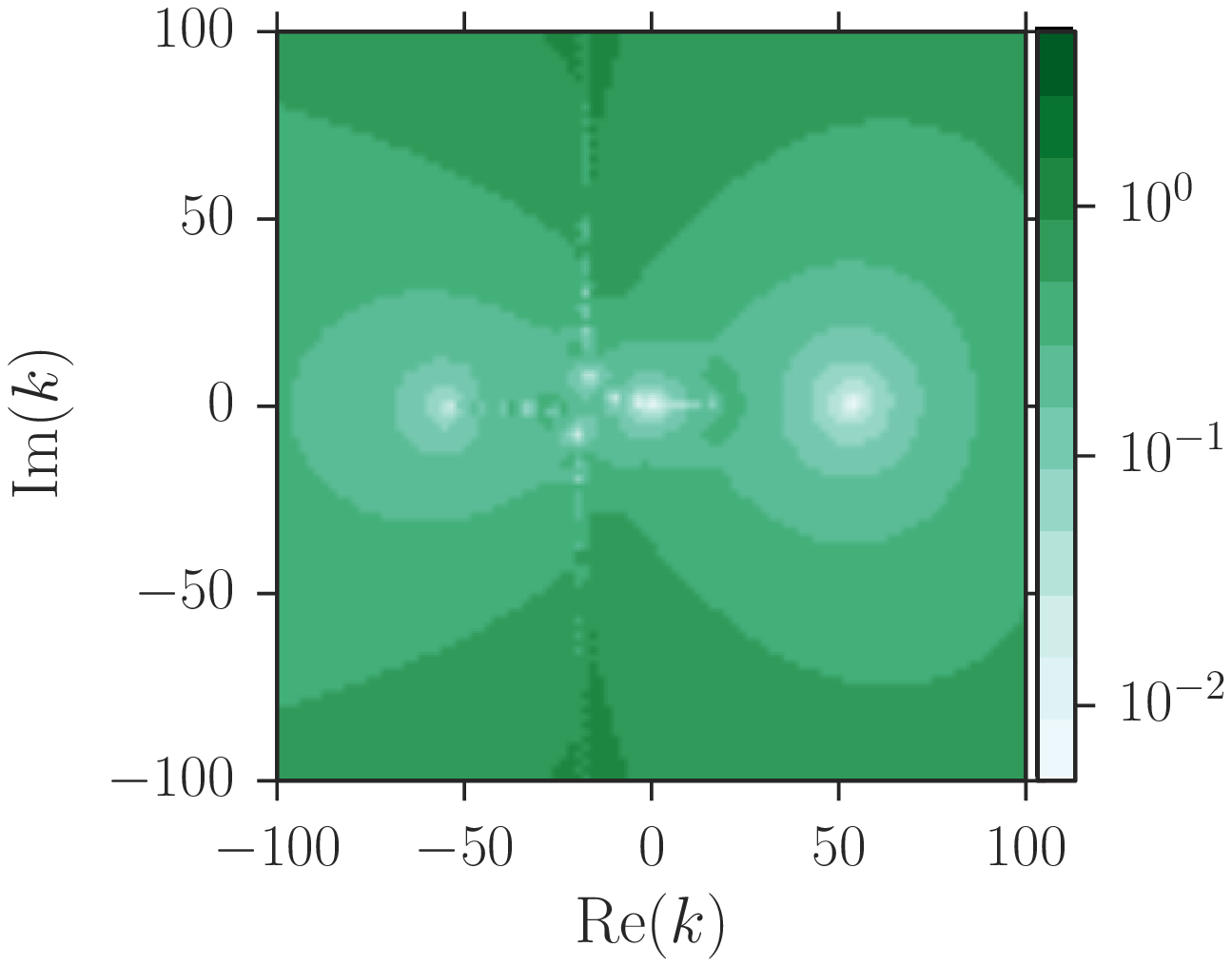}
    \label{fig:midfreqs2}
    \end{subfigure}%
    ~ 
    \begin{subfigure}[t]{0.33\textwidth}
        \centering
        \caption{}
        \includegraphics*[width=1.\textwidth]{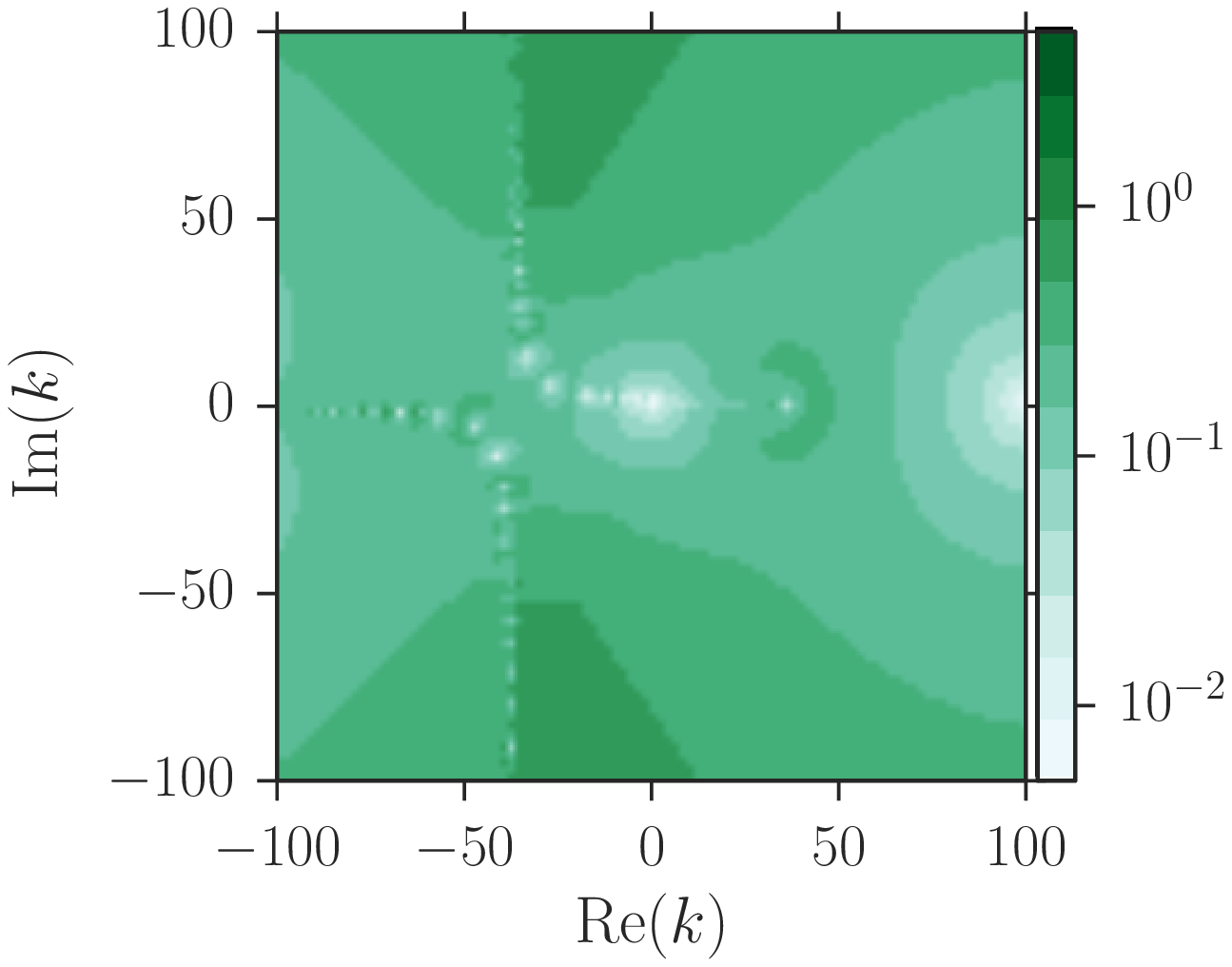}
    \label{fig:highfreqs2}
    \end{subfigure}%
    \caption{(a,b,c) viscous impedance error. (d,e,f) shear impedance error. (a,d) $\omega = 7+0.07\i$; (b,e) $\omega = 28+0.28\i$; (c,f) $\omega = 56+0.56\i$. Parameters are $m = 12$, $M = 0.5$, $\Ren = 1\times 10^{5}$, $\delta = 2\times 10^{-3}$. Base profiles as in \cref{baseflow}.} \label{fig:viscshearadmittanceerrs2}
\end{figure}

\Cref{fig:viscshearadmittanceerrs2} shows, for a thinner boundary layer $\delta = 2\times 10^{-3}$, the viscous errors become more prevalent than before, as can be seen by comparing \cref{fig:lowfreqv2} with \cref{fig:lowfreqs2}. At lower frequencies, the viscous error is significant throughout the wavenumber plane, not just in the anomalous region. \Cref{fig:lowfreqv2} shows large regions of the $k$-plane suffer from $\mathcal{O}(1)$ viscous errors, with a median error (calculated using all values in the plotted domain) of $0.5$ to one decimal place (compared to the median shear error of $1.3$ in \cref{fig:lowfreqs2}).  As the frequency increases, the viscous error becomes smaller, with the median error reducing to $0.05$ in \cref{fig:highfreqv2} (compared to the median shear error of $0.27$ in \cref{fig:highfreqs2}).  This could be attributed to the dependence of the acoustic boundary layer thickness $\delta_{\mathrm{ac}}$ on frequency: a larger ratio $\delta_{\mathrm{ac}}/\delta$ occurs at lower frequencies, hence viscous effects are stronger.

\begin{figure}
\captionsetup[subfigure]{aboveskip=1pt}
    \centering
    \begin{subfigure}[t]{0.33\textwidth}
        \centering
        \caption{}
        \includegraphics*[width=1.\textwidth]{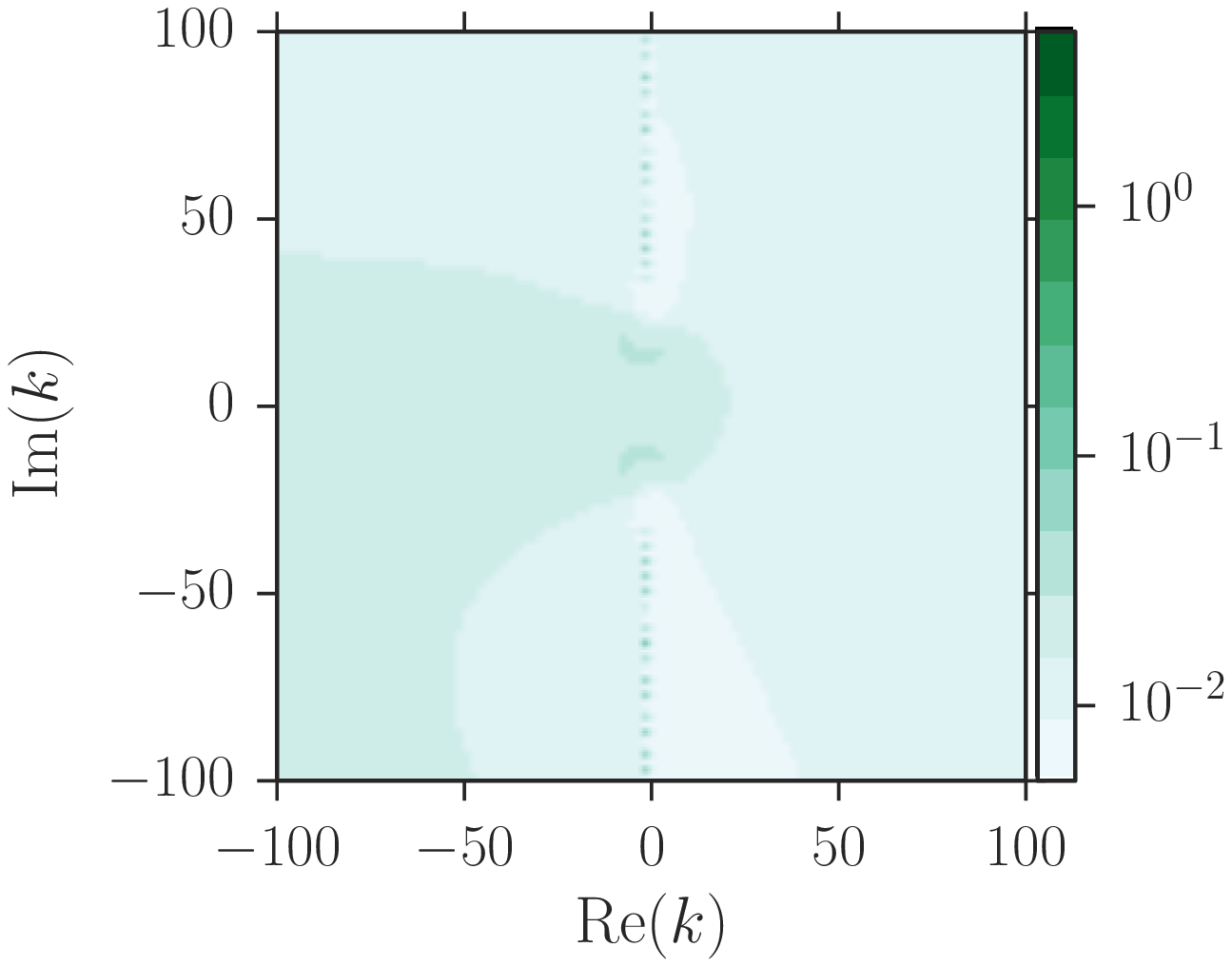}
    \label{fig:vM1}
    \end{subfigure}%
    ~ 
    \begin{subfigure}[t]{0.33\textwidth}
        \centering
        \caption{}
        \includegraphics*[width=1.\textwidth]{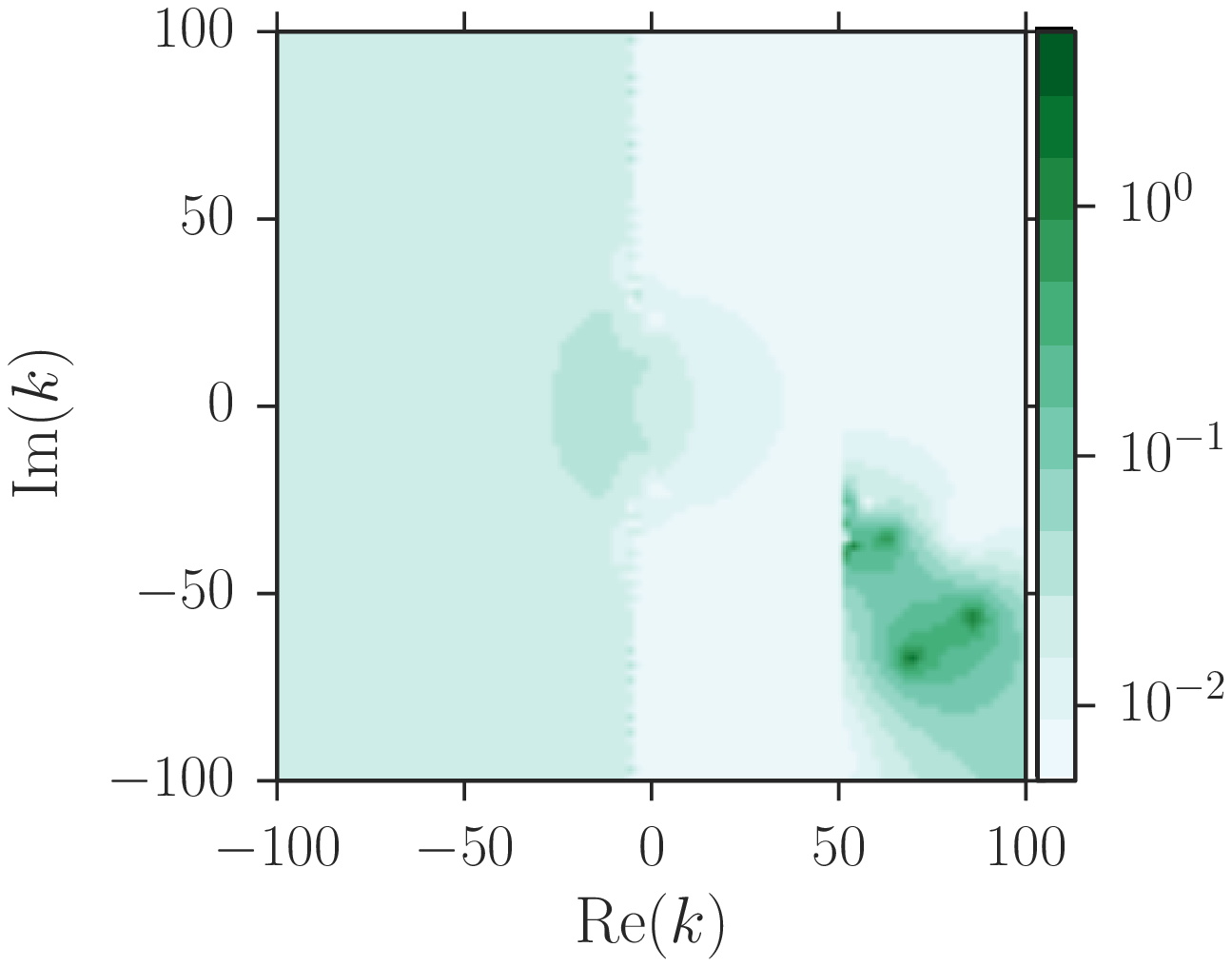}
    \label{fig:vM3}
    \end{subfigure}%
    ~
    \begin{subfigure}[t]{0.33\textwidth}
        \centering
        \caption{}
        \includegraphics*[width=1.\textwidth]{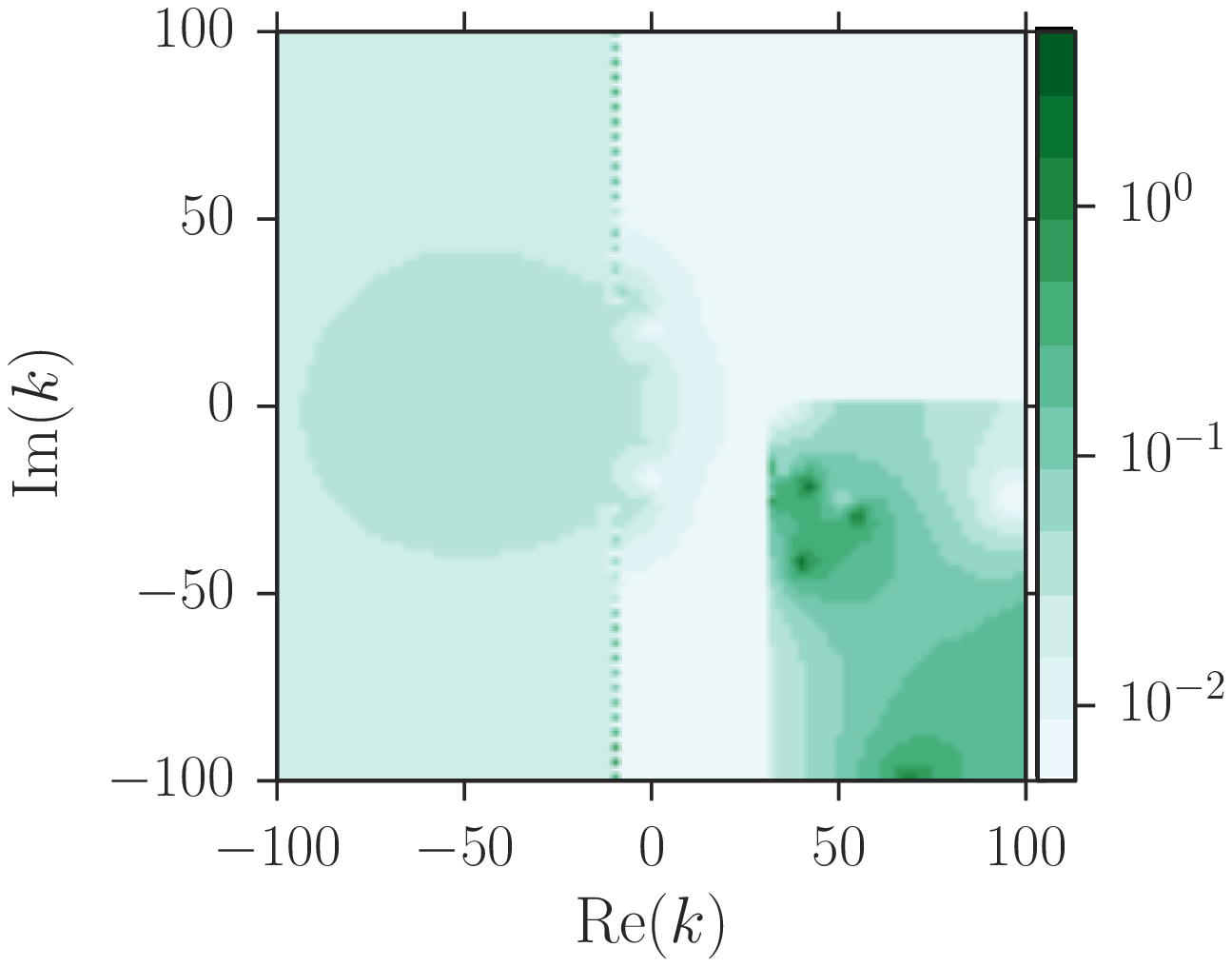}
    \label{fig:vM5}
    \end{subfigure}%

    \begin{subfigure}[t]{0.33\textwidth}
        \centering
        \caption{}
        \includegraphics*[width=1.\textwidth]{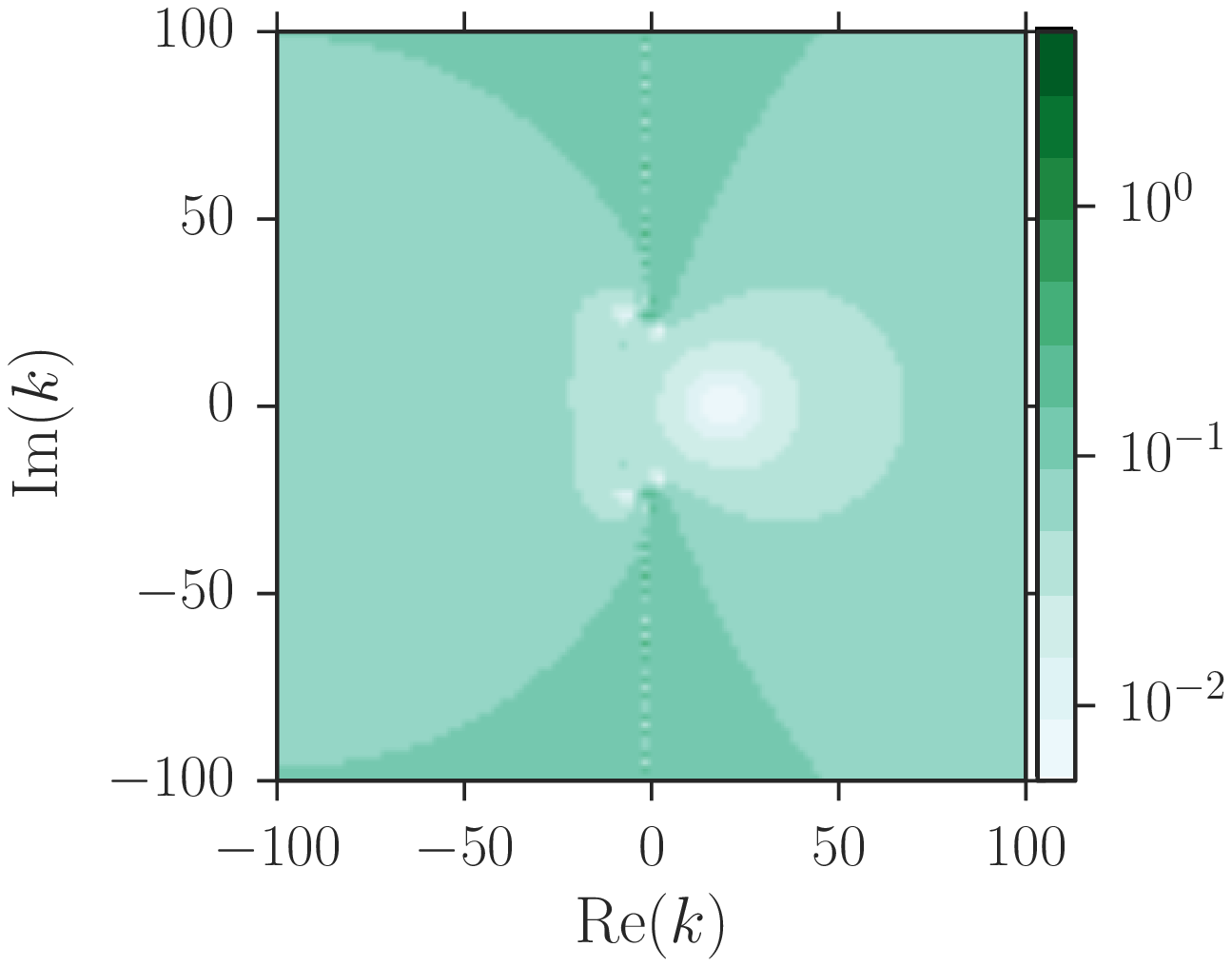}
    \label{fig:sM1}
    \end{subfigure}%
    ~ 
    \begin{subfigure}[t]{0.33\textwidth}
        \centering
        \caption{}
        \includegraphics*[width=1.\textwidth]{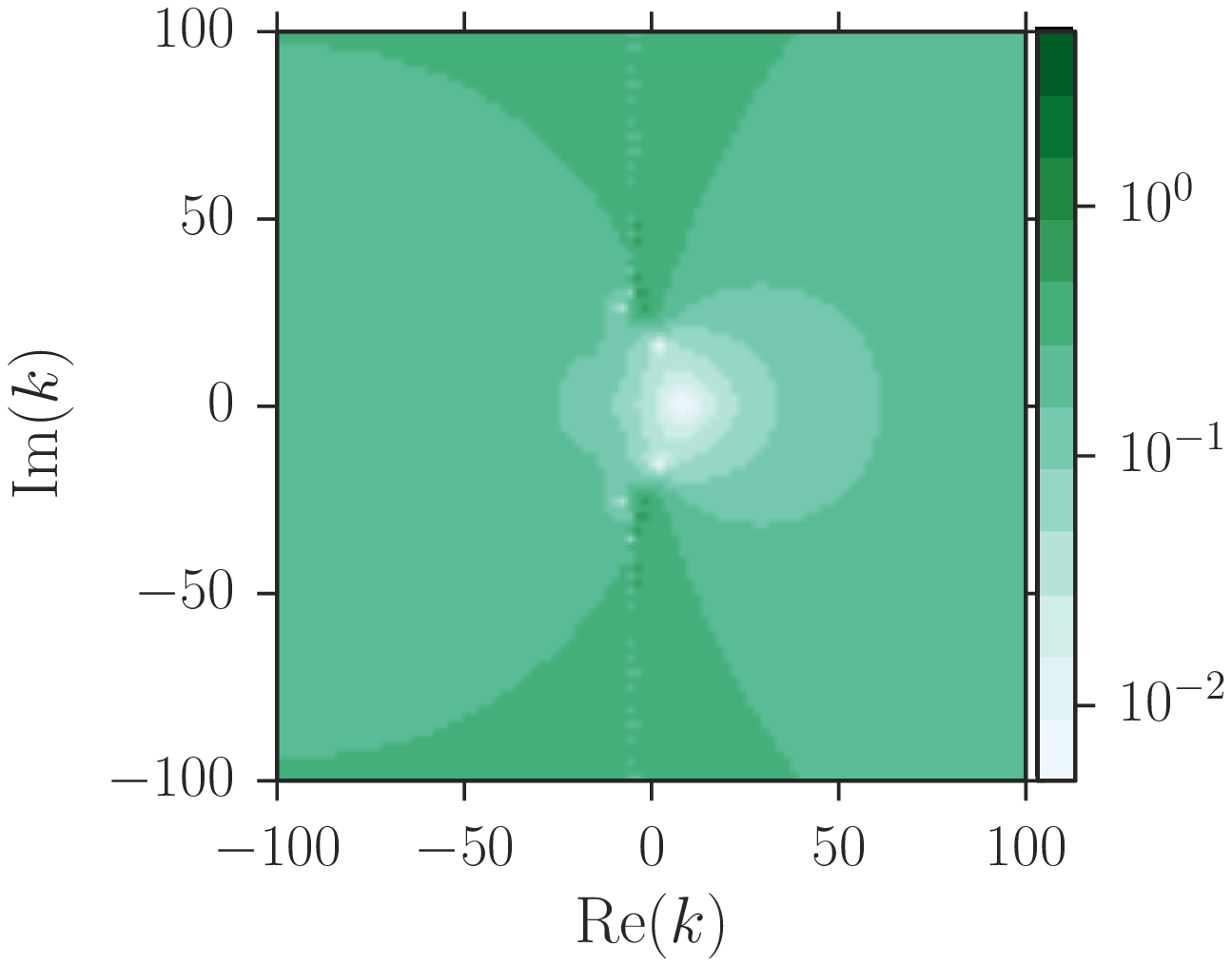}
    \label{fig:sM3}
    \end{subfigure}%
    ~
    \begin{subfigure}[t]{0.33\textwidth}
        \centering
        \caption{}
        \includegraphics*[width=1.\textwidth]{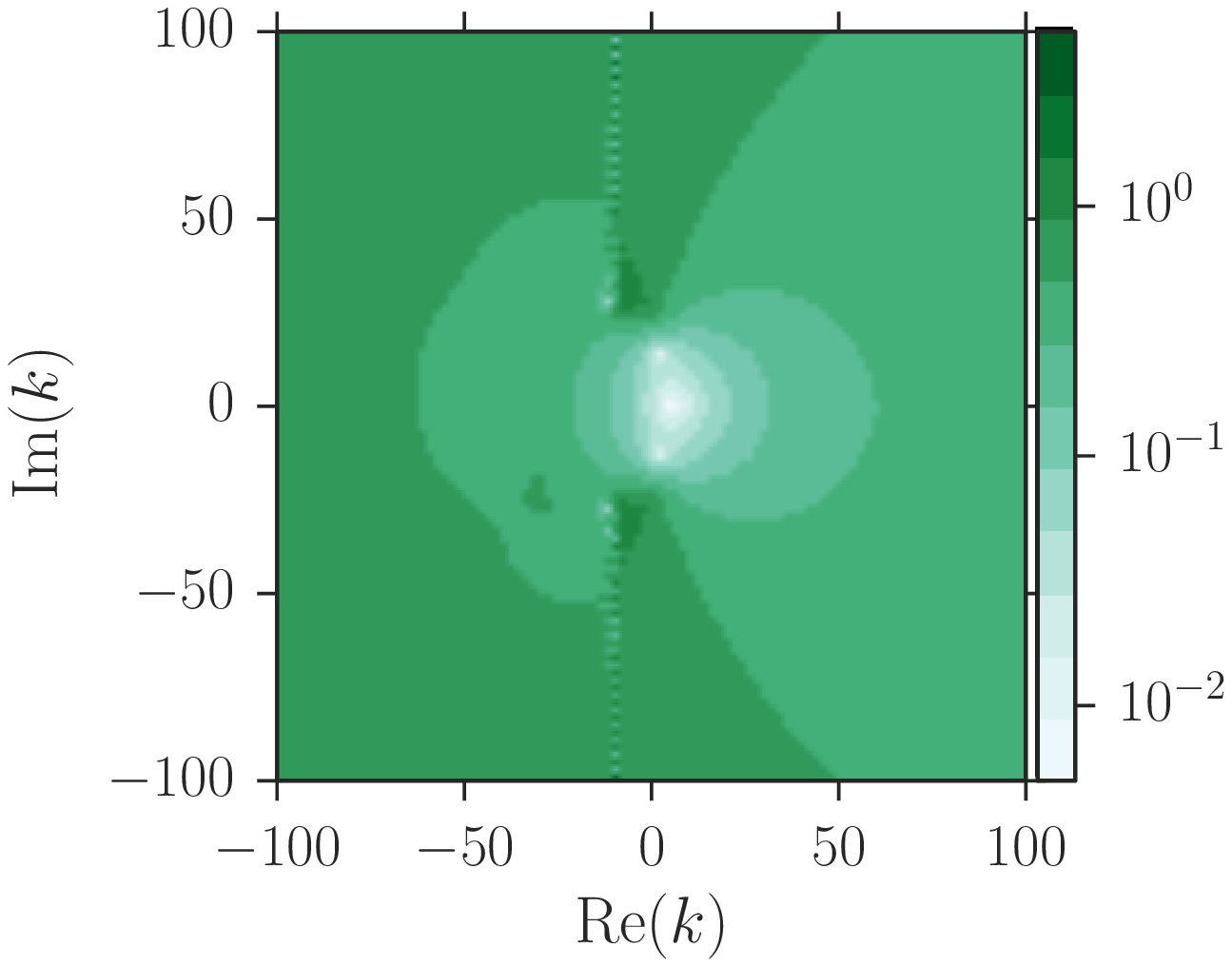}
    \label{fig:sM5}
    \end{subfigure}%
    \caption{(a,b,c) viscous impedance error. (d,e,f) shear impedance error. (a,d) $M = 0.1$; (b,e) $M = 0.3$; (c,f) $M = 0.5$. Parameters are $\omega = 15+0.15\i$, $m = 24$, $\Ren = 1\times 10^{5}$, $\delta = 3\times 10^{-2}$. Base profiles as in \cref{baseflow}.} \label{fig:viscshearadmittanceerrs3}
\end{figure}

\begin{figure}
\captionsetup[subfigure]{aboveskip=1pt}
    \centering
    \begin{subfigure}[t]{0.33\textwidth}
        \centering
        \caption{}
        \includegraphics*[width=1.\textwidth]{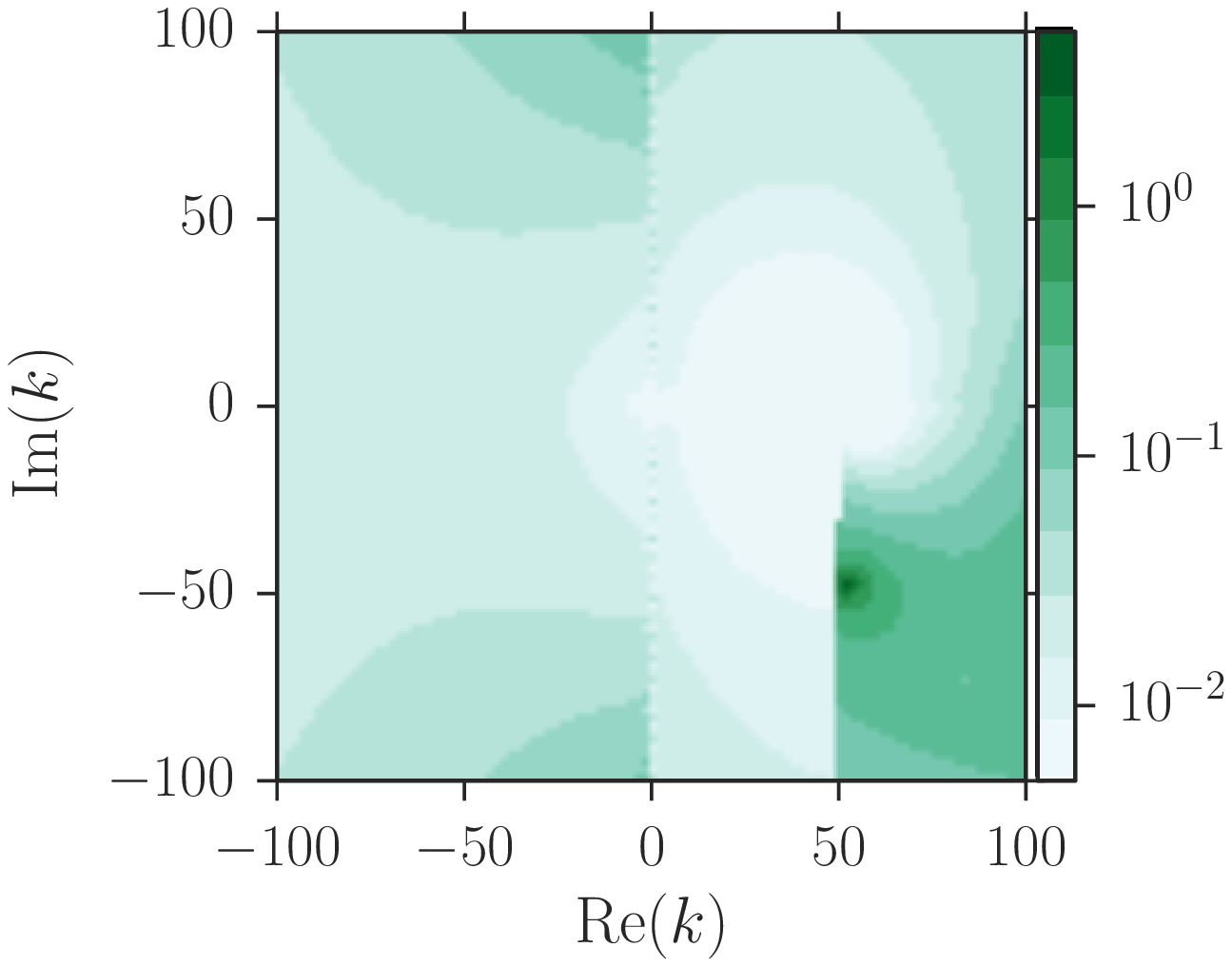}
    \label{fig:vM12}
    \end{subfigure}%
    ~ 
    \begin{subfigure}[t]{0.33\textwidth}
        \centering
        \caption{}
        \includegraphics*[width=1.\textwidth]{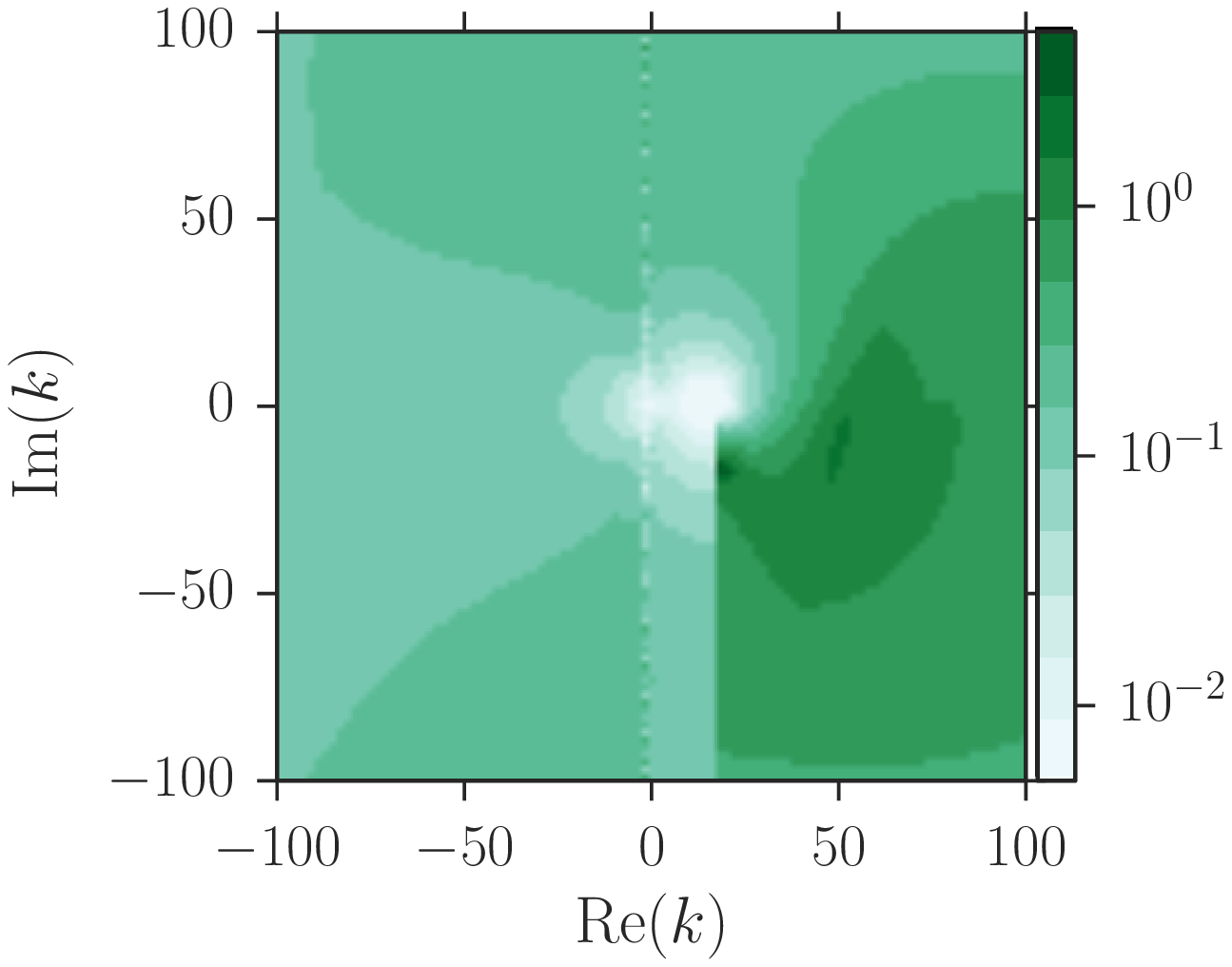}
    \label{fig:vM32}
    \end{subfigure}%
    ~
    \begin{subfigure}[t]{0.33\textwidth}
        \centering
        \caption{}
        \includegraphics*[width=1.\textwidth]{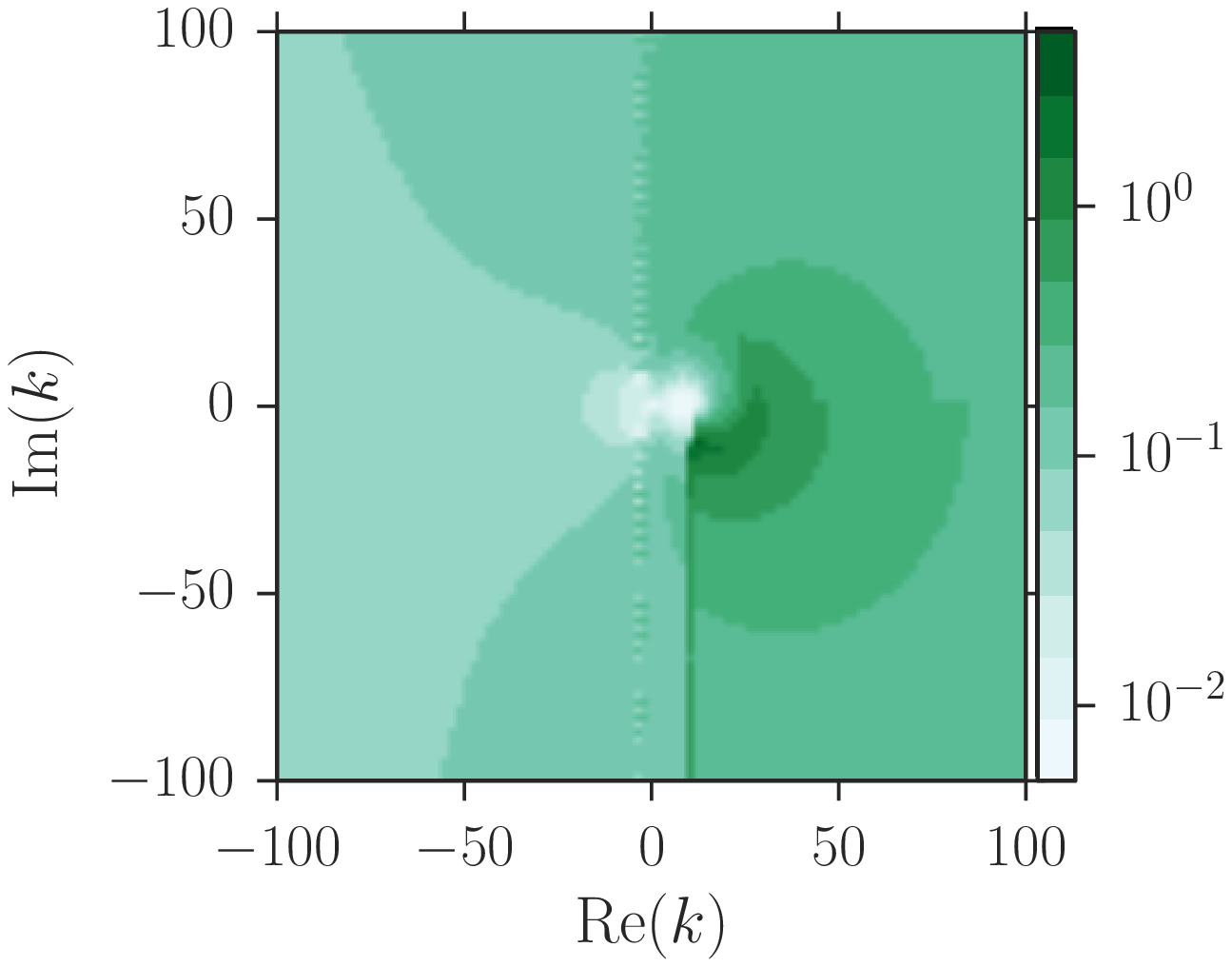}
    \label{fig:vM52}
    \end{subfigure}%

    \begin{subfigure}[t]{0.33\textwidth}
        \centering
        \caption{}
        \includegraphics*[width=1.\textwidth]{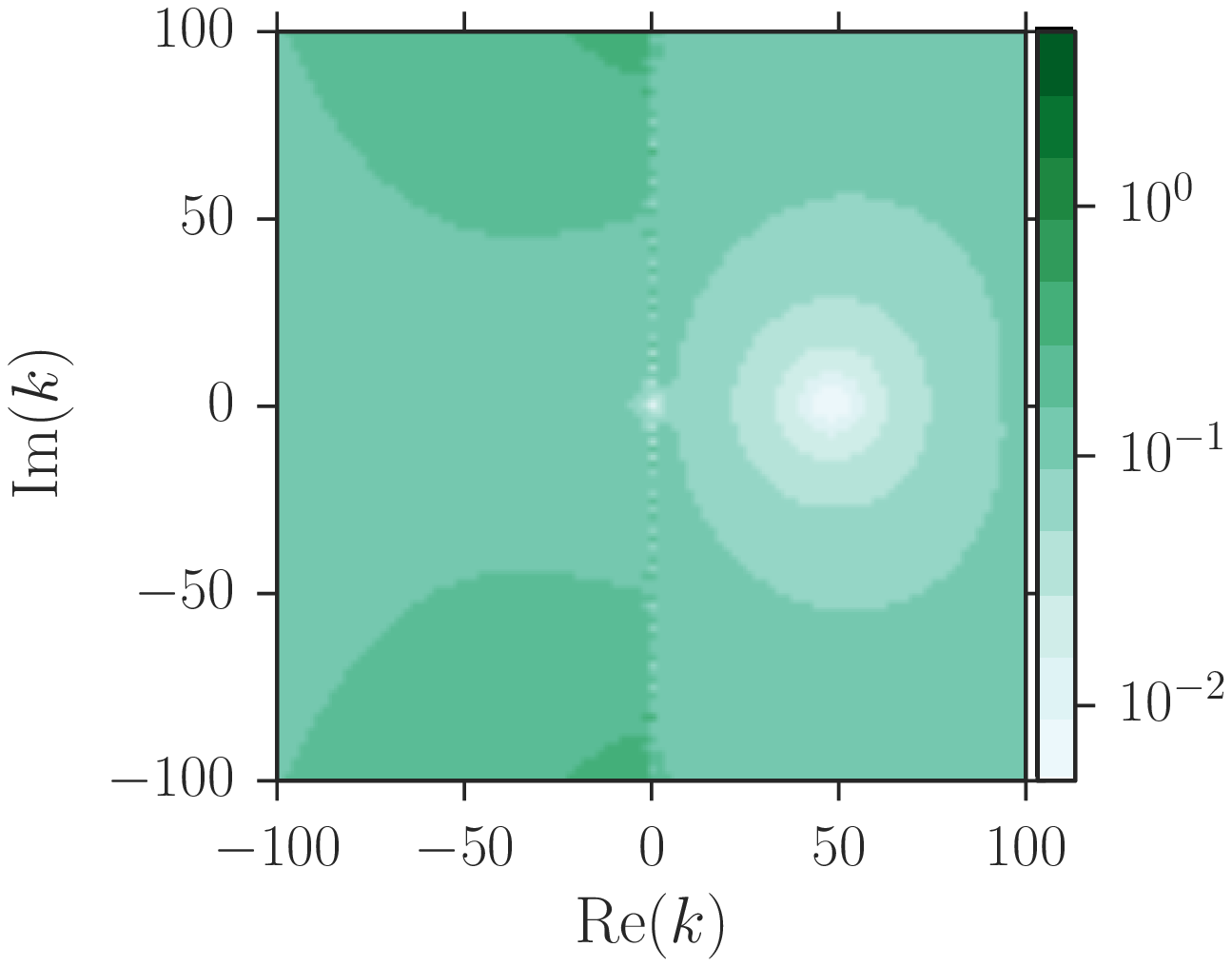}
    \label{fig:sM12}
    \end{subfigure}%
    ~ 
    \begin{subfigure}[t]{0.33\textwidth}
        \centering
        \caption{}
        \includegraphics*[width=1.\textwidth]{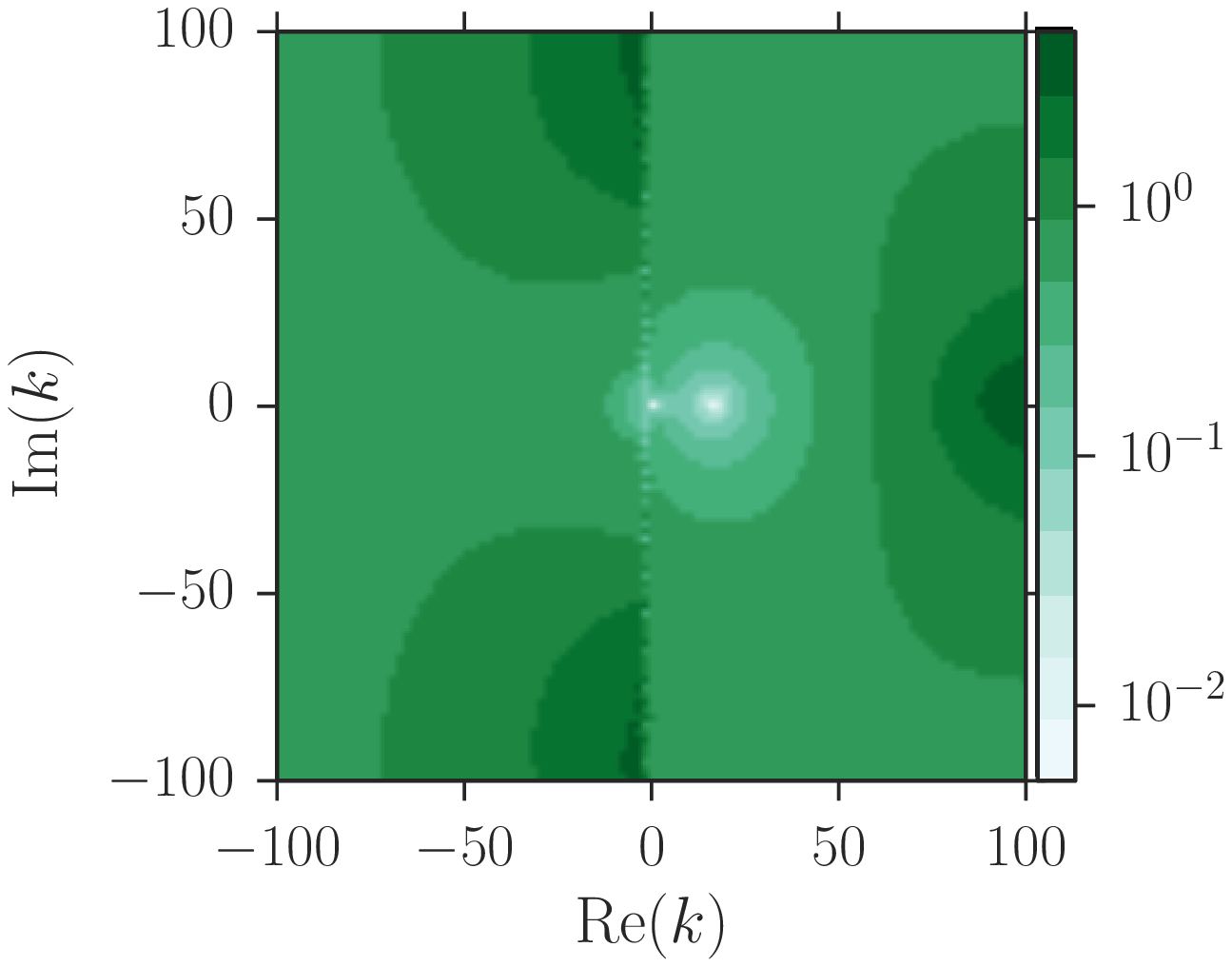}
    \label{fig:sM32}
    \end{subfigure}%
    ~
    \begin{subfigure}[t]{0.33\textwidth}
        \centering
        \caption{}
        \includegraphics*[width=1.\textwidth]{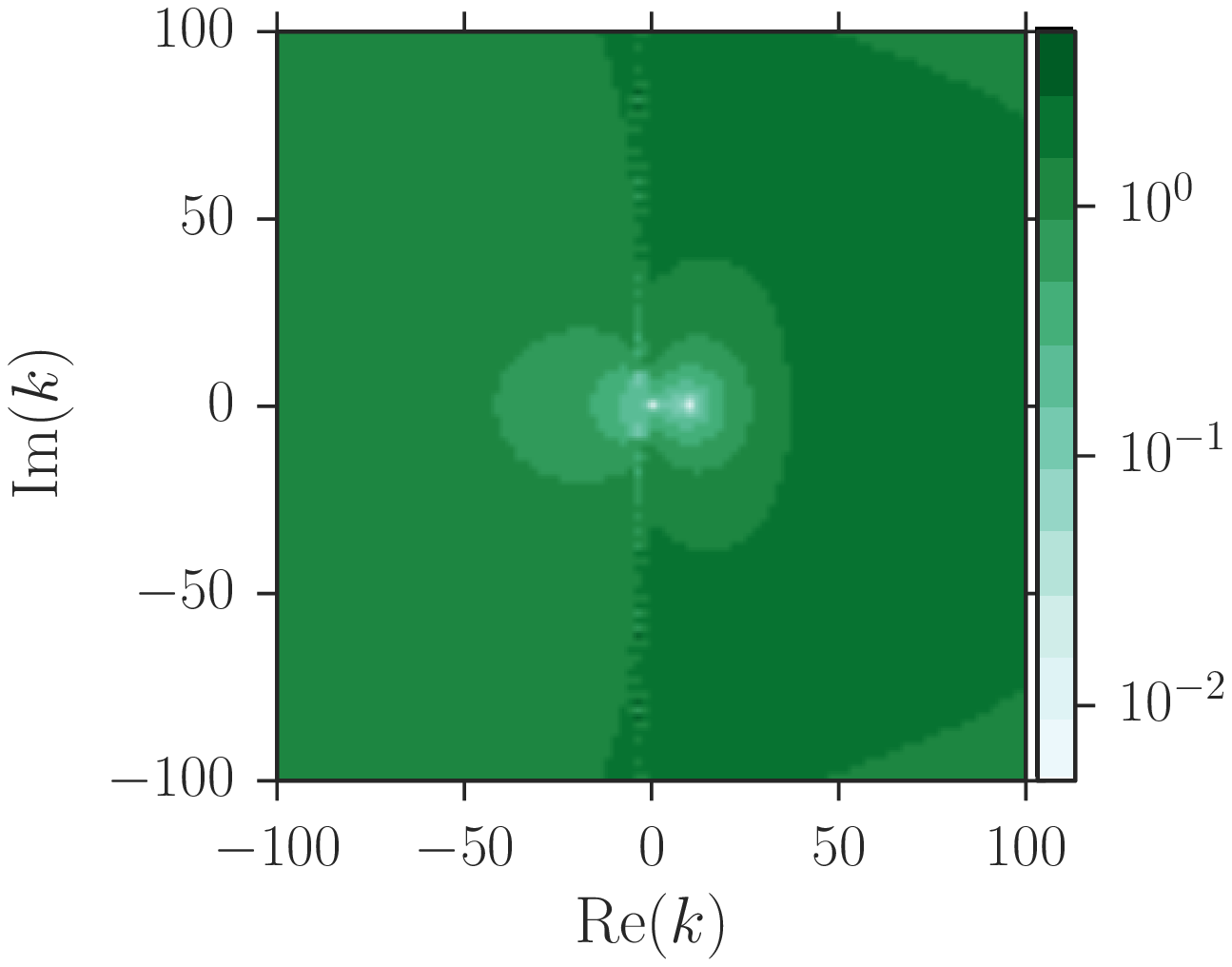}
    \label{fig:sM52}
    \end{subfigure}%
    \caption{(a,b,c) viscous impedance error. (d,e,f) shear impedance error. (a,d) $M = 0.1$; (b,e) $M = 0.3$; (c,f) $M = 0.5$. Parameters are $\omega = 5+0.05\i$, $m = 7$, $\Ren = 5\times 10^{6}$, $\delta = 2\times 10^{-3}$. Base profiles as in \cref{baseflow}.} \label{fig:viscshearadmittanceerrs4}
\end{figure}

\Cref{fig:viscshearadmittanceerrs3,fig:viscshearadmittanceerrs4} show the effect of increasing Mach number on the error, with the Reynolds number held fixed%
\footnote{Recall that $\Ren$ is defined here with respect to the sound speed, $\Ren = c_0^*l^*\!\rho_0^*/\mu_0^*$, rather than with respect to the flow speed, $\overline{\Ren} = U_0^*l^*\!\rho_0^*/\mu_0^* = M\Ren$.  Hence, \cref{fig:viscshearadmittanceerrs3,fig:viscshearadmittanceerrs4} show that, for a given fluid with fixed $\mu_0^*$ and $c_0^*$, increasing the flow speed $U_0^*$ leads to increasing $\overline{\Ren}$ but, confusingly, larger viscous error, justifying our previous choice of  $\Ren$ as the Reynolds number.}%
.  As the Mach number increases, so too does the shear error, as might be anticipated since a higher Mach number means neglecting larger velocity gradients in the uniform flow case.  However, it is seen (particularly in \cref{fig:vM12,fig:vM32,fig:vM52}) that increasing the Mach number leads to larger viscous errors throughout the $k$-plane and not just inside the anomalous region, with the median error increasing from $0.02$ in \cref{fig:vM12} to $0.14$ in \cref{fig:vM52}. 

\subsection{Accuracy of modes in the $k$-plane} \label{sec:kmodes}

As described in the introduction, the Ingard--Myers~\citep{ingard1959,myers1980} (or Myers) boundary condition corresponds to the limit of a sheared inviscid boundary layer with a vanishing thickness~\citep{eversman&beckemeyer1972,tester1973a}, and for the situation considered here may be written as
\begin{align}
\omega\t{v}(1) &= (\omega - Mk)\t{p}(1)/Z &
&\Rightarrow&
Z_{\mathrm{eff}} &= \frac{\omega}{\omega - Mk}Z, \label{Zeffmyers}
\end{align}
where $Z_\mathrm{eff}$ is the effective impedance for which the Myers boundary condition is $\t{p}(1)/\t{v}(1) = Z_\mathrm{eff}$.  We use the Myers condition here to find $k$-plane modes under the uniform inviscid assumption.  For the sheared viscous and sheared inviscid results, the numerics of \cref{sec:numericalmethod} are used along with the dispersion relation \cref{dispersion}. Unless stated, the impedance boundary model \cref{msdZ} is used in the following computations.

\begin{figure}
\captionsetup[subfigure]{aboveskip=0pt}
    \centering
    \begin{subfigure}[t]{0.45\textwidth}
        \centering        
        \includegraphics*[width=1.\textwidth]{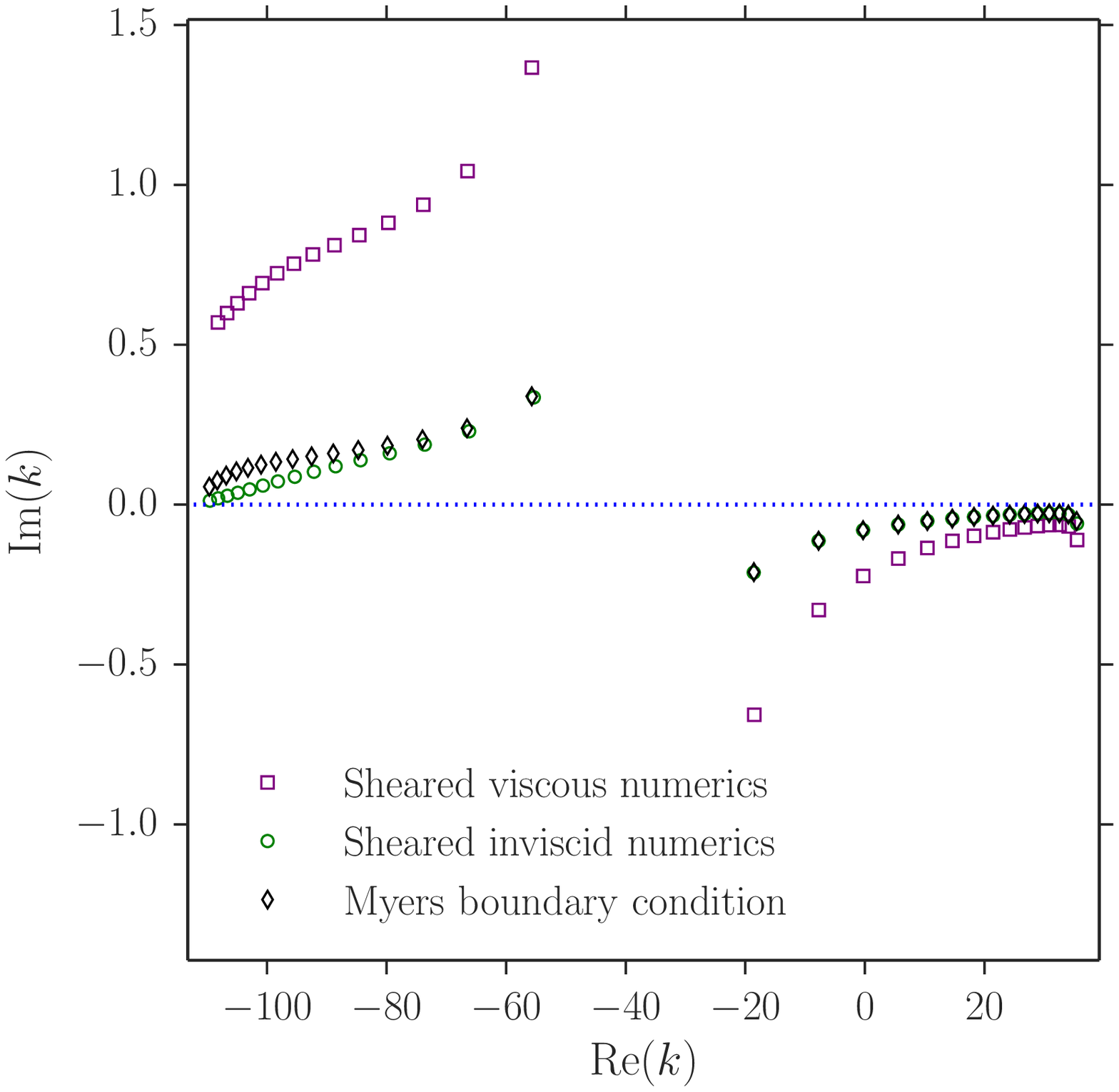}
    \caption{}
    \label{fig:leastcutoff}
    \end{subfigure}%
    ~
    \begin{subfigure}[t]{0.45\textwidth}
        \centering
        \includegraphics*[width=1.\textwidth]{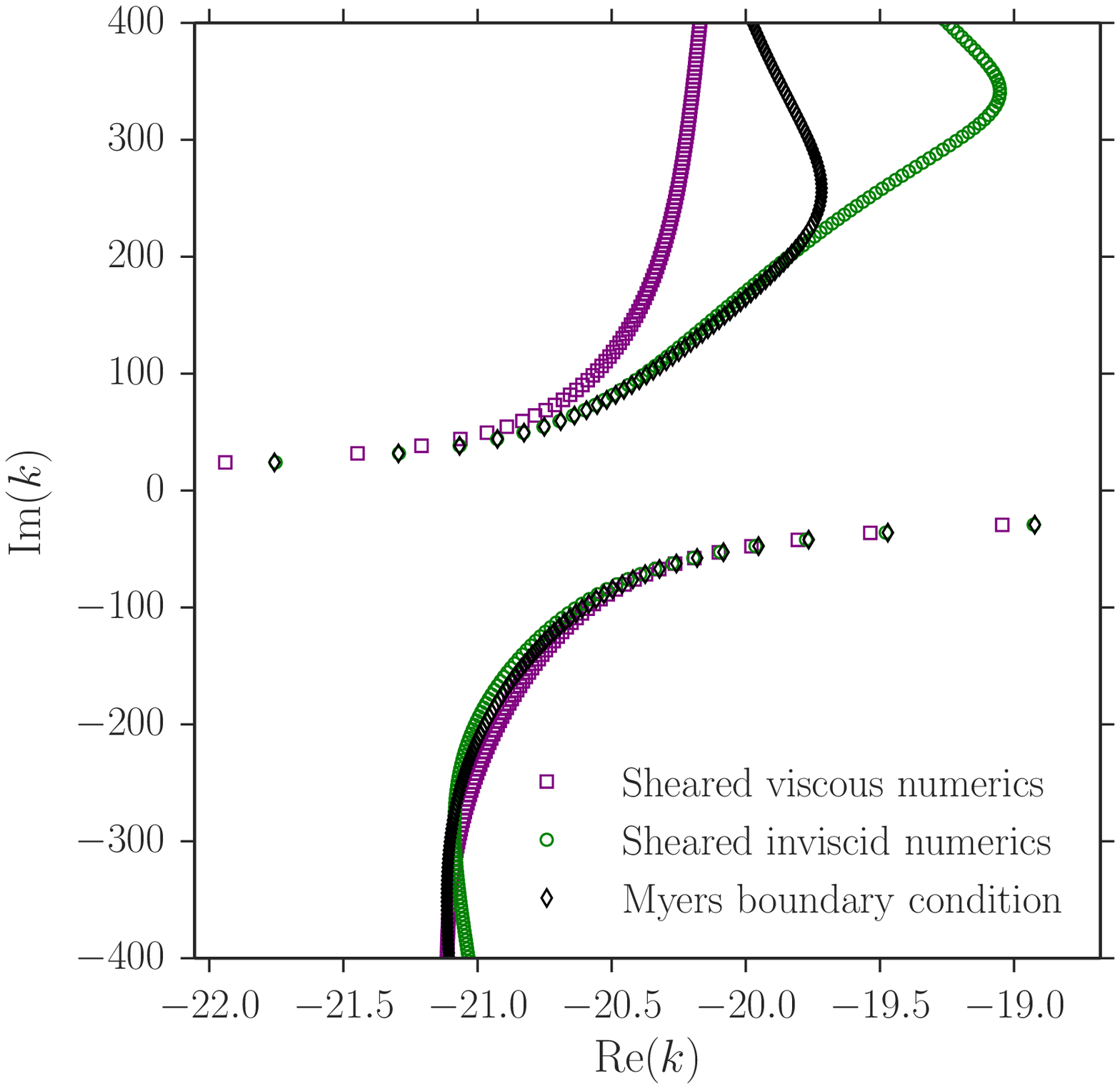}
    \caption{}
    \label{fig:cutoffmodes}
    \end{subfigure}%
    \caption{(a) Cut-on modes of the Myers boundary condition, sheared inviscid numerics (LEE), and sheared viscous numerics (LNSE), for $\omega = 56$, $m=12$, $M=0.5$, $\delta = 2\times 10^{-3}$, $\Ren = 1\times10^{5}$ with a mass--spring--damper boundary impedance~\cref{msdZ} with $R = 3$, $d = 0.15$ and $b = 1.15$ giving $Z = 3 + 8.38\i$. (b) Cut-off modes for $\omega = 31$, $m=24$, $M=0.5$, $\delta = 2\times 10^{-4}$, $\Ren = 2.5\times10^{7}$ with a boundary impedance of $Z = 2+0.6\i$. In both (a) and (b) the hyperbolic boundary layer profiles in \cref{baseflow} are used. } \label{fig:kmodes1}
\end{figure}

\Cref{fig:kmodes1} plots the solutions to the dispersion relations~\cref{dispersion} for the sheared viscous and sheared inviscid numerics, together with the solutions to the dispersion relation~\cref{Zeffmyers} for the uniform inviscid solution together with the Myers boundary condition.  The modes near the real axis may be considered propagating (cut on), with $|\Im(k)|$ giving the axial decay rate of the mode due to the lined wall; $|\Im(k)|$ is therefore extremely important in aeroengine design, as it predicts how much of the engine noise is absorbed by the liner and how much is available to propagate to the far field.  One effect of a thin sheared boundary layer is to change the impedance of the wall as seen by the acoustics outside the boundary layer~\citep{brambley2011b}, thus changing the amount by which the nearly propagating modes are damped.  \Cref{fig:leastcutoff} shows that viscosity can also play a vital role in determining the damping rate of the modes, even at the high frequency of $\omega = 56$ used in this case, and that inviscid calculations underestimate the decay rate of these cut-on modes.  This may also explain the result in \citet{boyeretal2011} where the growth rate of the surface wave was overestimated by inviscid computations.

\Cref{fig:cutoffmodes} shows viscosity has less of an effect on the well cut-off modes, although the agreement is parameter dependent.  Accurate prediction of these cut-off modes is far less important in aeroengine design than that of the nearly cut-on modes, since all models predict the cut-off modes to decay extremely fast along the axis of the duct.
   
    \subsection{Surface waves} \label{sec:numsurfacemodes}
    
Surface waves~\citep{rienstra2003} are an important consideration when investigating lined surfaces, as certain surface waves may represent a hydrodynamic instability of flow over the surface~\citep{rienstra2003,brambley2011a}.  Asymptotic analysis has shown that an inviscid finite thickness boundary layer can support a maximum of six modes localised near the boundary~\citep{brambley2013} while a vanishingly thin boundary layer can support only up to four~\citep{rienstra2003}.  Here, we investigate whether the inclusion of viscosity changes the number or character of the surface wave modes, by tracking the modes as viscosity is turned off in the computations ($1/\Ren\to0$).
    
\begin{figure}
\captionsetup[subfigure]{aboveskip=0pt}
    \centering
    \begin{subfigure}[t]{0.45\textwidth}
        \centering        
        \includegraphics*[width=1.\textwidth]{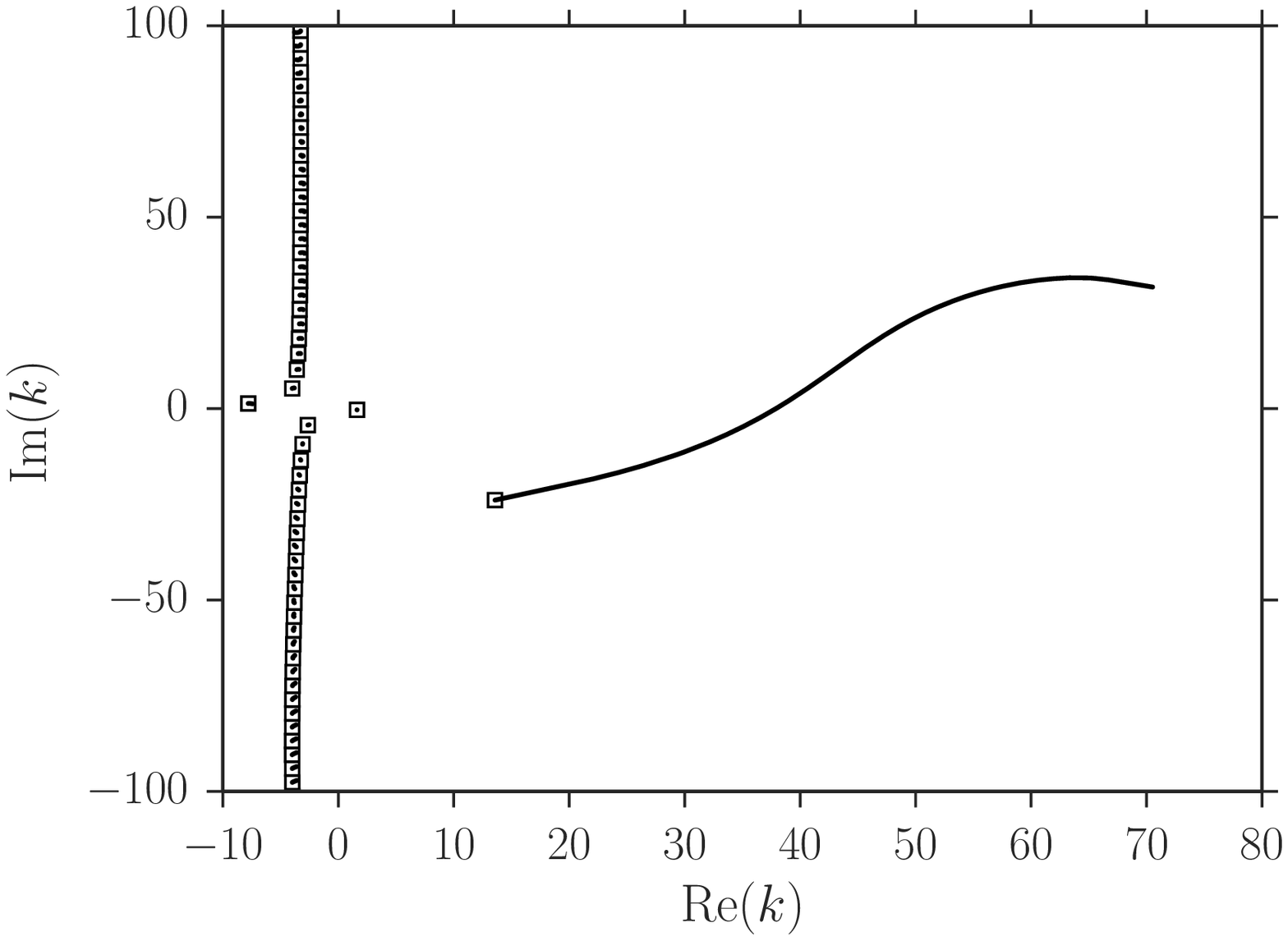}
    \caption{}
    \label{fig:sm1}
    \end{subfigure}%
    ~
    \begin{subfigure}[t]{0.45\textwidth}
        \centering
        \includegraphics*[width=1.\textwidth]{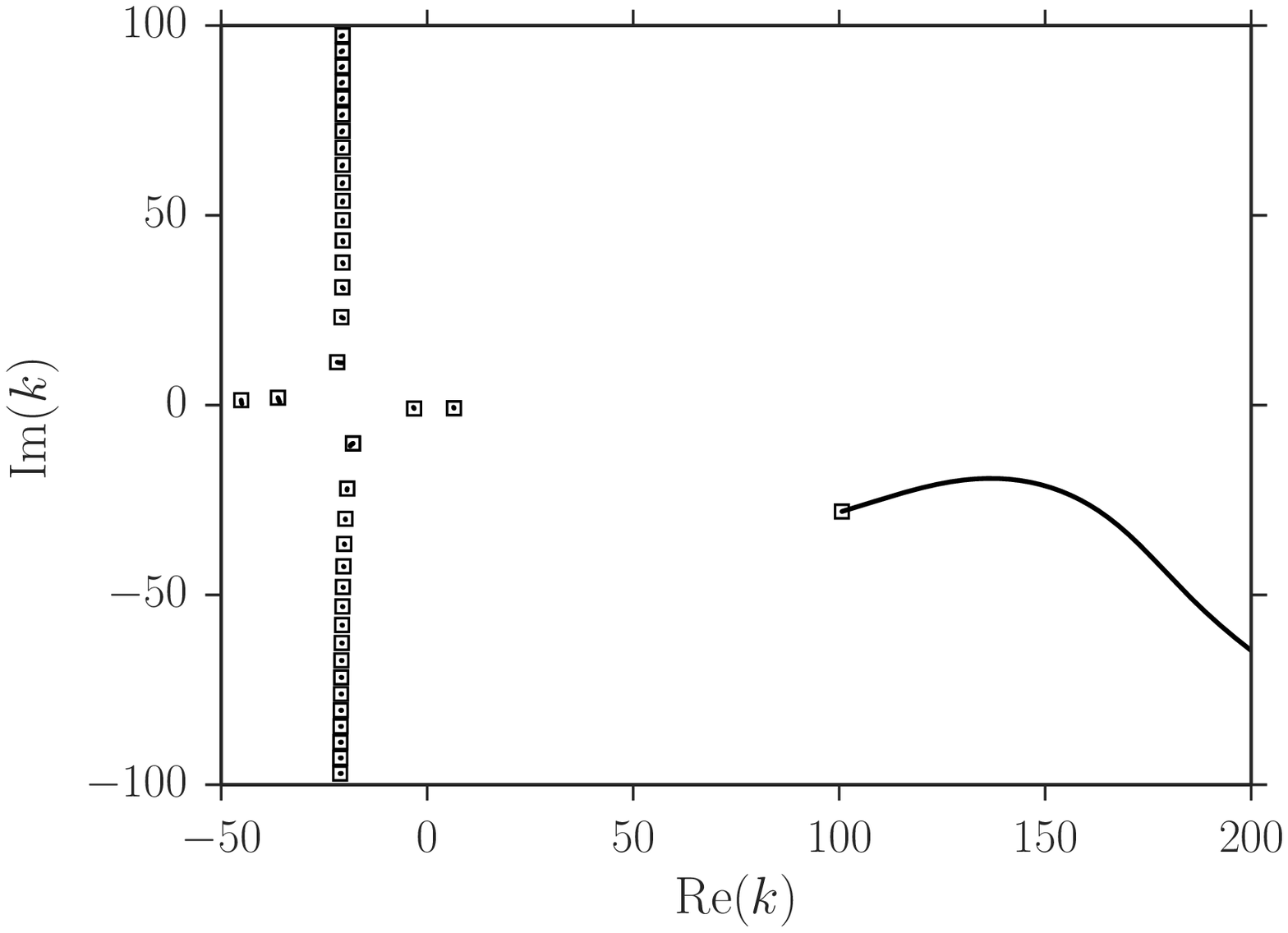}
    \caption{}
    \label{fig:sm2}
    \end{subfigure}%
    
    \begin{subfigure}[t]{0.9\textwidth}
        \centering
        \includegraphics*[width=1.\textwidth]{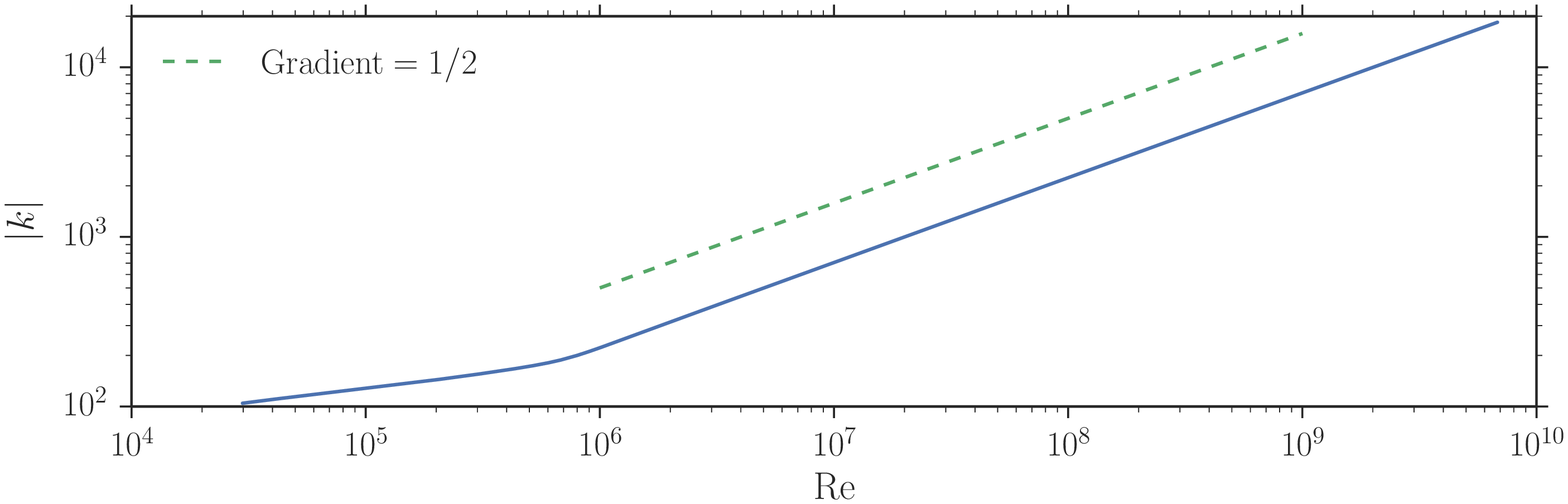}
    \caption{}
    \label{fig:sminf}
    \end{subfigure}%
    \caption{In (a) and (b), markers show modes in the $k$-plane for the viscous sheared numerics, tracks follow surface wave modes as $\Ren$ is increased.  At the end of the tracks the computations are purely inviscid. Parameters are: (a) $\omega = 5$, $m=0$, $M=0.5$, $\delta = 2\times 10^{-3}$, $Z=3+0.52\i$ calculated using \cref{msdZ}, markers at $\Ren = 2.5\times 10^{5}$; (b) $\omega = 31$, $m=24$, $M=0.5$, $\delta = 5.8\times 10^{-3}$, $Z=3+4.61\i$ calculated using \cref{msdZ}, markers at $\Ren = 3.0\times 10^{4}$. In both (a) and (b) the hyperbolic boundary layer profiles in \cref{baseflow} are used. (c) shows the behaviour of the surface mode $k_{\mathrm{sm}}$ on the right side of (b) as $\Ren$ is increased on a log--log scale, demonstrating that $\abs{k_{\mathrm{sm}}}$ is tending to infinity as $1/\Ren \to 0$.} \label{fig:ns-pb_sm}
\end{figure}
    
\Cref{fig:ns-pb_sm} shows the behaviour of modes of the viscous linearised Navier--Stokes equations as the Reynolds number is increased. The markers signify the most viscous point (lowest $\Ren$) considered, and the lines end where $1/\Ren = 0$ (inviscid). In both \cref{fig:sm1} and \cref{fig:sm2} the change in the acoustic modes is small compared to that in the surface wave modes.  \Cref{fig:sm1} is an example where the viscous surface wave mode (marker in the bottom right quadrant) moves substantially as $\Ren$ is increased, crossing the real axis at $\Ren \simeq 1.35\times 10^{6}$ and therefore changing in character from being exponentially decaying as $x$ increases to being exponentially growing.  \Cref{fig:sm2} shows a viscous surface wave mode originating in the lower right quadrant and tending to infinity as $1/\Ren\to 0$, as confirmed in \cref{fig:sminf}.  This mode therefore has no inviscid equivalent, and hence the inclusion of viscosity in the boundary layer is seen to support a greater number of surface waves modes than a purely inviscid boundary layer.

    \subsection{Stability} \label{sec:temporalstability}

\begin{figure}
    \centering      
        \includegraphics*[width=0.9\textwidth]{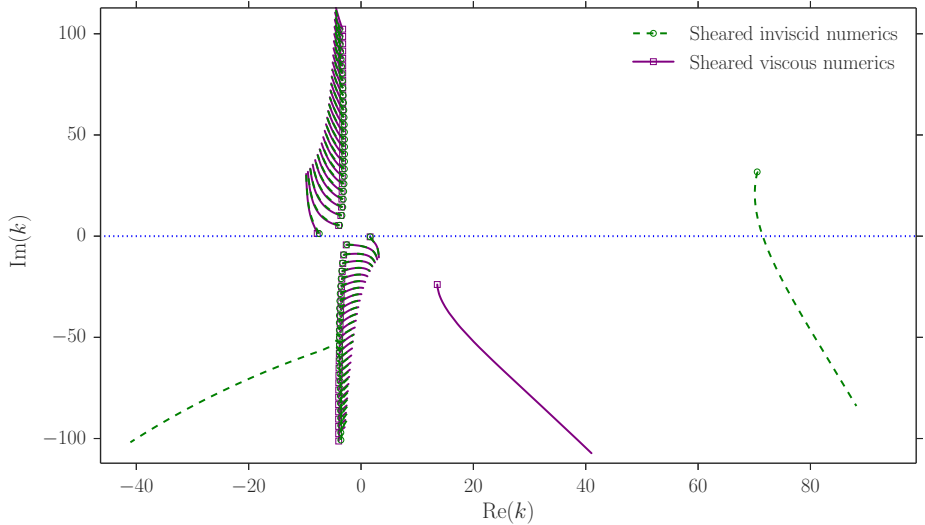}
    \caption{Briggs--Bers trajectories in the $k$-plane for inviscid and viscous sheared numerics. Markers at $\omega = 5$, $Z=3+0.52\i$; lines showing trajectories as $\Im(\omega )$ is reduced from zero to $-15$ with $\Re(\omega)=5$ held constant. The boundary impedance evolves as in \cref{msdZ}, with $R = 3$, $d = 0.15$, $b = 1.15$. Parameters are $m=0$, $M=0.5$, $\delta = 2\times 10^{-3}$, $\Ren = 2.5\times 10^{5}$.  The hyperbolic base profiles \cref{baseflow} are used.} \label{fig:briggs-bers}
\end{figure}

Viscosity is intrinsically linked to stability in shear flow.  For example, it was reported by \citet{brambley2011b} that viscous effects change the growth rate of the Myers vortex sheet instability from having a $k^{1/2}$ to a $k^{1/3}$ wavenumber dependence.
In the previous section, a surface wave mode was found as the Reynolds number was increased to switch from exponential decay to exponential growth as $x$ increases.  We further investigate stability here by performing a Briggs--Bers~\citep{briggs1964,bers1983} stability analysis, reducing $\Im(\omega )$ from zero with $\Re(\omega)$ held fixed, for a mass--spring--damper impedance~\cref{msdZ} with
$R = 3$, $d = 0.15$, and $b = 1.15$. The resulting Briggs--Bers trajectories for the $k$-plane modes are shown in \cref{fig:briggs-bers}.  All of the viscous modes are stable for the plotted parameters, since the trajectories do not cross the $\Re(k)$ axis. All but one of the inviscid modes are stable, with the surface wave in the right half plane crossing the $\Re(k)$ axis as $\Im(\omega )$ is reduced from zero, indicating the mode to be a right-running convective instability. Importantly, this mode is found to stabilise as the Reynolds number is decreased past a critical value $\Ren \simeq 1.35\times 10^{6}$, well within the normal operating range of an aircraft engine.

\begin{figure}
    \centering      
        \includegraphics*[width=0.9\textwidth]{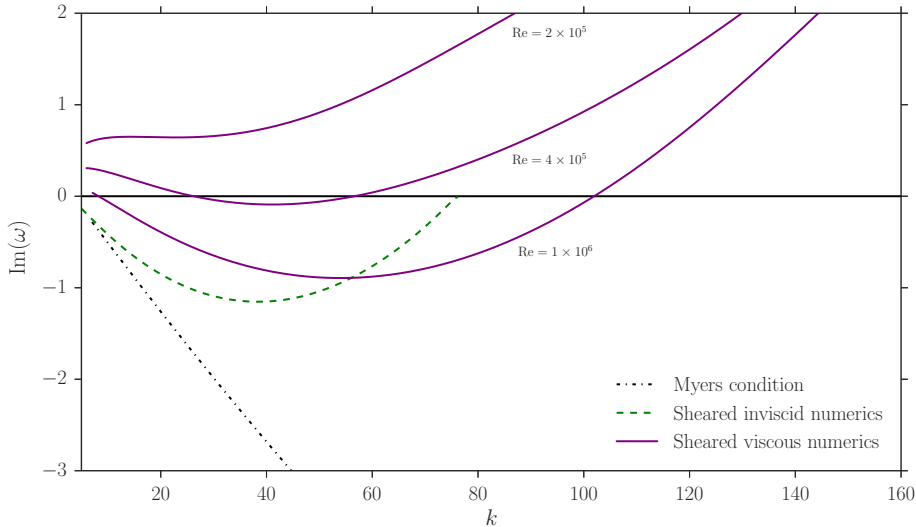}
    \caption{The decay rate $\Im(\omega)$ of the unstable mode is plotted as the real wavenumber $k$ increases. The mode stabilises as $\Ren$ is decreased past $\Ren = 4 \times 10^{5}$. Parameters are $m=0$, $M=0.5$, $\delta = 5\times 10^{-3}$, with the mass--spring--damper liner model of \cref{msdZ}, with $R = 3$, $d = 0.15$, $b = 1.15$. The hyperbolic base profiles \cref{baseflow} are used.} \label{fig:w_unstablemodes}
\end{figure}

A temporal stability analysis may also be performed by choosing a real wavenumber $k$ and solving the dispersion relation~\cref{dispersion} for the complex frequency $\omega (k)$, with $-\Im(\omega(k))$ then giving the growth rate at that wavenumber.  \Cref{fig:w_unstablemodes} shows the behaviour of the growth rate of the unstable surface wave as the (real) wavenumber is increased. The Myers boundary condition displays the well-known instability, the growth rate of which is unbounded in $k$.  The inviscid sheared numerics have an instability for all real $k$, with $\Im(\omega)$ asymptoting to zero but remaining negative. Results for the viscous sheared numerics are shown for three Reynolds numbers. At the highest value, $\Ren = 1\times 10^{6}$, it can be seen that: viscosity stabilises short wavelengths; the most unstable wavelength is altered from the inviscid value; and the maximum growth rate is reduced in the viscous case, although there are possibly some wavenumbers where the viscous system gives faster growing instability than the inviscid system.  As the Reynolds number is reduced (below $\Ren = 4\times 10^{5}$ for these parameters) the flow becomes stable for all real $k$ (as shown by the stable $\Ren = 2\times 10^{5}$ mode).  This may explain the apparent difficulty, expressed in the literature, of observing this instability experimentally.

It is important to remember that absolute, as well as convective, instabilities could be present. Absolute instabilities occur when two surface modes, one originating in the upper--half $k$-plane for $\Im(\omega)=0$ and one originating in the lower--half plane, collide for $\Im(\omega)<0$, causing a pinch at the resulting double root in the $k$-Fourier inversion contour~\citep{briggs1964,bers1983}.  These absolute instabilities would dominate any convective instability at large times. \Citet{brambley2013} showed asymptotically and numerically that for an inviscid sheared flow, the boundary layer must be extremely thin ($\delta \sim 10^{-4}$) for an absolute instability to arise.  In the viscous case, we conjecture that absolute instabilities require a large $\Ren$ (i.e. weak viscosity) coupled with a thin boundary layer; however, this is purely a conjecture, and in the present work we do not investigate absolute instabilities.

\section{Asymptotic analysis} \label{sec:asymptotics}

Several simplified boundary conditions have been proposed which take account of near-wall effects on the wall impedance.  These include: models based on an inviscid fluid with a vanishingly thin shear layer~\citep{myers1980}, meaning $\mathcal{O}(\delta)$ quantities are neglected; an inviscid fluid with a finite-thickness shear layer~\citep{myers&chuang1984,brambley2011a}, meaning $\mathcal{O}(\delta)$ quantities are included; a viscous fluid with a vanishingly thin shear layer~\citep{brambley2011b}, meaning $\mathcal{O}(\delta)$ quantities are neglected; or other restrictive simplifying assumptions~\citep[e.g.][]{aureganetal2001,nayfeh1974}.  In this section we derive a simplified boundary layer model capable of reproducing the important effects of both shear and viscosity seen above, both including finite-thickness shear by including $\mathcal{O}(\delta)$ quantities and including viscosity.

We first present asymptotics based on a reasonably straightforward rescaling in~\S\ref{sec:blasymp}.  As it turns out these result in equations that still need to be solved numerically, an alternative asymptotic solution in the high frequency limit is presented in~\S\ref{sec:highfreq} that yields tractable equations with analytic solutions.  Both asymptotic solutions are subsequently compared with full LNSE solutions in~\S\ref{sec:asympresults}.

\subsection{Boundary layer asymptotics} \label{sec:blasymp}

We analyse asymptotically the near-wall behaviour of the linearised Navier--Stokes equations \cref{lns} by rescaling into the boundary layer,
\begin{align}
r&=1-\delta y, \qquad \qquad \mathrm{with} \qquad \qquad \mu = \frac{\H}{\Ren} = \H \xi \delta^{2}, \label{blscaling}
\end{align}
where $\xi $ measures the magnitude of the molecular viscosity $\mu$ compared with that expected from a Blasius boundary layer of thickness $\delta$.  We assume here $\xi \le \mathcal{O}(1)$, with $\xi = \mathcal{O}(1)$ for a Blasius boundary layer, $\xi \ll 1$ for a turbulent boundary layer, and $\xi = 0$ for an inviscid boundary layer.  These scalings are supplemented by $\h{u} = \delta \t{u}$ and $\h{T} = \delta \t{T}$, which are required to balance the viscous with the inertial terms at leading order in the axial momentum, energy, and continuity equations~\citep{brambley2011b}. These scalings lead to a system of ordinary differential equations in the boundary layer variable $y$ and in powers of the boundary layer thickness $\delta $. \Citet{brambley2011b} keeps only the leading order terms. As a direct extension of that work, and in order to model the effects of both the shear and the viscosity, we work here to $\mathcal{O}(\delta )$.  We find
\begin{subequations}
\begin{gather}
\i (\omega - Uk)\h{T} + \i kT\h{u} + T^{2}\!\left(\frac{\t{v}}{T}\right)_{\!y}\! = \delta\!\left[\gamma \i (\omega - Uk)T\t{p} + T\t{v} - \i m T\t{w}\right], \label{BL_mass_eq}\\
\begin{multlined}[c][.9\displaywidth]\i (\omega - Uk)\h{u} - U_{y}\t{v} - \xi (\gamma -1)^{2}T(T\h{u}_{y} + U_{y}\h{T})_{y} =\\\delta\!\left[\i (\gamma -1)kT\t{p} - \xi (\gamma -1)^{2}T(T\h{u}_{y} + U_{y}\h{T})\right],\end{multlined} \label{BL_xmom_eq}\\
\begin{multlined}[c][.9\displaywidth]\t{p}_{y} = \delta \Big[\i \rho (\omega - Uk)\t{v} - \xi (2+\beta) (\gamma -1)(T\t{v}_{y})_{y}\\ - \i \xi \beta k (\gamma -1) (T\h{u})_{y} - \i \xi k (\gamma - 1)(T\h{u}_{y} + U_{y}\h{T})\Big],\end{multlined} \label{BL_rmom_eq}\\
\xi (T\t{w}_{y})_{y} - \frac{\i (\omega - Uk)}{(\gamma - 1)^{2}T}\t{w} + \frac{\i m}{\gamma - 1}\t{p} = \mathcal{O}(\delta ), \label{BL_thetamom_eq}\\
\begin{multlined}[c][.9\displaywidth]\i (\omega - Uk)\h{T} - T_{y}\t{v} - \frac{1}{\mathrm{Pr}}\xi (\gamma -1)^{2}T(T\h{T})_{yy} - \xi (\gamma -1)^{2}T(U_{y}^{2}\h{T} + 2TU_{y}\h{u}_{y}) =\\
\delta\!\left[(\gamma - 1)\i(\omega - Uk)T\t{p} - \frac{1}{\mathrm{Pr}}\xi (\gamma -1)^{2}T(T\h{T})_{y}\right].\end{multlined} \label{BL_energy_eq}
\end{gather}
\label{BL_governing_eqs}%
\end{subequations}
Note that the azimuthal momentum equation \cref{BL_thetamom_eq} is written to leading order as the azimuthal acoustic velocity $\t{w}$ appears only in the first order forcing in the continuity equations \cref{BL_mass_eq}. If the parameter $\xi $ were set to zero in \cref{BL_governing_eqs}, an inviscid system would be recovered which, when solved, would lead to the modified Myers condition as derived by \citet{brambley2011a}. Immediately deducible from \cref{BL_rmom_eq} is that, in contrast with the leading order viscous model of \citet{brambley2011b}, the pressure is not constant across the boundary layer; instead, variation in the pressure appears at first order as an integral across the boundary layer. The system \cref{BL_governing_eqs} may be solved asymptotically assuming expansions of the acoustic quantities of the form $q = q_{0} + \delta q_{1} + \mathcal{O}(\delta ^{2})$.

The acoustic axial and azimuthal velocities satisfy no slip at the lining $r=1$, $y=0$, to all orders, and the acoustic temperature satisfies the isothermal wall condition to all orders.  The leading order pressure is a constant, and our chosen normalisation therefore dictates that $\t{p}_{0} \equiv 1$ and $\t{p}_{1}(0) = 0$.  Similarly, we choose for the impedance boundary condition to be satisfied exactly, such that $\t{v}_{0}(0) = 1/Z$ and $\t{v}_{1}(0) = 0$.  Prohibiting exponentially growing solutions as $y \to \infty$ gives further boundary conditions, as described in \cref{sec:A}.  This leads to
\begin{equation}
    \begin{aligned}
\h{u}_{0}(0) &=   0,& \h{T}_{0}(0) &=   0,& \t{w}_{0}(0) &=   0,& &\t{p}_{0} \equiv 1,& \t{v}_{0}(0) &= \frac{1}{Z},\\
\h{u}_{0}(y) &\to 0,& \h{T}_{0}(y) &\to 0,& \t{w}_{0}(y) &\to \frac{m}{\omega - Mk} && \text{as } y \to \infty,
    \end{aligned}
\label{bcs0}%
\end{equation}
at $\mathcal{O}(1)$, and
\begin{equation}
    \begin{aligned}
\h{u}_{1}(0) &= 0,& \h{T}_{1}(0) &= 0,& &\t{p}_{1}(0) = 0,& \t{v}_{1}(0) &= 0, \\
\h{u}_{1}(y) &\to \frac{k}{\omega - M k},&  \h{T}_{1}(y) &\to 1, & &\text{as } y \to \infty, & &
    \end{aligned}
\label{bcs1}%
\end{equation}
at $\mathcal{O}(\delta )$.  The limit $y\to\infty$ of this boundary layer solution must match to the uniform inviscid acoustics outside the boundary layer, \cref{uniform}.  Defining $\pinf$ and $\vinf$ as the values of pressure and normal velocity of the uniform inviscid solution at the lining $r=1$, we expand $\pinf = \pinf^{(0)} + \delta \pinf^{(1)}$ and similarly for $\vinf$, so that the uniform inviscid solutions \cref{uniform} close to the lining may be expanded as
\begin{subequations}
\begin{align}
p_{\mathrm{ui}}(1-\delta y) =&\, \pinf^{(0)} + \delta \pinf^{(1)} + \delta y \i (\omega + Mk)\vinf^{(0)} + \mathcal{O}(\delta ^{2}), \\
v_{\mathrm{ui}}(1-\delta y) =&\, \vinf^{(0)} + \delta \vinf^{(1)} + \delta y \left(\vinf^{(0)} - \frac{(\omega - M k)^{2} - k^{2} - m^{2}}{\i (\omega - Mk)}\pinf^{(0)}\right) + \mathcal{O}(\delta ^{2}).
\end{align}
\label{outer2}%
\end{subequations}
In practice, the system \cref{BL_governing_eqs}, \cref{bcs0,bcs1} is solved across the boundary layer for a finite range $y\in[0,Y]$, with the values of $\t{p}_{0}$, $\t{p}_{1}$, $\t{v}_{0}$, $\t{v}_{1}$ at $y=Y$ extrapolated to infinity (see \cref{sec:A}) and matched with the relations \cref{outer2} to find $\pinf^{(j)}$ and $\vinf^{(j)}$; the effective impedance $Z_{\mathrm{eff}} = (\pinf^{(0)} + \delta \pinf^{(1)})/(\vinf^{(0)} + \delta \vinf^{(1)})$ may then be formed.

The system \cref{BL_governing_eqs}, \cref{bcs0,bcs1} must in general be solved numerically.  It does not, therefore, suggest an easily applicable closed-form boundary condition capable of capturing the behaviour of the acoustics in a sheared viscothermal boundary layer. With this in mind, we now consider the high frequency limit of the LNSE.

    \subsection{High frequency asymptotics} \label{sec:highfreq}

We now consider the limits $\omega \gg 1$ and $\delta \ll 1$ with $\omega \delta \sim \ee \ll 1$ (where $\ee$ is not to be confused with the acoustic amplitude $\epsilon_{a}$ used earlier).  If we were to expand the outer solutions \cref{uniform} near the wall $r=1-\delta y$ in powers of $\omega $ and $\delta $, then at order $n$ (in $\delta $) the largest term would be of the form $(\omega \delta)^{n}$.  Thus, for a useful outer expansion, we need $\delta \sim 1/\omega ^{a}$ with $a>1$. We choose here the distinguished scaling $\ee = 1/\sqrt{\omega}$ (informed by the expansion of the outer solution near the boundary, \cref{highfreqouter} below), and hence the two small parameters are related by $\delta = \ee ^{3} \dd$ where $\dd=\mathcal{O}(1)$. This scaling agrees well with reported parameters for a turbofan intake~\citep{gabard2013}, with a blade passing frequency $\omega = 28$ and upstream boundary layer thickness $\delta = 7\times 10^{-3}$ giving $\dd\approx 1.04$.  With the above scaling choices, the outer solutions expand as
\begin{subequations}
\begin{align}
p_{\mathrm{ui}}(1-\delta y) =&\, \pinf + \i \ee \bar{\delta }(1 - ML) y \vinf + \frac{1}{2}\ee^{2} \bar{\delta }^2 (N^2 - \bar{\alpha }^2) y^2 \pinf + \mathcal{O}(\ee^{3}), \label{phighfreqouter}\\
v_{\mathrm{ui}}(1-\delta y) =&\, \vinf + \i \ee \bar{\delta } \frac{\bar{\alpha }^{2} - N^{2}}{1 - ML} y \pinf + \frac{1}{2}\ee^{2} \bar{\delta }^2 (N^2 - \bar{\alpha }^{2}) y^2 \vinf + \mathcal{O}(\ee^{3}), \label{vhighfreqouter}
\end{align}
\label{highfreqouter}%
\end{subequations}
where
\begin{equation}
\bar{\alpha }^{2} = (1 - ML)^{2} - L^{2}, \label{alphabar}
\end{equation}
and $L = k/\omega $, $N = m/\omega $ with $L$, $N$ assumed to be $\mathcal{O}(1)$.

To find the inner solution, we follow \citet{brambley2011b} in introducing a multiple scales WKB ansatz for the acoustic quantities,
\begin{align}
\frac{\mathrm{d}}{\mathrm{d}y} &= \frac{\partial}{\partial\bar{y}} + \frac{1}{\ee }\eta(\bar{y})\frac{\partial}{\partial\theta },& \label{multiplescales}
&\text{with}&
\bar{y}(y) &= y,&
\theta (y) &= \frac{1}{\ee} \int_{0}^{y}\eta (y') \mathrm{d}y',
\end{align}
then relabel $\bar{y}$ to $y$. The function $\eta (y)$ is a combination of base flow quantities,
\begin{equation}
\eta ^{2}(y) = \frac{\i (1-U(y)L)}{\xi (\gamma - 1)^{2}T^{2}(y)}, \label{eta}
\end{equation}
with $\Ren\{\eta(y)\} > 0$ as $y\to \infty$, and represents the viscous decay rate of vorticity away from the boundary.  The acoustic quantities are assumed to vary over both the short length-scale $\theta $ and the long length-scale $y$. The base flow quantities vary only over the long length-scale $y$. The short length-scale $\theta$ can be thought of as equivalent to the classical acoustic boundary layer scaling $r = 1 - \delta\theta/\sqrt{\omega}\eta$~\citep[][chap.~2, pg.~11]{ingard2010}.  From the system \cref{lns}, we make the pre-emptive scalings
\begin{align}
\t{u} &= \frac{\hh{u}}{\omega \delta } ,& \t{T} &= \frac{\hh{T}}{\omega \delta}, \label{hfutscaling}
\end{align}
and expand in powers of $\ee$,
\begin{subequations}
\begin{gather}
\begin{multlined}[c][.85\displaywidth]\hh{u}_{\theta \theta} - \hh{u} = \frac{\i U_{y}\t{v}}{1-UL} - \frac{\ee \dd L}{\rho (1-UL)}\t{p} - \frac{\ee}{\eta ^{2}T}\Big[(\eta T \hh{u}_{\theta})_{y} + \eta T \hh{u}_{\theta y} + \eta U_{y}\hh{T}_{\theta }\Big] \\
- \frac{\ee ^{2}}{\eta ^{2}T}\Big[(T\hh{u}_{y})_{y} + (U_{y}\hh{T})_{y}\Big] - \ee ^{3}\Big[\dd^{2}(1+\beta )\frac{\i L}{\eta }\t{v}_{\theta }\Big] + \mathcal{O}(\ee ^{4}),\end{multlined} \label{xmom_hf}\\
\begin{multlined}[c][.85\displaywidth]\frac{1}{\Pr}\hh{T}_{\theta \theta} - \hh{T} = \frac{\i T_{y}\t{v}}{1-UL} - \ee \frac{\dd}{\rho }\t{p} - \frac{\ee}{\eta ^{2}T}\Big[\frac{1}{\Pr} (\eta T \hh{T}_{\theta })_{y} + \frac{1}{\Pr} \eta (T\hh{T}_{\theta})_{y} + 2\eta TU_{y}\hh{u}_{\theta}\Big] \\
- \frac{\ee ^{2}}{\eta ^{2}T}\Big[\frac{1}{Pr}(T\hh{T})_{yy} + 2TU_{y}\hh{u}_{y} + U_{y}^{2}\hh{T}\Big] + \mathcal{O}(\ee ^{4}), \end{multlined} \label{energy_hf}\\
\t{v}_{\theta } = - \ee \Big[\!\frac{\i (1-UL)}{\eta T}\hh{T} + \frac{\i L}{\eta }\hh{u} + \frac{T}{\eta }\Big(\frac{\t{v}}{T}\Big)_{\!y}\Big]\! + \ee ^{2}\Big[\i (1-UL)\frac{\dd \gamma}{\eta }\t{p} - \frac{\i \dd N}{\eta }\t{w}\Big] + \mathcal{O}(\ee ^{4}), \label{mass_hf}\\
\begin{multlined}[c][.85\displaywidth]\t{p}_{\theta} = -\ee \frac{\t{p}_{y}}{\eta } - \frac{\dd \i \rho (1-UL)}{\eta}\bigg\{\ee^{2}\Big[(2+\beta)\t{v}_{\theta \theta} - \t{v}\Big]\\
+ \ee^{3}\Big[\frac{(2+\beta)}{{\eta ^{2}T}}\Big((\eta T\t{v}_{\theta})_{y}+ \eta T\t{v}_{\theta y}\Big) + (1+\beta)\frac{\i L}{\eta }\hh{u}_{\theta}\Big]\bigg\} + \mathcal{O}(\ee ^{4}), \end{multlined} \label{rmom_hf}\\
\begin{multlined}[c][.85\displaywidth]\t{w}_{\theta \theta} - \t{w} = -\frac{N}{\rho (1-UL)}\t{p} - \frac{\ee}{\eta ^{2}T}\Big[(\eta T \t{w}_{\theta})_{y} + \eta T \t{w}_{\theta y}\Big]\\
- \frac{\ee ^{2}}{\eta ^{2}T}\Big[(T\t{w}_{y})_{y} + \i \dd N(1+\beta )\eta T\t{v}_{\theta}\Big] + \mathcal{O}(\ee ^{3}). \end{multlined} \label{thetamom_hf}
\end{gather}
\label{hf_eqs}%
\end{subequations}
The equations \cref{hf_eqs} are not quite a high frequency expansion of the boundary layer equations \cref{BL_governing_eqs}; the high frequency has caused some terms to jump order, and consequently we retain some terms that are absent in the standard $\mathcal{O}(\delta)$ analysis in \cref{sec:blasymp}.  Also note that in contrast with the high frequency asymptotics of \citet{brambley2011b}, the model proposed here has variation in the acoustic pressure at $\mathcal{O}(\ee)$, and `finite thickness shear' terms (\i.e. first order in the boundary layer thickness $\delta $) appearing at $\mathcal{O}(\ee^{2})$.

Solving the system \cref{hf_eqs} for the inner solutions leads to the acoustic pressure and radial velocity
\begin{subequations}
\begin{align}
\t{p}(y,\theta) = &\, F_{0}(y) + \ee F_{1}(y) + \ee^{2} F_{2}(y), \label{hf_p} \\
\t{v}(y,\theta) = &\, A_{0}(y) + \ee \Big[A_{1}(y) + \frac{\i L}{\eta(y) } B_{0}(y)\mathrm{e}^{-\theta } + \frac{\i (1-U(y)L)}{\sigma \eta(y) T(y)}D_{0}(y)\mathrm{e}^{-\sigma \theta }\Big] \notag \\
&\,+ \ee^{2}\Big[A_{2}(y) + \frac{\i L}{\eta }B_{1}(y)\mathrm{e}^{-\theta } + a_{0}(y)\mathrm{e}^{-\theta } + \frac{\i (1-U(y)L)}{\sigma \eta (y) T(y)}D_{1}(y)\mathrm{e}^{-\sigma \theta }\Big]. \label{hf_v}
\end{align}
\label{hf_pv}%
\end{subequations}
where $\sigma^2 = \Pr$.  The functions $F_{j}(y)$, $A_{j}(y)$, $B_{j}(y)$, $D_{j}(y)$ and $a_{0}(y)$ are determined by boundary, secularity and matching conditions, as described in \cref{sec:B}.  We note that $\t{p}$ does not vary on the short length scale $\theta$ until $\mathcal{O}(\ee ^{3})$, which is beyond the order of solution we present here.  We asymptotically match \cref{hf_pv} with the outer solutions \cref{highfreqouter} in the limit $y\to \infty$ (see \cref{sec:B1}), which, using the definitions $Z = \t{p}(0)/\t{v}(0)$ and $Z_{\mathrm{eff}} = \pinf/\vinf$, leads to the effective impedance
\begin{equation}
Z_{\mathrm{eff}} = \frac{1}{\omega - M k}\frac{\omega Z - \dfrac{k U_{y}(0)}{\sqrt{\omega}\eta (0)}Z - \i \delta I_{0}(\omega - M k)^{2} + \omega\mathcal{B} Z}{1 + \i \omega Z\delta I_{1}\dfrac{k^{2} + m^{2}}{(\omega - M k)^{2}} + \mathcal{A} + \mathcal{C}Z}, \label{Zeff_hf}
\end{equation}
where
\begin{subequations}
\begin{gather}
\mathcal{A} = (\delta I_{0}\delta I_{1} + \delta^2I_{11} - \delta^2I_{01})(k^{2} + m^{2}) - \delta^2 I_{2}\big((\omega - M k)^{2} - k^{2} - m^{2}\big), \\
\begin{multlined}[c][.85\displaywidth]\mathcal{B} = (\delta I_{0}\delta I_{1} + \delta^2I_{3} - \delta^2I_{10})(k^{2} + m^{2}) - \delta^2 I_{00}\big((\omega - M k)^{2} - k^{2} - m^{2}\big) \\
-\i\frac{(\gamma - 1)^{2}}{\omega\Ren}\left[ \frac{I_{\mu}}{\delta^2} + 2\frac{\sigma}{1 + \sigma}T(1)U_r(1)^{2} - \frac{5k^{2}}{4\omega^2}T(1)^2U_r(1)^{2}\right],\end{multlined}\\
\mathcal{C} = \frac{(\gamma-1)T(1)}{\sqrt{\i \omega\Ren}}\left[\i k U_r(1)\delta I_{1}\frac{k^{2} + m^{2}}{(\omega - M k)^{2}} + \frac{\i}{\omega}(k^{2} + m^{2})(\gamma - 1)T(1) + \frac{\i \omega}{\sigma}(\gamma - 1)\right],
\end{gather}
\label{hf_coefficients}%
\end{subequations}
and $I_{j}$ are the integrals
\begin{gather}
\delta I_{0} = \int_{0}^{1}\chi_{0}\,\mathrm{d}r, \qquad\qquad \delta I_{1} = \int_{0}^{1}\chi_{1}\,\mathrm{d}r, \qquad\qquad \delta^2I_{2} = \int_{0}^{1}(1-r)\chi_{0}\,\mathrm{d}r, \nonumber\\
\delta^2I_{3} = \int_{0}^{1}(1-r)\chi_{1}\,\mathrm{d}r, \qquad\quad \frac{I_{\mu}}{\delta^2} = \int_{0}^{1}\frac{\chi_{\mu}}{\delta^3}\,\mathrm{d}r, \qquad\quad \delta^2I_{01} = \int_{0}^{1}\!\!\chi_{0}(r)\!\!\int_{r}^{1}\!\!\chi_{1}(r')\,\mathrm{d}r'\mathrm{d}r, \nonumber\\
\delta^2I_{10} = \int_{0}^{1}\!\!\chi_{1}(r)\!\!\int_{r}^{1}\!\!\chi_{0}(r')\,\mathrm{d}r'\mathrm{d}r, \qquad\qquad \delta^2I_{00} = \int_{0}^{1}\Big(\int_{r}^{1}\chi_{0}(r')\,\mathrm{d}r' - \delta I_{0}\Big)\mathrm{d}r, \nonumber\\
\delta^2I_{11} = \int_{0}^{1}\Big(\int_{r}^{1}\chi_{1}(r')\,\mathrm{d}r' - \delta I_{1}\Big)\mathrm{d}r,
\end{gather}
with
\begin{gather}
\chi_{0} = 1 - \frac{\rho (\omega - U k)^{2}}{(\omega - M k)^{2}}, \qquad\qquad \chi_{1} = 1 - \frac{(\omega - M k)^{2}}{\rho (\omega - U k)^{2}}\nonumber\\
\frac{\chi_{\mu}}{\delta^3} = \frac{-\omega}{\omega - U k}\Big[\frac{1}{2\sigma^2}(T^{2})_{rrr} + (TU_{r}^{2})_{r} + \frac{k T}{\omega - U k}(TU_{r})_{rr}\Big].
\end{gather}
At leading order, the boundary condition \cref{Zeff_hf} reduces to the Myers condition, \cref{Zeffmyers}. At $\mathcal{O}(\ee)$, and in the limit of a vanishingly thin shear layer $\delta \to 0$, \cref{Zeff_hf} reduces to the $\mathcal{O}(\ee)$ high frequency result presented in \citet{brambley2011b}, while at $\mathcal{O}(\ee)$ in the limit of infinite Reynolds number ($\xi=0$) \cref{Zeff_hf} reduces to the modified Myers boundary condition~\citep{brambley2011a}.

\section{Comparison of asymptotic and numerical results} \label{sec:asympresults}

\begin{figure}
\captionsetup[subfigure]{aboveskip=0pt}
    \centering
    \begin{subfigure}[t]{0.42\textwidth}
        \centering        
        \includegraphics*[width=1.\textwidth]{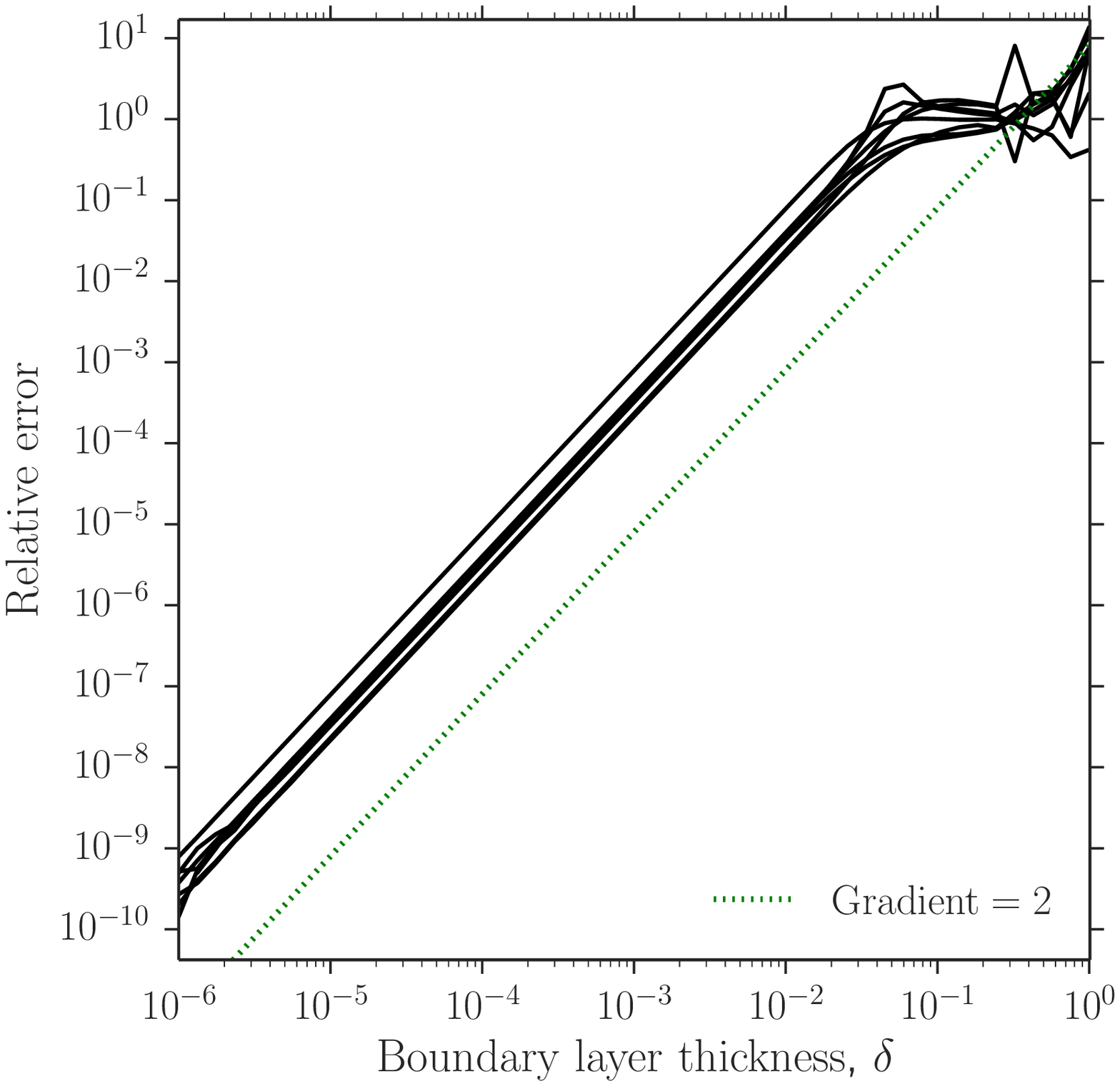}
    \caption{}
    \label{fig:vfasymerr}
    \end{subfigure}%
    ~
    \begin{subfigure}[t]{0.42\textwidth}
        \centering
        \includegraphics*[width=0.971\textwidth]{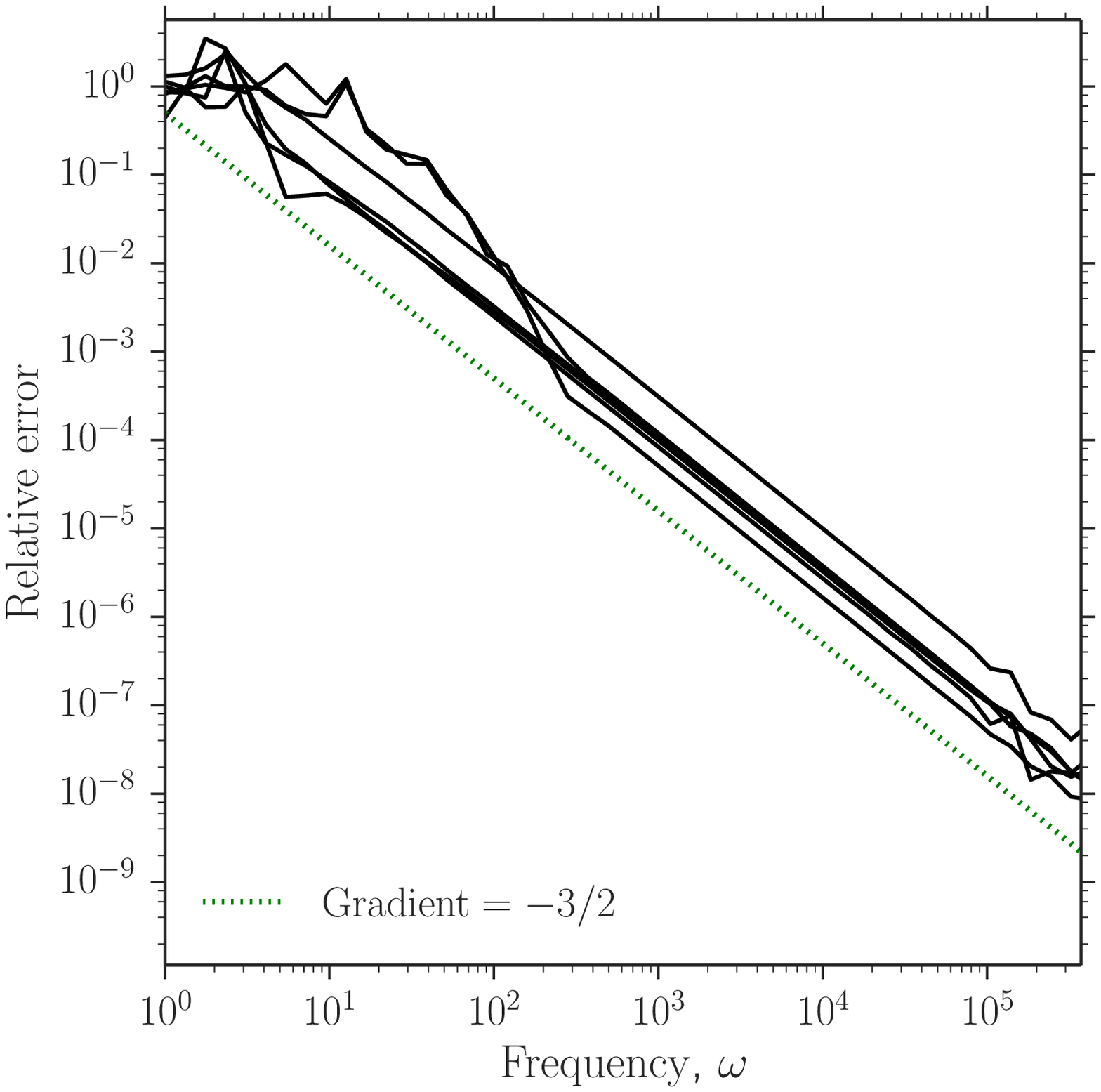}
    \caption{}
    \label{fig:hfasymerr}
    \end{subfigure}%
    \caption{Asymptotic accuracy plots with relative error defined as $\abs{Z_{\mathrm{eff}}(Z)/(\t{p}_{ui}(1)/\t{v}_{ui}(1)) - 1}$, where the function $Z_{\mathrm{eff}}(Z)$ is the asymptotic effective impedance (from either \cref{sec:blasymp} or \cref{sec:highfreq}) with the input boundary impedance $Z = \t{p}_{sv}(1)/\t{v}_{sv}(1)$. (a) accuracy of the reduced boundary layer model \cref{BL_governing_eqs}, \cref{bcs0,bcs1} with respect to $\delta $.  The first order solution has an error of $\mathcal{O}(\delta ^{2})$ (gradient $2$). Parameters are $\omega = 10$, $m=0$, $M=0.5$, $k=\pm1\pm\i$, $\pm1$, $\pm \i$, $\Ren = 1/\delta^{2}$. (b) accuracy of the high frequency effective impedance \cref{Zeff_hf} with respect to $\omega$; the error is $\mathcal{O}(\ee^{3})=\mathcal{O}(\omega ^{-3/2})$ (gradient $-3/2$). Parameters are $M=0.5$, $\delta = \omega ^{-3/2}$, $\Ren = 1/\delta^{2}$, $m/\mathrm{int}(\omega) = 1$, $k/\omega = \exp{\{\i \arg{(\tilde{k})}\}}$ where $\tilde{k}=\pm1\pm\i$, $1$, $-\i$. In both (a) and (b) the hyperbolic base flow profiles \cref{baseflow} are used.} \label{fig:asympaccuracy}
\end{figure}

The two asymptotic boundary layer models presented above are compared here against numerical solution of the linearised Navier--Stokes equations. \Cref{fig:asympaccuracy} shows that the two models are correct to their stated order of accuracy in their respective limits $\delta\to 0$ and $\omega\to\infty$.

\begin{figure}
\captionsetup[subfigure]{aboveskip=0pt}
    \centering
    \begin{subfigure}[t]{0.9\textwidth}
        \centering        
        \includegraphics*[width=1.\textwidth]{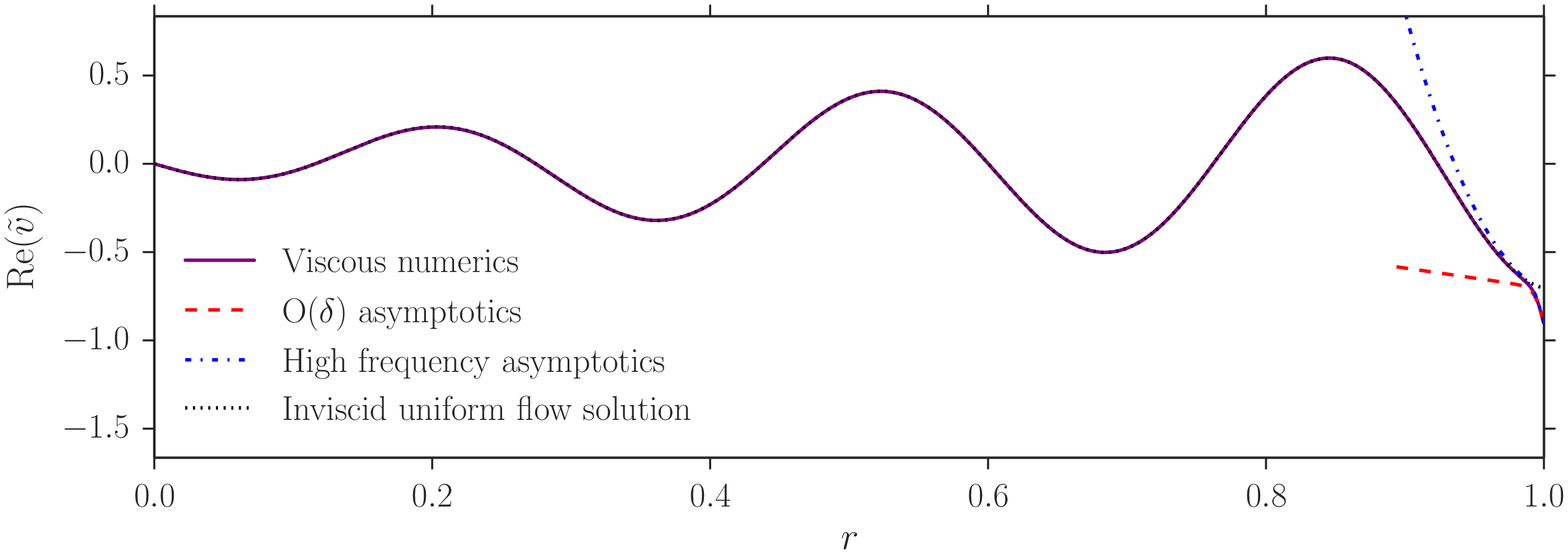}
    \caption{}
    \label{fig:fullduct}
    \end{subfigure}%
    
    \begin{subfigure}[t]{0.9\textwidth}
        \centering
        \includegraphics*[width=1.\textwidth]{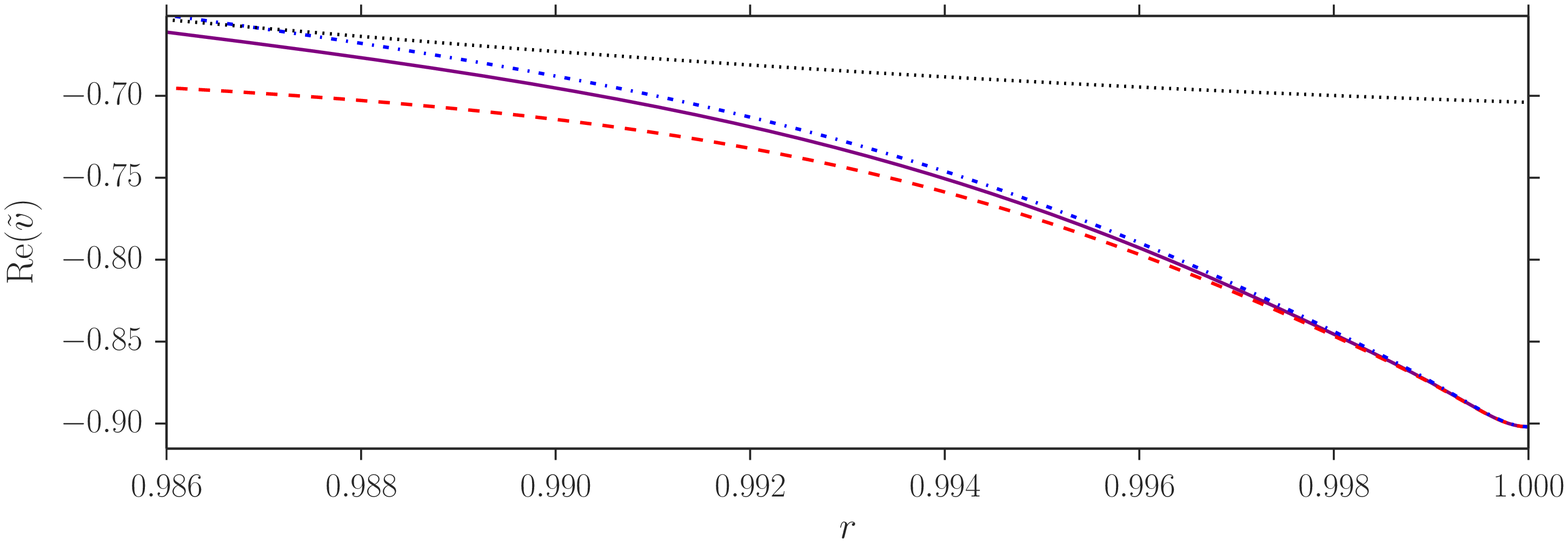}
    \caption{}
    \label{fig:blzoom}
    \end{subfigure}%
    \caption{Acoustic radial velocity mode shapes of the viscous numerics, the $\mathcal{O}(\delta)$ boundary layer asymptotics \cref{BL_governing_eqs}, the high frequency boundary layer asymptotics \cref{hf_v}, and the inviscid uniform outer solution \cref{uniform} to which the asymptotic solutions match in the limit $y\to \infty$. The asymptotic solutions are defined in the space of a boundary layer variable $y$, and plotted in the $r$ domain using $r = 1 - \delta y$. (a) shows the full duct, (b) shows the boundary layer. Parameters are $\omega = 31$, $k=14 + \i$, $m=2$, $M=0.5$, $\delta = 7\times 10^{-3}$, $\Ren = 5\times 10^{5}$; the hyperbolic base flow \cref{baseflow} is used.} \label{fig:asympmodeshapes}
\end{figure}

\Cref{fig:asympmodeshapes} shows the mode shapes of the acoustic radial velocity of the two asymptotic solutions, $\mathcal{O}(\delta)$ \cref{BL_governing_eqs} and high frequency \cref{hf_v}, compared with viscous numerics.  For reference, the inviscid uniform flow solution $\tilde{v}_\mathrm{ui}$ \cref{uniform}, to which the asymptotic solutions match, is also plotted.  Both models replicate the viscous mode shape well inside the boundary layer.  It appears in \cref{fig:blzoom} that the high frequency asymptotics outperform the $\mathcal{O}(\delta)$ asymptotics due to the high frequency ($\omega=31$) used in this case.

\subsection{$k$-plane modes} \label{sec:kplaneasym}

To find duct modes for the asymptotic models, a dispersion relation must be satisfied,
\begin{equation}
Z_{\mathrm{eff}}(Z) = \frac{\t{p}_{ui}(1)}{\t{v}_{ui}(1)}, \label{asym_disp}
\end{equation}
where the effective impedance $Z_{\mathrm{eff}}(Z)$ is the result of the asymptotic model (e.g. from \cref{Zeff_hf}) given the actual boundary impedance $Z$ as input (see \cref{sec:appendixNUM} for more information).  In this section, we choose a frequency $\omega $ and find complex $k(\omega)$ that satisfy \cref{asym_disp}.

\begin{figure}
\captionsetup[subfigure]{aboveskip=0pt}
    \centering
    \begin{subfigure}[t]{0.45\textwidth}
        \centering        
        \includegraphics*[width=0.99\textwidth]{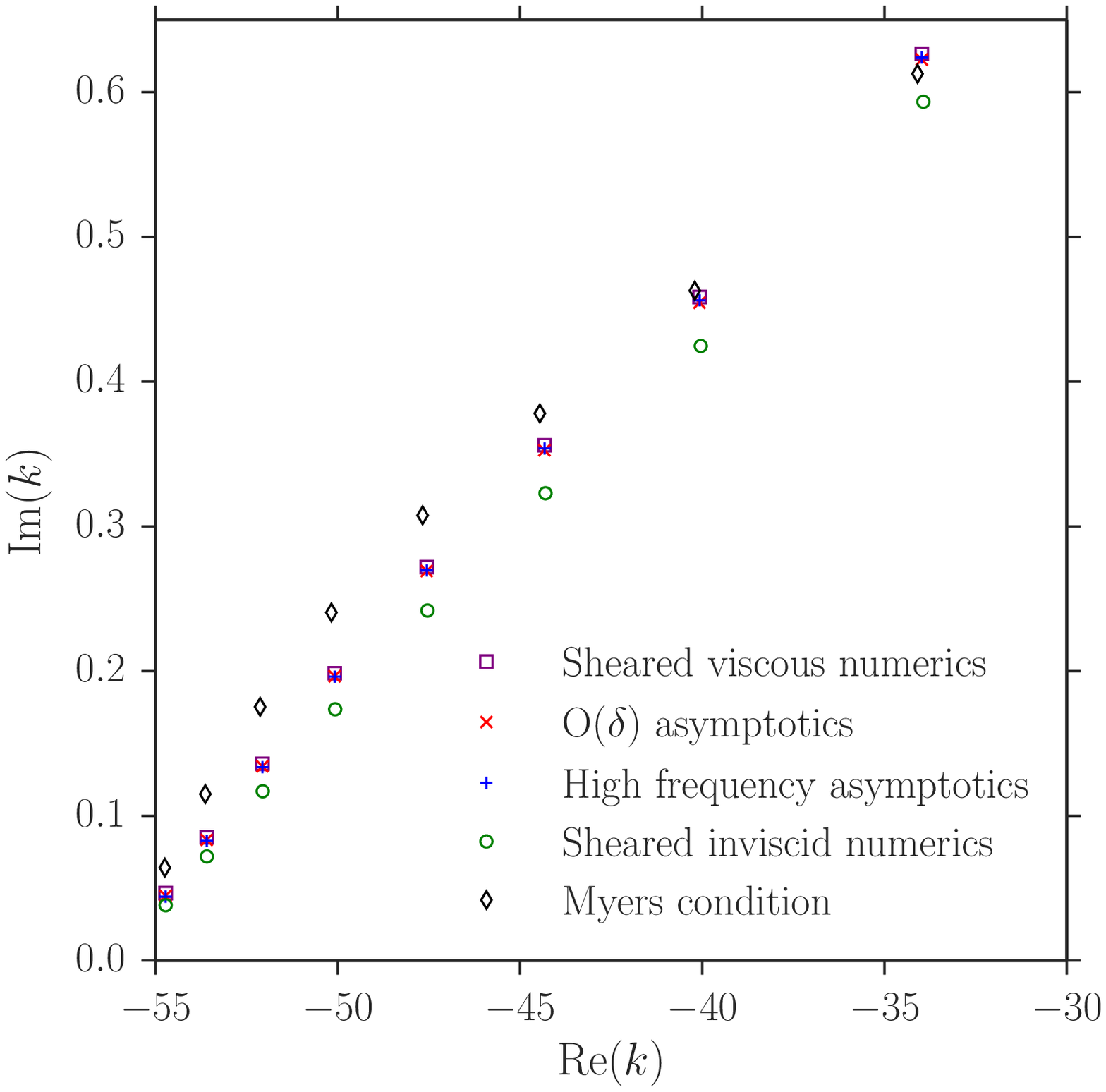}
    \caption{}
    \label{fig:leastcutoffasym}
    \end{subfigure}%
    ~
    \begin{subfigure}[t]{0.45\textwidth}
        \centering
        \includegraphics*[width=1.\textwidth]{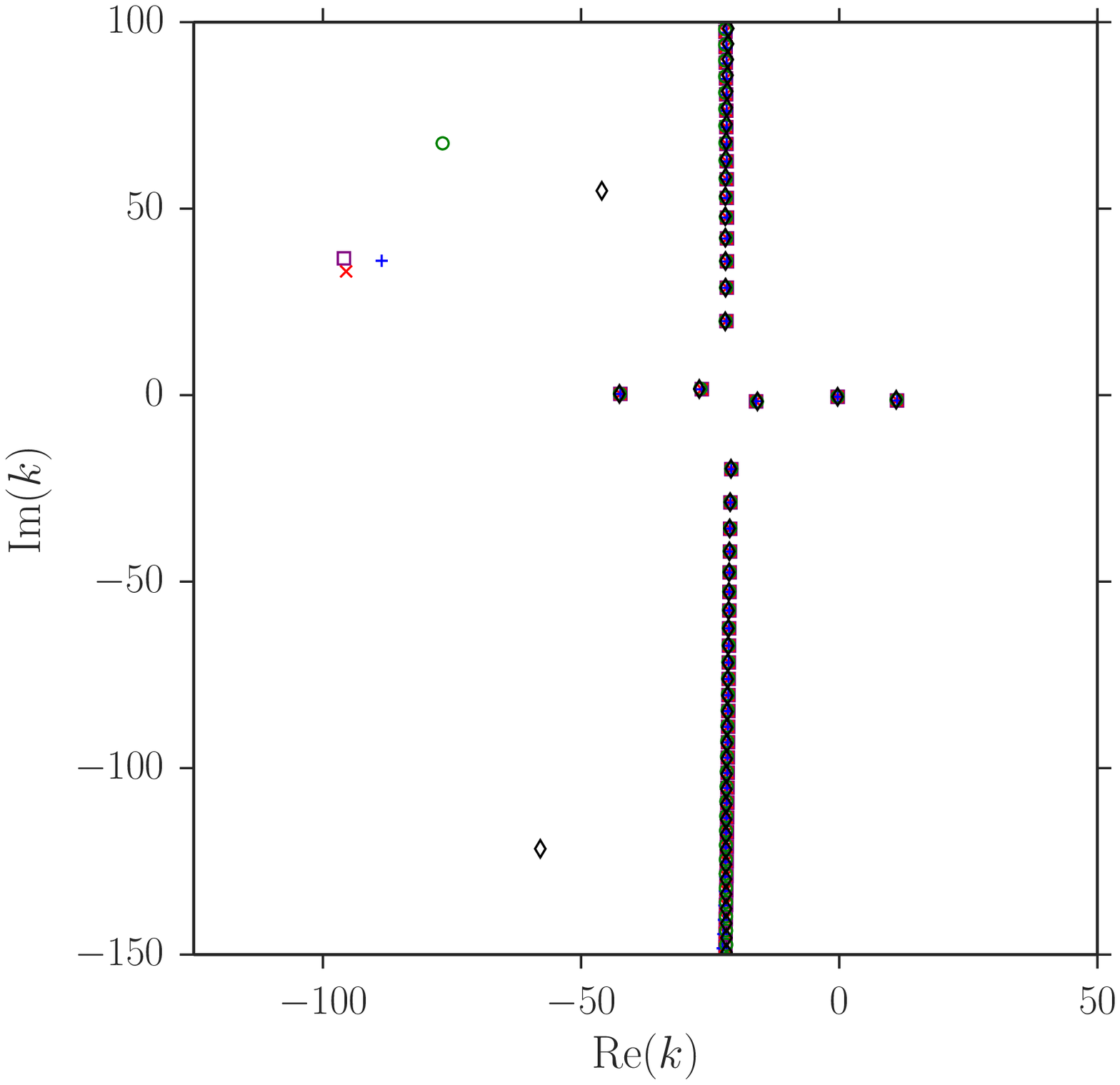}
    \caption{}
    \label{fig:surfaceasym}
    \end{subfigure}%
\caption{(a) Upstream cut-on modes of the two asymptotic models \cref{BL_governing_eqs} and \cref{Zeff_hf}, the viscous and inviscid numerics, and the Myers boundary condition, for $\omega = 28$, $m=0$, $M=0.5$, $\delta = 2\times 10^{-3}$, $\Ren = 5\times10^{6}$ with a boundary impedance of $Z = 3+4.16\i $ (calculated using the mass--spring--damper impedance \cref{msdZ}). (b) Mode spectra showing one surface wave mode in the upper left quadrant. Parameters are $\omega = 31$, $m=24$, $M=0.5$, $\delta = 1\times 10^{-3}$, $\Ren = 1\times10^{6}$, with a boundary impedance of $Z = 0.6-2\i $.  In both (a) and (b) the hyperbolic boundary layer profiles in \cref{baseflow} are used.} \label{fig:kmodesasym}
\end{figure}

The asymptotic models are seen in \cref{fig:kmodesasym} to replicate the $k$-plane modes of the LNSE well.  As with \cref{fig:kmodes1}, the attenuation (given by $\Im(k)$) of the nearly propagating upstream modes is badly predicted by the Myers condition. The effect of viscosity is to increase the attenuation of these cut-on modes, as seen in \cref{fig:leastcutoffasym}, while the effect of shear is to reduce attenuation (for modes travelling upstream).  Both asymptotic models perform well, suggesting the physics of both the shear and the viscosity have been correctly captured in both asymptotic formulations. In \cref{fig:surfaceasym}, the asymptotic boundary conditions accurately predict the viscous surface wave mode in the upper left quadrant, unlike either the inviscid numerics or the Myers condition.

\begin{figure}
\captionsetup[subfigure]{aboveskip=0pt}
    \centering        
        \includegraphics*[width=0.9\textwidth]{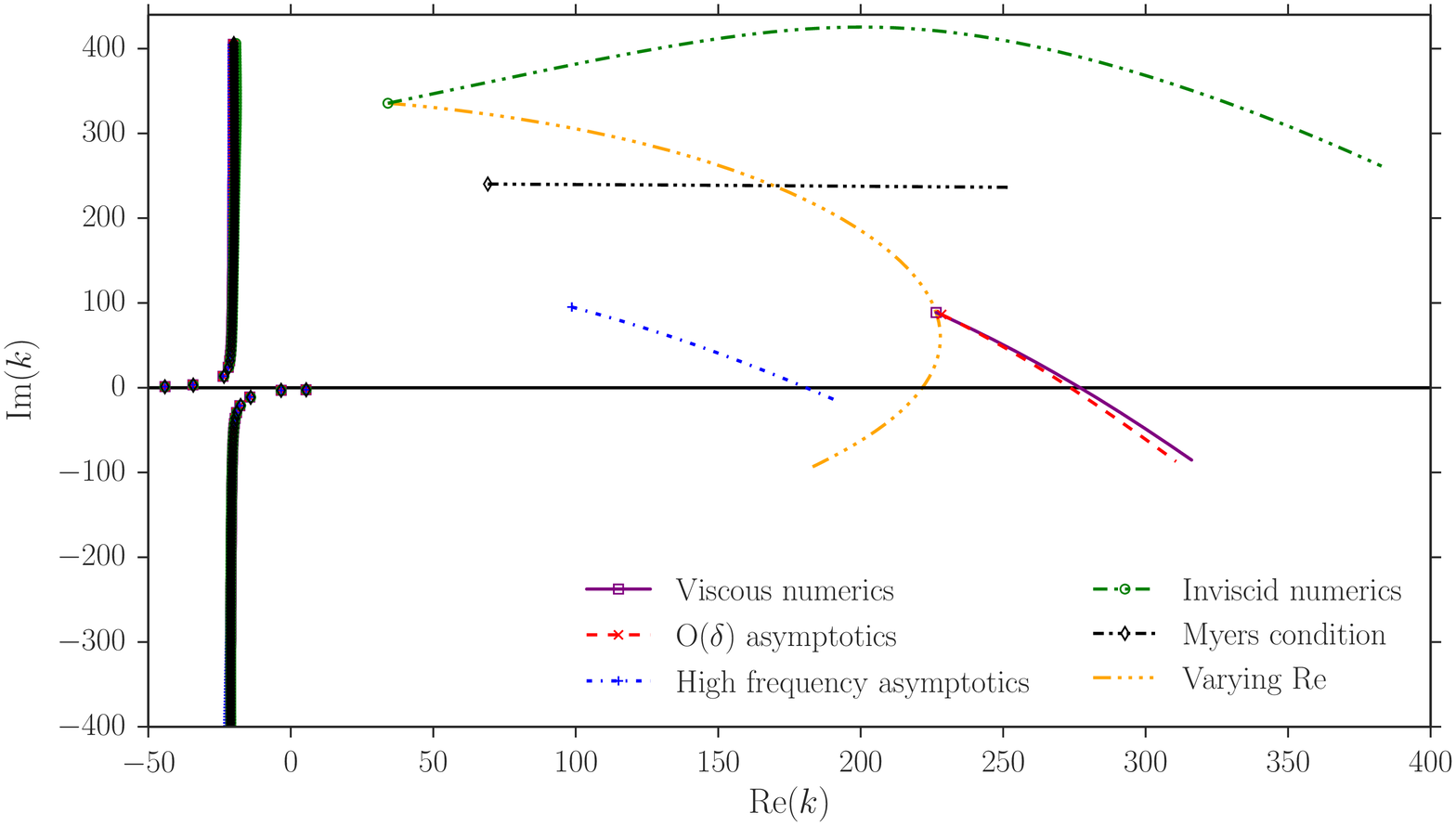}
\caption{Markers show modes in the $k$-plane of the two asymptotic models \cref{BL_governing_eqs} and \cref{Zeff_hf}, the viscous and inviscid numerics, and the Myers boundary condition, for $\omega = 31$, $m=24$, $M=0.5$, $\delta = 2\times 10^{-4}$, $\Ren = 2.5\times10^{7}$ with a boundary impedance of $Z = 2+0.6\i $.  Also shown are Briggs-Bers trajectories of the surface wave modes in the upper right quadrant as $\Im(\omega)$ is reduced from zero to around $-20$ with $\Re(\omega) = 31$ held fixed.  The impedance is governed by a mass--spring--damper model $Z(\omega ) = R + \i \omega d - \i b/\omega$, with $R = 2$, $d = 0.02$, $b = 0.62$, such that $Z = 2+0.6\i $ at $\omega = 31$.  The track labelled ``Varying $\Ren$'' follows the inviscid numerical surface wave mode as the Reynolds number is reduced from $\Ren=\infty$; it passes through the viscous mode at $\Ren = 2.5\times 10^{7}$ and crosses the real axis when $\Ren \simeq 1.04\times 10^{7}$. The hyperbolic boundary layer profiles in \cref{baseflow} are used. } \label{fig:briggsberssasym}
\end{figure}

In \cref{fig:briggsberssasym} the inviscid numerics and Myers boundary condition again predict a surface wave in a very different position to the viscous numerical surface wave (upper right quadrant). The dashed ``Varying $\Ren$'' line traces the movement of this surface wave mode as the Reynolds number is decreased from infinity; as both the inviscid and the viscous surface wave mode lie along this line, we identify the viscous surface wave mode as the viscous equivalent of the inviscid surface wave mode.  The $\mathcal{O}(\delta)$ asymptotics \cref{BL_governing_eqs} perform very well, while the high frequency asymptotics \cref{Zeff_hf} do not do so well in predicting the position of the LNSE surface wave mode.  This can be explained by the large value of the axial wavenumber at the LNSE surface wave mode being outside the range of validity of the high-frequency asymptotics, since $k \simeq 226+88\i $ for this mode gives $\abs{L} = 1.6/\ee$, contradicting the assumption of $L=k/\omega$ being $\mathcal{O}(1)$ following \cref{alphabar}.

Also shown in \cref{fig:briggsberssasym} are the Briggs-Bers trajectories of the surface wave modes as $\Im(\omega)$ is reduced from zero to around $-10$. The LEE mode (sheared inviscid numerics) remains far above the real $k$ axis as $\Im(\omega)$ is reduced, while the LNSE mode (viscous numerics) crosses the real axis, indicating a downstream-propagating convective instability.  The two asymptotic models predict the correct convective instability, although the high frequency asymptotics are inaccurate for the reasons discussed in the previous paragraph.

\subsection{$\omega $-plane modes}

The temporal stability properties of the new asymptotic models are investigated here. We choose a real $k$ and solve the dispersion relation \cref{asym_disp} to find complex frequency roots $w(k)$.  The exponential factor $\exp{\{ \i \omega t\}}$ implies that the temporal growth rate of a mode is given by $-\Im(\omega)$.

\begin{figure}
\captionsetup[subfigure]{aboveskip=0pt}
    \centering        
        \includegraphics*[width=0.9\textwidth]{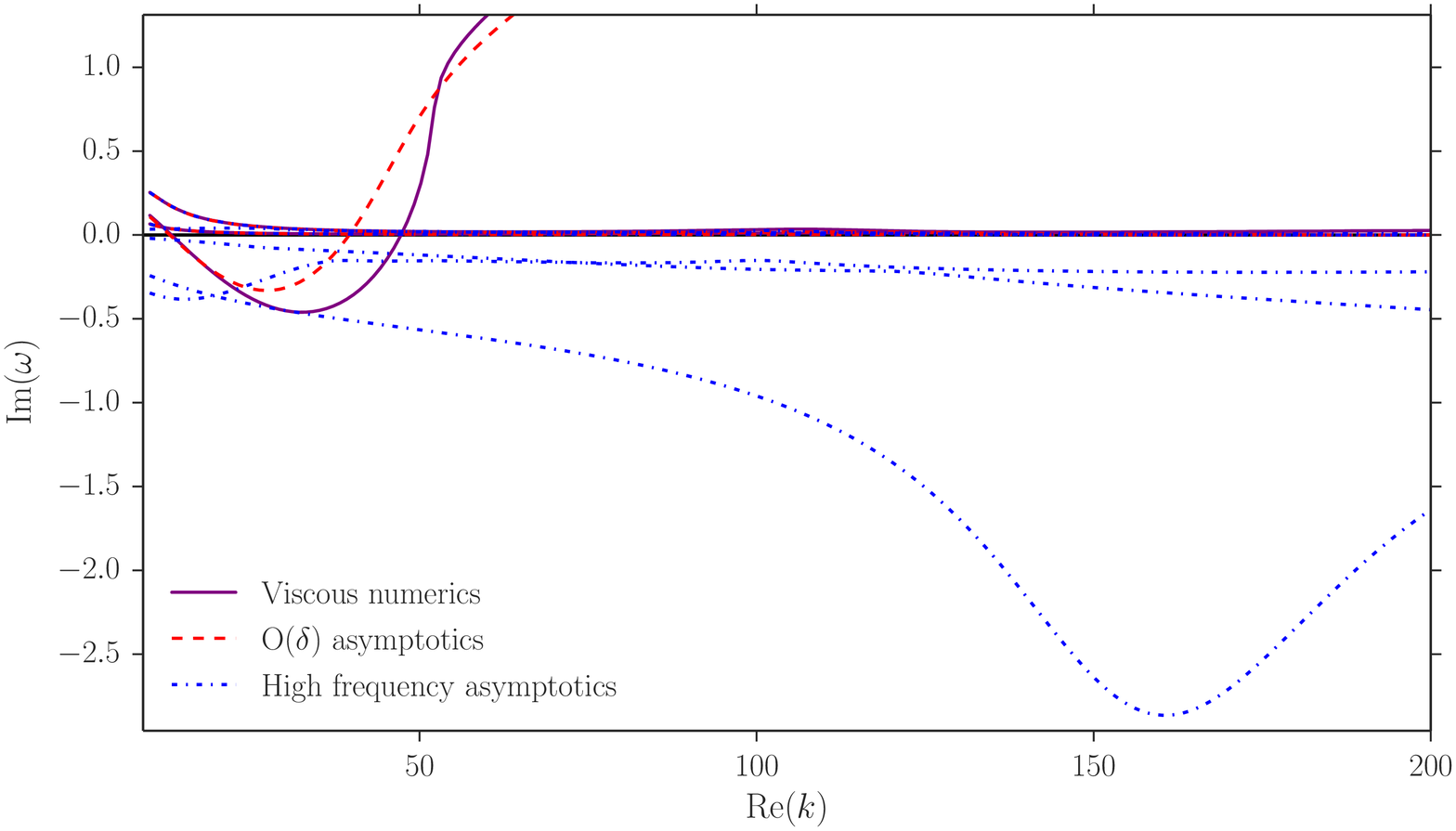}
\caption{Temporal growth rates, $-\Im(\omega)$, of modes as $k$, real, is increased. A mode below the $k$ axis is a growing instability.  Plotted are modes for the LNSE, $\mathcal{O}(\delta)$ asymptotics \cref{BL_governing_eqs}, and high frequency asymptotics \cref{Zeff_hf}. Parameters are $m = 0$, $M = 0.5$, $\Ren = 3\times10^{6}$, $\delta = 7\times10^{-3}$ with the hyperbolic base flow \cref{baseflow}. The boundary impedance is modelled as a mass--spring--damper \cref{msdZ} with a mass $d = 0.01$, spring $b = 10$ and damping $R = 0.75$.} \label{fig:stability_asym}
\end{figure}

\Cref{fig:stability_asym} compares the behaviour of the $\omega$ modes as $k$, real, is increased for the LNSE, $\mathcal{O}(\delta)$ asymptotics, and high frequency asymptotics.  As in \cref{sec:temporalstability}, the LNSE displays an instability that has a well-defined maximum growth rate and restabilises at a finite $k$.  The $\mathcal{O}(\delta)$ asymptotic model reproduces this behaviour: the growth rate of the instability is bounded (due to the regularising effect of a finite thickness shear layer), and the mode restabilises for small enough wavelength (due to the small-scale damping by viscosity); this model is well-posed.  Recall that neither the Myers boundary condition~\citep{ingard1959,myers1980} nor the leading order viscous boundary condition~\citep{brambley2011b} lead to a bounded growth rate, while the first order inviscid boundary correction~\citep{brambley2011a} gives a bounded growth rate that remains unstable for all $k$.
\Cref{fig:stability_asym} also shows that the high frequency viscous asymptotics of \cref{sec:highfreq} perform poorly with regard to temporal instability, so there is no guarantee of well-posedness; this is because the temporal instability occurs either for low frequencies or for $|k/\omega|\gg 1$, which are both outside the region of asymptotic validity of the high frequency boundary condition \cref{Zeff_hf}.

\subsection{Accuracy of high frequency asymptotics at lower frequencies} \label{sec:lowhighfreq}

In \cref{sec:kplaneasym} it was shown that the high frequency asymptotics \cref{Zeff_hf} are efficient in predicting cut-off and cut-on acoustic modes at high frequencies, but that the model can fail relative to the $\mathcal{O}(\delta)$ asymptotics in its prediction of surface waves.  Here, we investigate the accuracy of the high frequency asymptotics at moderate to low frequencies for the cut-off and cut-on acoustic modes.

For the nearly cut-on acoustic modes, the parameter of most interest is the rate of attenuation per axial distance travelled.  The accuracy of the asymptotic models with respect to the LNSE numerics can be expressed as the difference in the predicted attenuation rate, given in decibels per duct radius as
\begin{equation}
\Delta \mathrm{dB} = 20\log_{10}\left[\frac{\Im(k)}{\Im(k_{\mathrm{LNSE}})}\right]. \label{deltadb}
\end{equation}
\Cref{tab:cutonw10w5w2}
\begin{table}%
\centering%
\begin{tabular}{ c|r@{}l r|r@{}l r|r@{}l r }
          &     \multicolumn{2}{|c}{$k$ ($\omega = 10$)}  &  \multicolumn{1}{c|}{$\Delta\mathrm{dB}$} & \multicolumn{2}{|c}{$k$ ($\omega = 5$)} &  \multicolumn{1}{c|}{$\Delta\mathrm{dB}$} & \multicolumn{2}{|c}{$k$ ($\omega = 2$)} &  \multicolumn{1}{c}{$\Delta\mathrm{dB}$} \\
    \hline
    LNSE  &      5.9042&\phantom{+}\llap{--\,}0.4802\i    &                  &     2.6441&\phantom{+}\llap{--\,}0.6808\i     &              &    -0.9533&\phantom{+}\llap{--\,}1.0094\i     &                 \\
    --    &    -18.6182&\phantom{+}\llap{+}0.1565\i       &                  &    -6.4055&\phantom{+}\llap{+}2.2000\i        &              &    -3.3209&\phantom{+}\llap{+}1.9036\i        &                 \\
    Myers &      5.9456&\phantom{+}\llap{--\,}0.4706\i    &    -0.1756       &     2.6555&\phantom{+}\llap{--\,}0.6043\i     &    -1.0354   &    -0.8293&\phantom{+}\llap{--\,}0.8665\i     &    -1.3261      \\
    --    &    -18.6321&\phantom{+}\llap{+}0.1259\i       &    -1.8948       &    -6.5494&\phantom{+}\llap{+}1.7036\i        &    -2.2212   &    -3.1805&\phantom{+}\llap{+}3.1690\i        &     4.4269      \\
    HF    &      5.9046&\phantom{+}\llap{--\,}0.4748\i    &    -0.0992       &     2.6309&\phantom{+}\llap{--\,}0.6817\i     &     0.0115   &    -1.0158&\phantom{+}\llap{--\,}0.9550\i     &    -0.4815      \\
    --    &    -18.6189&\phantom{+}\llap{+}0.1451\i       &    -0.6587       &    -6.3715&\phantom{+}\llap{+}2.2413\i        &     0.1616   &    -2.9887&\phantom{+}\llap{+}1.7280\i        &    -0.8407      \\
    OD    &      5.9043&\phantom{+}\llap{--\,}0.4790\i    &    -0.0219       &     2.6422&\phantom{+}\llap{--\,}0.6781\i     &    -0.0336   &    -0.9675&\phantom{+}\llap{--\,}1.0079\i     &    -0.0127      \\
    --    &    -18.6186&\phantom{+}\llap{+}0.1472\i       &    -0.5317       &    -6.4053&\phantom{+}\llap{+}2.2004\i        &     0.0016   &    -3.3079&\phantom{+}\llap{+}1.9001\i        &    -0.0159      \\
\end{tabular}%
\caption{Wavenumbers of the most cut-on modes, using the LNSE numerics, the Myers boundary condition, the high frequency (HF) asymptotics \cref{Zeff_hf} and the $\mathcal{O}(\delta)$ (OD) asymptotics \cref{BL_governing_eqs}.  Parameters are $\omega = 5$, $m = 2$, $\Ren = 1\times 10^{5}$, $\delta = 4\times 10^{-3}$. The boundary impedance $Z$ is found using \cref{msdZ} with $d = 0.08$, $b = 6$ and $R = 1.6$.  Base profiles as in \cref{baseflow}. The same parameters were used for \cref{fig:cutoffw5}.  The errors in the attenuation predicted by the approximate models are expressed in decibels per radius in the $\Delta\mathrm{dB}$ columns, calculated using \cref{deltadb}.}%
\label{tab:cutonw10w5w2}%
\end{table}%
shows that the attenuation rate of the cut-on modes is well predicted by the high frequency asymptotics \cref{Zeff_hf}, and even at the low frequency of $\omega = 5$ the $\mathcal{O}(\delta)$ asymptotics are only marginally more accurate.\footnote{For a duct of radius $l^{*} = 1\mathrm{m}$, a dimensionless frequency $\omega = 5$ corresponds to a sound frequency of $f^{*} \approx 270\mathrm{Hz}$. The value $\omega = 31$ gives $f^{*} \approx 1.6\mathrm{kHz}$.}  For $\omega = 2$, the $\mathcal{O}(\delta)$ asymptotics are significantly more accurate than the high frequency asymptotics, but the high frequency asymptotics still predict the attenuation of the two cut-on modes to within $1\mathrm{dB}$ per duct radius travelled --- that is, with much greater accuracy than the Myers boundary condition.  A similar situation is shown in \cref{tab:cutonw15w10w5}
\begin{table}%
\centering%
\begin{tabular}{ c|l r|l r|l r }  
          &     \multicolumn{1}{|c}{$k$ ($\omega = 15$)} &  \multicolumn{1}{c|}{$\Delta\mathrm{dB}$} & \multicolumn{1}{|c}{$k$ ($\omega = 10$)} &  \multicolumn{1}{c|}{$\Delta\mathrm{dB}$}   &   \multicolumn{1}{|c}{$k$ ($\omega = 5$)} &  \multicolumn{1}{c}{$\Delta\mathrm{dB}$} \\
    \hline
    LNSE  &    -26.5036+0.2031\i       &                  &    -13.8388+1.0301\i        &               &    -0.5166-3.4788\i        &             \\
    Myers &    -26.5169+0.2312\i       &    1.1256        &    -13.8435+1.0375\i        &    0.0618     &    -0.5106-3.4081\i        &   -0.1783   \\
    HF    &    -26.5039+0.2026\i       &   -0.0214        &    -13.8393+1.0295\i        &   -0.0054     &    -0.5166-3.4783\i        &   -0.0012   \\
    OD    &    -26.5077+0.2070\i       &    0.1677        &    -13.8420+1.0351\i        &    0.0416     &    -0.5211-3.4781\i        &   -0.0016   \\
\end{tabular}
\caption{As for \cref{tab:cutonw10w5w2}, but with $m = 6$, $\Ren = 8\times 10^{6}$, $\delta = 5\times 10^{-3}$.}%
\label{tab:cutonw15w10w5}%
\end{table}%
for different parameters. We see that the high frequency asymptotics can achieve impressive accuracy even at low frequencies, and may even out-perform the $\mathcal{O}(\delta)$ asymptotics.

\begin{figure}%
\centering\includegraphics[width=0.6\textwidth]{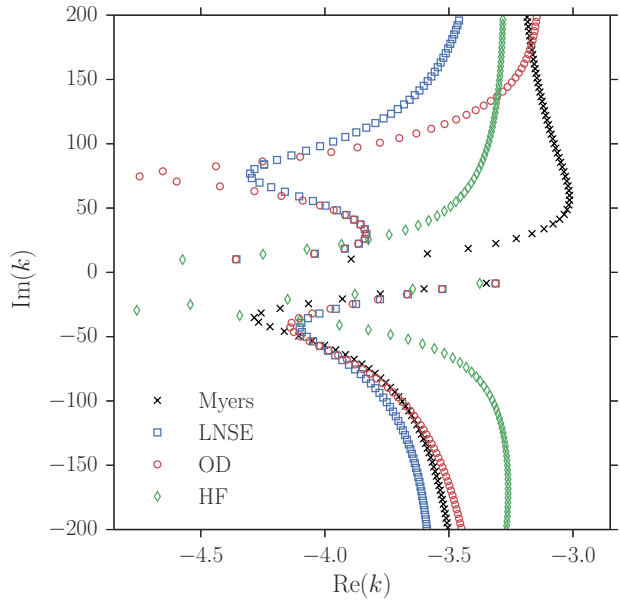}%
\caption{Cutoff modes in the $k$-plane of the Myers boundary condition, the LNSE numerics, the high frequency (HF) asymptotics \cref{Zeff_hf} and the $\mathcal{O}(\delta)$ (OD) asymptotics \cref{BL_governing_eqs} for $\omega = 5$, $m = 2$, $\Ren = 1\times 10^{5}$, $\delta = 4\times 10^{-3}$. The boundary impedance $Z$ is found using \cref{msdZ} with $d = 0.08$, $b = 6$ and $R = 1.6$.  Base profiles as in \cref{baseflow}.}%
\label{fig:cutoffw5}%
\end{figure}

In \cref{fig:cutoffw5},
the spectra of cut-off acoustic modes in the $k$-plane is plotted for the LNSE numerics, the Myers boundary condition and the two new asymptotic models, for the relatively low frequency $\omega = 5$.  The accuracy of the high frequency asymptotics is seen to have dropped with this reduction in frequency, although the behaviour of these cut-off modes is far less important than that of the cut-on modes considered above.

\subsection{The ratio $Z_{\mathrm{eff}}/Z$}

In this section the effective impedance $Z_{\mathrm{eff}}$ predicted by the high frequency asymptotics \cref{Zeff_hf} -- which accounts for a sheared and viscous boundary layer over an acoustic lining -- is compared with the boundary impedance $Z$.  This is done by considering the values of $\abs{Z_{\mathrm{eff}}/Z}$ and $\mathrm{arg}(Z_{\mathrm{eff}}/Z)$ over the complex $k$-plane for a given set of parameters.  The parameters $(Z, \omega, M, \Ren, \delta)$ are chosen to correspond to typical experimental facilities~\citep[e.g.][]{marxetal2010,jonesetal2005,auregan&leroux2008,renou&auregan2011}. Due to our scheme of nondimensionalisation, in which the frequency is scaled by the ratio of the speed of sound and the radius of the duct, $c_{0}^{*}/l^{*}$, the small ducts typically used in such facilities ($\sim 1\mathrm{cm}$ wide) lead to dimensionless frequencies that are too small to satisfy the asymptotic regimes assumed in the derivation of \cref{Zeff_hf}.  Therefore, we choose to scale the system up to the size of a typical aeroengine, where the fan diameters are typically $2$--$3.5\mathrm{m}$. The mean flow profiles used to evaluate \cref{Zeff_hf} are boundary layer expansions of the hyperbolic profiles \cref{baseflow},
\begin{align}
U(y) = M \tanh(y), & & T(y) = T_{0} + \tau \left(\cosh(y)\right)^{-1}, \label{ybaseflow}
\end{align}
where $y\in[0,16]$ is sufficient to capture the boundary layer, and we identify $\delta $ with the momentum thicknesses of the experimentally determined (fully turbulent) profiles.

\begin{figure}
\captionsetup[subfigure]{aboveskip=1pt}
    \centering
    \begin{subfigure}[t]{0.43\textwidth}
        \centering        
        \includegraphics*[width=1.\textwidth]{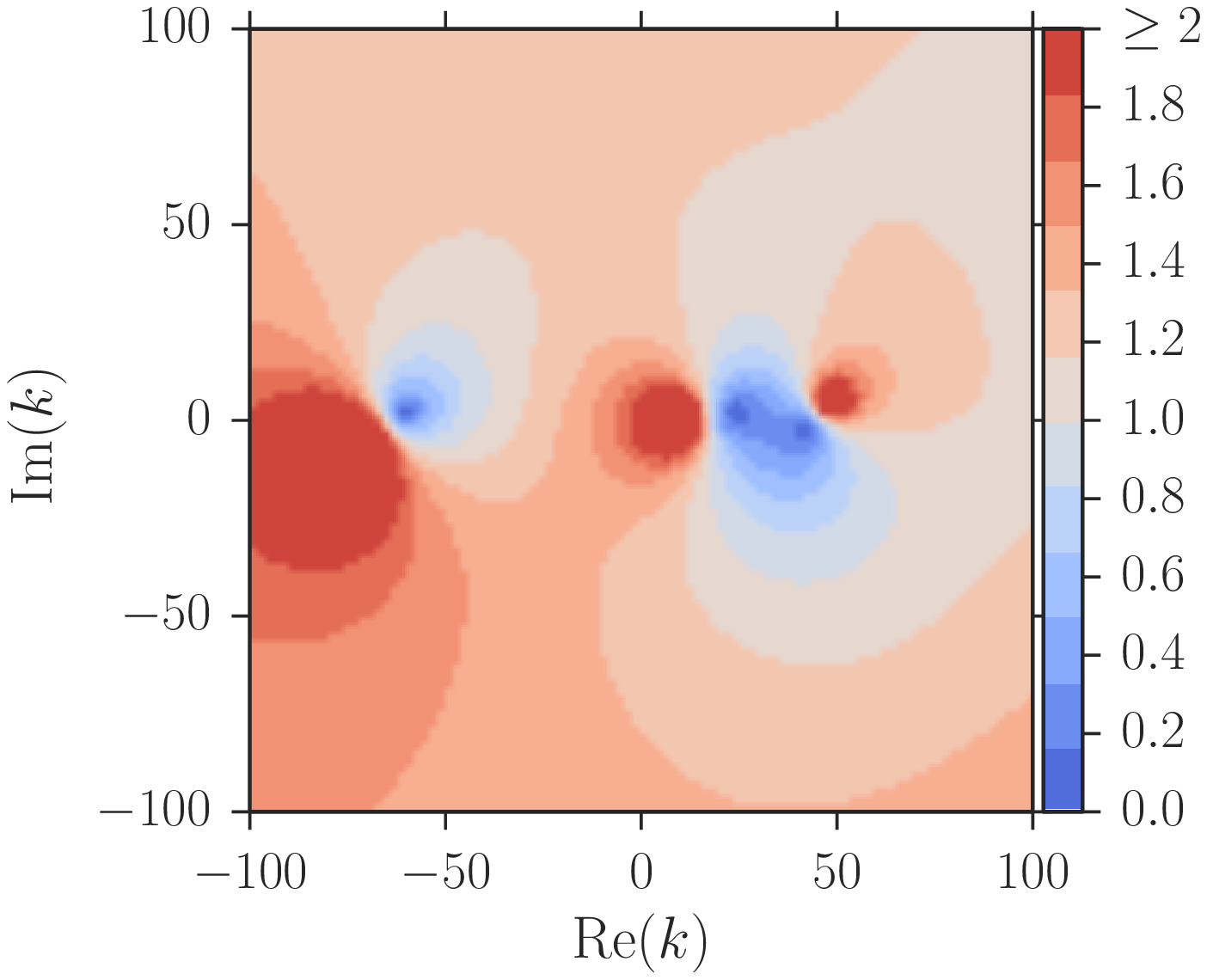}
        \caption{}
    \label{fig:11}
    \end{subfigure}%
    ~ 
    \begin{subfigure}[t]{0.45\textwidth}
        \centering
        \includegraphics*[width=1.\textwidth]{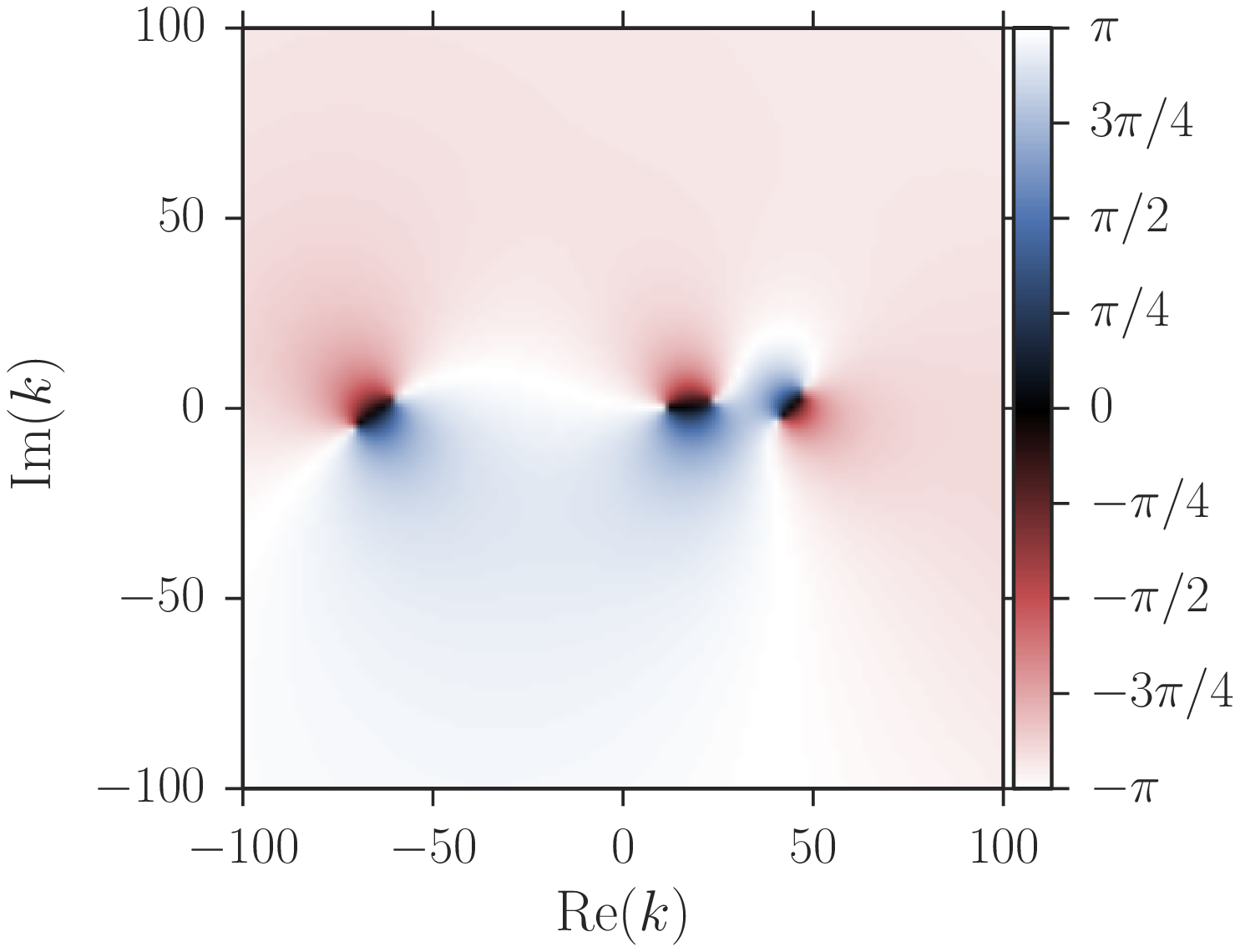}
        \caption{}
    \label{fig:12}
    \end{subfigure}%
        \caption{Contour plots in the $k$-plane of (a) $\abs{Z_{\mathrm{eff}}/Z}$, (b) $\mathrm{arg}(Z_{\mathrm{eff}}/Z)$ for the experimental setup of \citet{jonesetal2005}, with $f^{*} = 2500\mathrm{Hz}$, $Z = 0.93 - 1.43\i$, $M=0.335$.  For a duct radius $l^{*} = 1.5\mathrm{m}$ and sound speed $c_{0}^{*} = 340\mathrm{ms}^{-1}$, our dimensionless parameters are $\omega = 69.3$, $m = 0$, $\Ren = 3.4\times 10^{7}$, $\delta = 9\%$.  Base profiles as in \cref{ybaseflow}.} \label{fig:1}
\end{figure}

\begin{figure}
\captionsetup[subfigure]{aboveskip=1pt}
    \centering
    \begin{subfigure}[t]{0.43\textwidth}
        \centering        
        \includegraphics*[width=1.\textwidth]{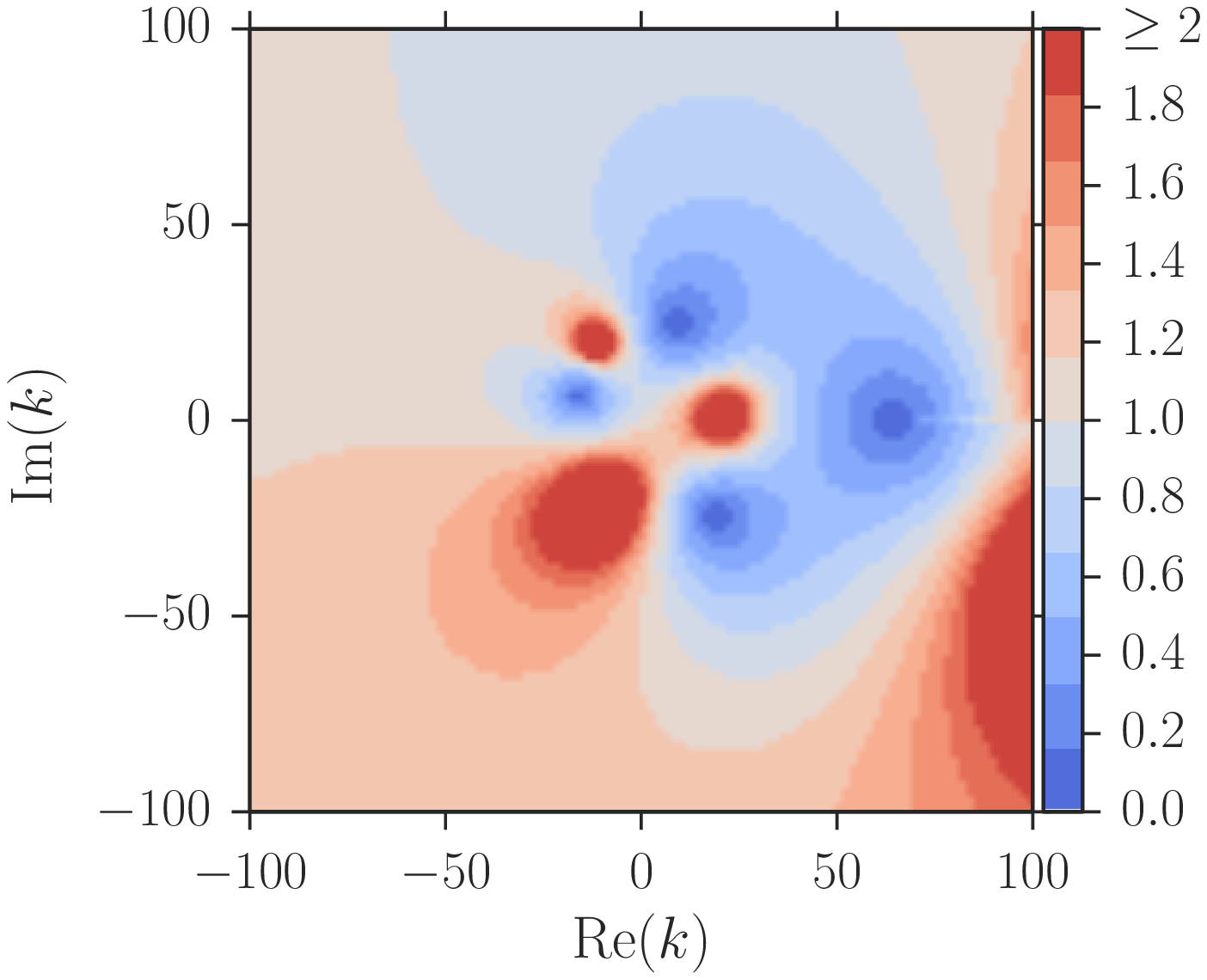}
        \caption{}
    \label{fig:21}
    \end{subfigure}%
    ~ 
    \begin{subfigure}[t]{0.45\textwidth}
        \centering
        \includegraphics*[width=1.\textwidth]{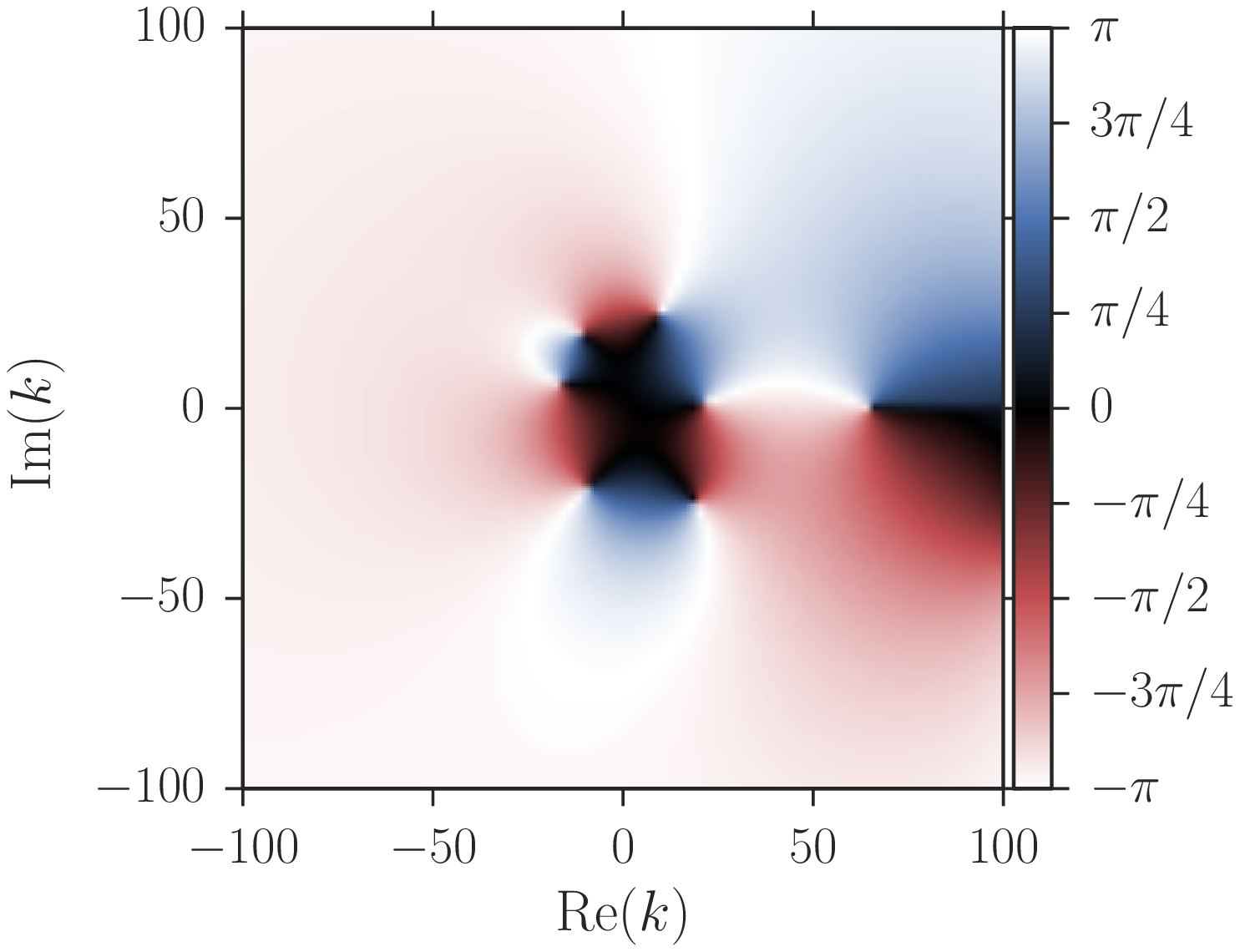}
        \caption{}
    \label{fig:22}
    \end{subfigure}%
    \caption{Contour plots in the $k$-plane of (a) $\abs{Z_{\mathrm{eff}}/Z}$, (b) $\mathrm{arg}(Z_{\mathrm{eff}}/Z)$ for the experimental setup of \citet{marxetal2010}, with $f^{*} = 1200\mathrm{Hz}$, $Z = 0.25 - 0.39\i$ and $M = 0.32$. For a duct radius $l^{*} = 1\mathrm{m}$ and sound speed $c_{0}^{*} = 363\mathrm{ms}^{-1}$, dimensionless parameters are $\omega = 20.8$, $m = 0$, $\Ren = 2.4\times 10^{7}$, $\delta = 5\%$.  Base profiles as in \cref{ybaseflow}.} \label{fig:2}
\end{figure}

\begin{figure}
\captionsetup[subfigure]{aboveskip=1pt}
    \centering
    \begin{subfigure}[t]{0.43\textwidth}
        \centering        
        \includegraphics*[width=1.\textwidth]{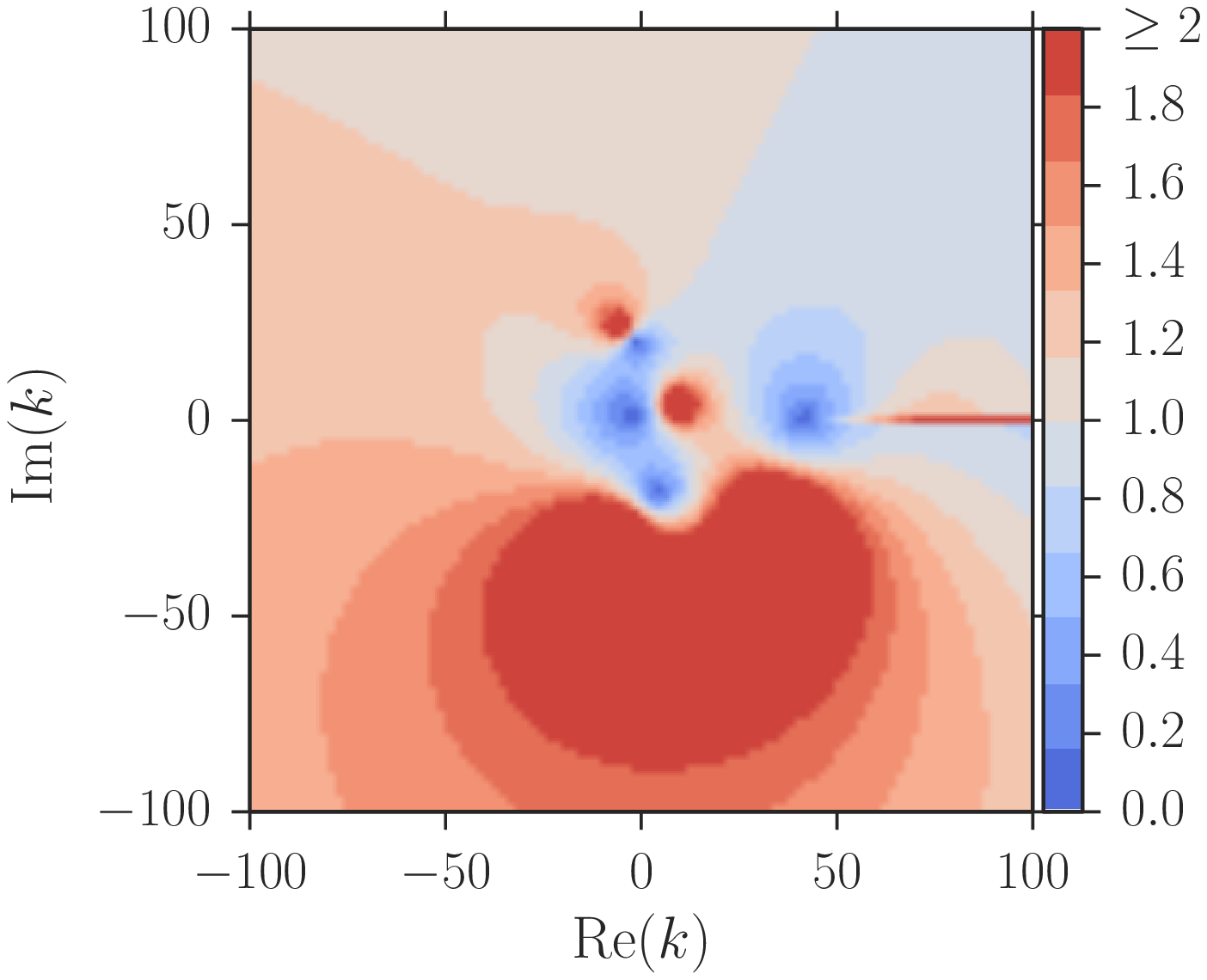}
        \caption{}
    \label{fig:31}
    \end{subfigure}%
    ~ 
    \begin{subfigure}[t]{0.45\textwidth}
        \centering
        \includegraphics*[width=1.\textwidth]{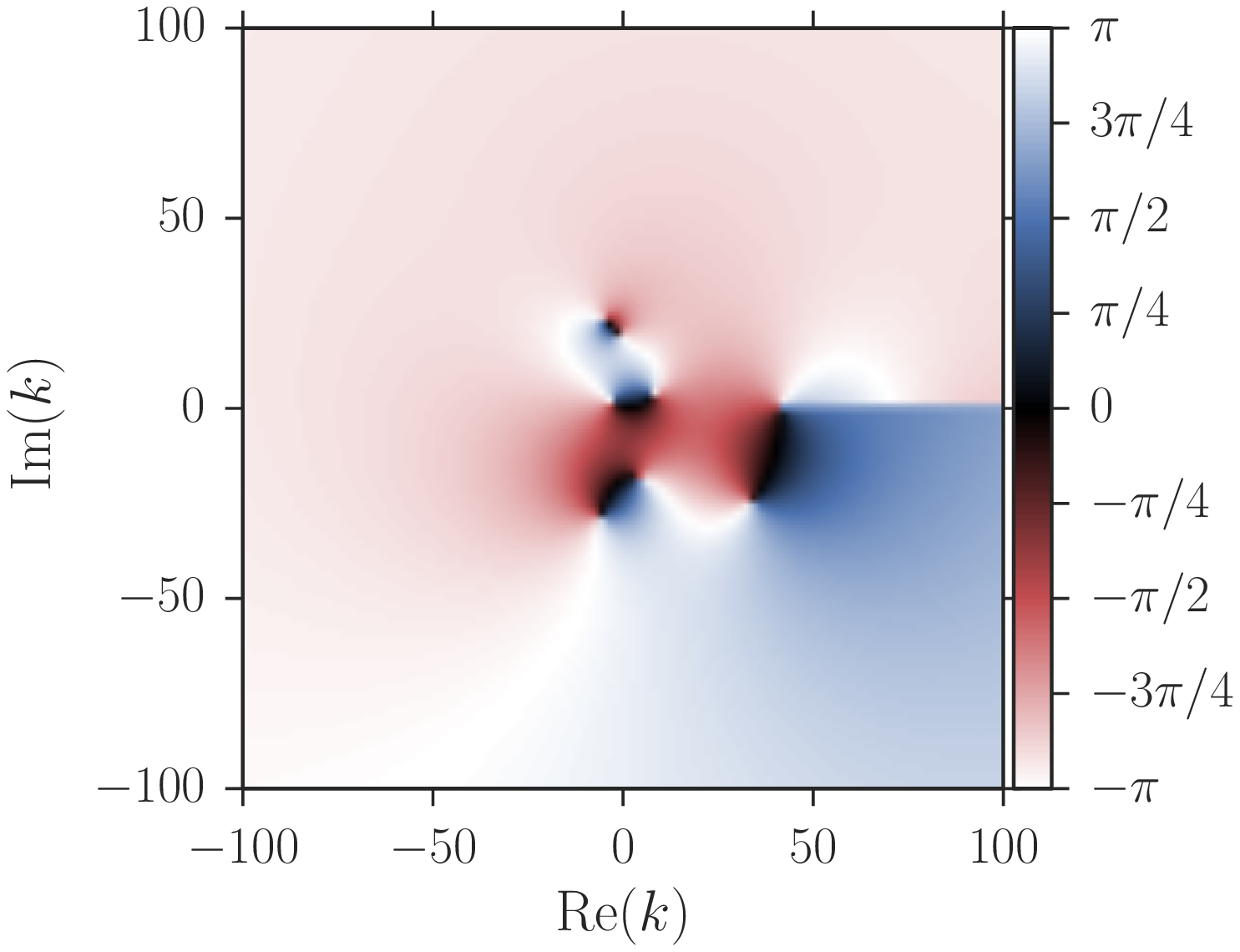}
        \caption{}
    \label{fig:32}
    \end{subfigure}%
        \caption{Contour plots in the $k$-plane of (a) $\abs{Z_{\mathrm{eff}}/Z}$, (b) $\mathrm{arg}(Z_{\mathrm{eff}}/Z)$ for the experimental setup of \citet{jonesetal2005}, with $f^{*} = 500\mathrm{Hz}$, $Z = 0.61 - 0.59\i$ and $M = 0.335$. For a duct radius $l^{*} = 1.5\mathrm{m}$ and sound speed $c_{0}^{*} = 340\mathrm{ms}^{-1}$, dimensionless parameters are $\omega = 13.9$, $m = 24$, $\Ren = 3.4\times 10^{7}$, $\delta = 9\%$.  Base profiles as in \cref{ybaseflow}.} \label{fig:3}
\end{figure}

\begin{figure}
\captionsetup[subfigure]{aboveskip=1pt}
    \centering
    \begin{subfigure}[t]{0.43\textwidth}
        \centering        
        \includegraphics*[width=1.\textwidth]{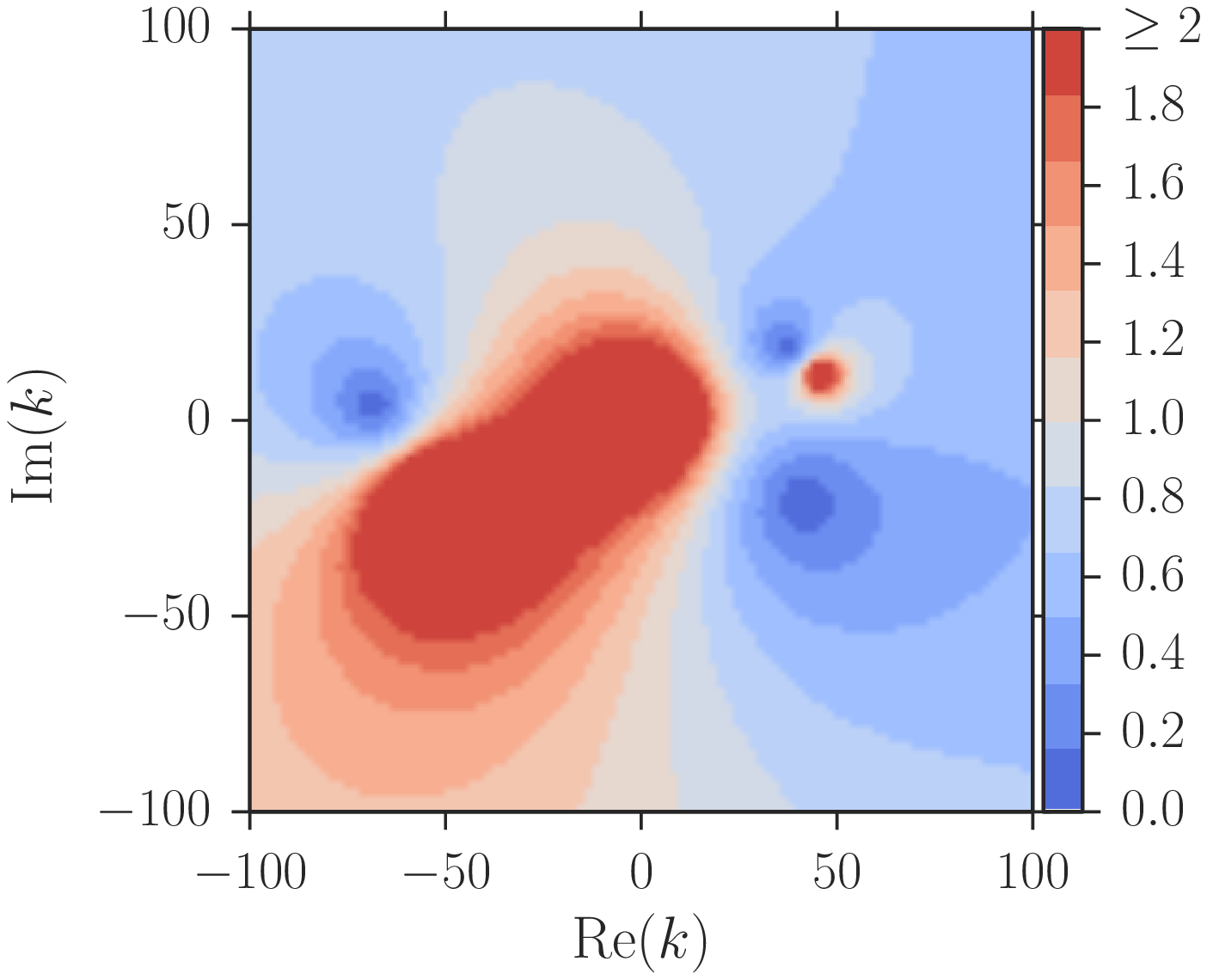}
        \caption{}
    \label{fig:41}
    \end{subfigure}%
    ~ 
    \begin{subfigure}[t]{0.45\textwidth}
        \centering
        \includegraphics*[width=1.\textwidth]{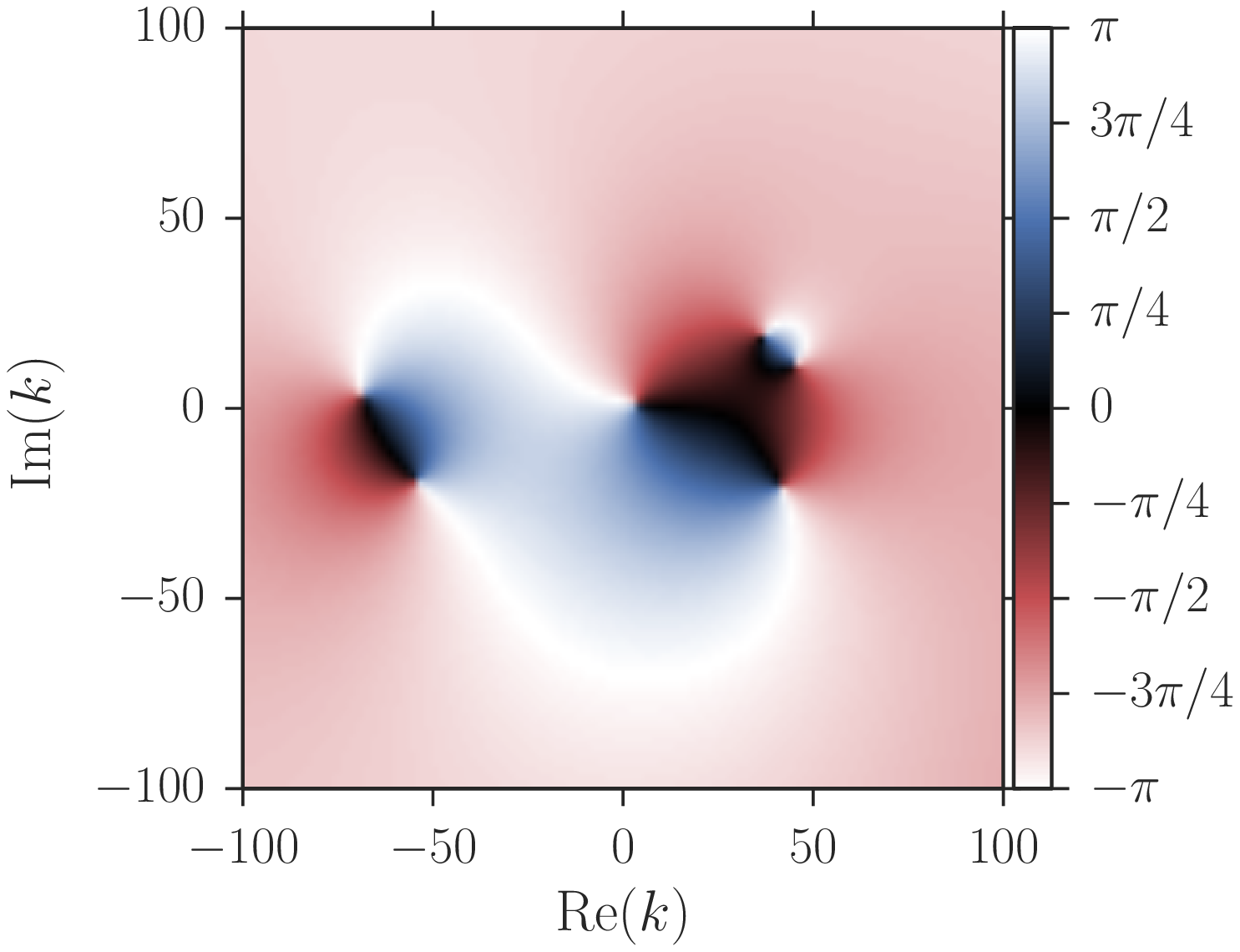}
        \caption{}
    \label{fig:42}
    \end{subfigure}%
    \caption{Contour plots in the $k$-plane of (a) $\abs{Z_{\mathrm{eff}}/Z}$, (b) $\mathrm{arg}(Z_{\mathrm{eff}}/Z)$ for the experimental setup of \citet{marxetal2010}, with $f^{*} = 5000\mathrm{Hz}$ and $M = 0.32$. The impedance $Z = 3.33 + 1\i$ was found by extrapolating to a higher frequency the expression of \citet{auregan&leroux2008}: $Z = -\i\breve{a} \cot(\breve{b}\omega^{*} + (1-\i)\breve{c} \sqrt{\omega^{*}})$, where $\breve{a} = 1.25$, $\breve{b} = 1.85\times10^{-4}$ and $\breve{c} = 2\times10^{-3}$.  For a duct radius $l^{*} = 1\mathrm{m}$ and sound speed $c_{0}^{*} = 363\mathrm{ms}^{-1}$, dimensionless parameters are $\omega = 86.5$, $m = 24$, $\Ren = 2.4\times 10^{7}$, $\delta = 5\%$.  Base profiles as in \cref{ybaseflow}.} \label{fig:4}
\end{figure}

\Cref{fig:1,fig:2,fig:3,fig:4} display results for two different experimental setups, with two parameter sets for each (details are given in the figure captions). \Cref{fig:11,fig:21} show that for the zeroth-order azimuthal mode ($m=0$) there can occur large areas of the complex $k$-plane where $\abs{Z_{\mathrm{eff}}/Z}$ lies close to unity. However, the corresponding plots of $\mathrm{arg}(Z_{\mathrm{eff}}/Z)$, \cref{fig:12,fig:22}, show that $Z_{\mathrm{eff}}$ and $Z$ commonly lie in different quadrants of the complex plane, hence their close relative magnitudes belie their disparity.  For the higher azimuthal order shown in \cref{fig:31,fig:41}, $\abs{Z_{\mathrm{eff}}/Z}\gtrsim2$ over large sections of the $k$-plane. This suggests that modes that rapidly vary in the azimuthal direction -- $m=24$ is indeed rapidly varying, but typical of rotor-alone noise at take-off~\citep{mcalpine2006} -- interact differently with the coupled boundary layer--acoustic lining system to a plane wave, say, and hence see an appreciably different effective impedance.

\section{Conclusion}

Although it is not always feasible to include viscous and thermal conductive effects in all aeroacoustic computations, we believe there should be more careful thought about the situations in which they should be included, and an understanding of the size of the errors introduced by omitting them.  Certainly, at low frequencies and for thin boundary layers the acoustics can be significantly affected by viscosity, with errors being introduced by its neglect that are of the same order of magnitude as the errors introduced by neglecting shear (e.g.\ \cref{fig:lowfreqv2}).  In general, the damping of upstream propagating well cut-on modes is found to be poorly predicted by inviscid numerics (see \cref{fig:kmodes1,fig:kmodesasym}), showing that small errors can lead to significant variations in important small quantities.  Viscosity is also indispensable when investigating the physical onset of instability, and in particular the growth rate of the instability is strongly dependent on viscosity (see \cref{fig:w_unstablemodes}).  The flow appears to be totally stabilized if viscosity is strong enough, which for the parameters considered here (see \cref{fig:w_unstablemodes}) occurred at a Reynolds numbers around $\Ren=4\times 10^5$ .  The existence of the anomalous region~\citep{brambley2011b}, located for $\Re(k) > \omega/M$, $\Im(k) < 0$, can also lead to significant errors in inviscid computations.  

When full viscous numerics are not practical, it should be possible to make use of the asymptotic models presented here in frequency domain computations.  The boundary layer model in \cref{sec:blasymp} may be seen as an extension of the viscous Myers model of \citet{brambley2011b} in order to account for a nonzero-thickness viscous shear layer by including $\mathcal{O}(\delta)$ terms, or indeed as an extension of the inviscid modified Myers model of \citet{brambley2011a} to include viscothermal terms --- in effect, \cref{sec:blasymp} gives a viscous modified Myers condition.  Although no closed-form solution for the effective impedance can be found for such a model, the reduced governing equations could be incorporated into an inviscid code as a boundary solver. The relatively few assumptions that are made in the derivation of the model \cref{BL_governing_eqs} mean the resulting boundary condition performs well for a wide range of parameters, and also yields the correct stability properties.  If one is concerned with high frequency sound (such as is common in aeroacoustics), the analytical effective impedance boundary condition \cref{Zeff_hf} may be easily applied in the frequency domain at the wall of an inviscid uniform flow.  It is shown in \cref{sec:lowhighfreq} that the high frequency asymptotics predict the cut-on modes of the linearised compressible Navier--Stokes equations well even at relatively low frequencies down to $\omega \sim \mathcal{O}(1)$. 

Analytical modelling of sound propagation in lined ducts with uniform flow, where the acoustic pressure modes may be written in terms of Bessel functions, requires an impedance boundary condition to be applied at the duct wall in order to form a dispersion relation.  By applying the high frequency boundary condition \cref{Zeff_hf}, it would be possible to form analytical mode shapes for acoustics that account for a thin-but-finite-thickness viscothermal shear layer above the lining.

The success of commonly-used impedance eduction techniques, which connect the ``far-field'' acoustic response of a liner to its on-surface impedance, depends upon the quality of the liner model employed. The accuracy with which the closed-form high-frequency liner model \cref{Zeff_hf} predicts cut-on modes (see \cref{fig:leastcutoffasym}) suggests that this model would be a useful tool in such impedance eduction methods, if the asymptotic regimes are respected. Moreover, the high-frequency model has more degrees of freedom (the $\delta I_{j}$ integrals) than previous viscous models (such as that of \citet{aureganetal2001}), allowing a better fit to the data to be achieved.

The temporal stability properties of the linearised compressible Navier--Stokes equations are well-approximated by the $\mathcal{O}(\delta)$ boundary layer model \cref{BL_governing_eqs}: the model is wellposed and has a well-defined maximum growth rate of instability, and captures the damping of small wavelengths by viscosity.  The high frequency asymptotic model derived in \cref{sec:highfreq} is not well-suited to temporal stability analysis due to the scaling assumptions $\omega \gg 1$ and $k/\omega \lesssim \mathcal{O}(1)$ made in its derivation; this suggests that it may not perform well if adapted to time domain applications, but should be suitable for use in the frequency domain where temporal instability is excluded.

We present results assuming a homogeneous impedance boundary, which may not be achieved in practice.  Ongoing numerical investigations~\citep{tametal2014,zhang&bodony2016} may help in this regard, and illumine the important effects of inhomogeneities that could be included in theoretical studies. Moreover, we assume a thin boundary layer with no axial variation. Comparison of these results to DNS and experiments~\citep[such as][]{alomar&auregan2016} would help validate these assumptions.  Measurements of parameter values from aeroengines in flight (in particular boundary layer thicknesses) would be useful both to inform the relevant asymptotic regimes of interest in future theoretical work, and to predict the impact of the current theoretical work on aeroengine noise.

One aspect lacking from the current study is the effect of turbulence within a turbulent boundary layer on the acoustics.  Since in most cases high Reynolds number boundary layers are turbulent, this would be an interesting avenue of future research.  One way to address this would be by altering the choice of viscous function $\H$ in \cref{visctherm} to contain some radially dependent eddy viscosity, which could be inserted into the current work with no great difficulty.  Another worthwhile extension to this study would be an investigation of the absolute stability of viscous flow over a liner. An absolute instability would dominate a convective instability for large times; gaining an understanding of the effect of viscosity on the absolute stability properties may therefore be valuable, although our expectation is that absolute instabilities will be confined to extremely thin boundary layers at extremely high Reynolds numbers~\citep{brambley2013}.

\section*{Acknowledgements}

E.J.B. gratefully acknowledges support from a Royal Society University
Research Fellowship, and from a college lectureship from Gonville~\&
Caius College, Cambridge.
D.K. was supported by an EPSRC grant.

\appendix

\section{Numerical boundary conditions and extrapolating to infinity} \label{sec:A}

Equation~\cref{BL_governing_eqs} gives the governing equations for the acoustics in a thin boundary layer $y>0$, $r=1-\delta y$.  Outside this boundary layer, as $y \to \infty$, the mean flow is considered uniform and the outer inviscid acoustic solution is given in \cref{uniform}.  Here we consider matching the two.  To aid this matching, we assume that the mean flow varies within the boundary layer only for $y<Y$, within which region \cref{BL_governing_eqs} must be solved numerically.  For $y>Y$, however, the governing equations may be solved analytically. The solutions may be used to extend the numerical solutions found in $y\in[0,Y]$ and extrapolate them in the limit $y\to\infty$ in order to match with the outer inviscid acoustic solution \cref{uniform}.  For $y>Y$, the governing equations \cref{BL_governing_eqs} reduce to 
\begin{subequations}
\begin{gather}
\xi \h{u}_{yy} - \i(\omega - M k)\h{u} = \delta \left[\xi \h{u}_{y} - \i k\t{p}\right], \label{oBL_xmom_eq} \\
\frac{\xi}{\mathrm{Pr}} \h{T}_{yy} - \i (\omega - Mk)\h{T} = \delta \left[\frac{\xi}{\mathrm{Pr}}\h{T}_{y} - \i(\omega - M k)\t{p}\right], \label{oBL_energy_eq} \\
\xi \t{w}_{yy} - \i (\omega - M k)\t{w} = -\i m\t{p}, \label{oBL_thetamom_eq}\\
\t{v}_{y} = -\i (\omega - M k)(\gamma - 1)\h{T} - \i k\h{u} + \delta \left[\i \gamma (\omega - M k)\t{p} + \t{v} - \i m\t{w}\right], \label{oBL_mass_eq}\\
\t{p}_{y} = \delta \left[\i (\omega - M k)\t{v} - \xi (2 + \beta)\t{v}_{yy} - \i k \xi (1 + \beta) \h{u}_{y}\right], \label{oBL_rmom_eq}
\end{gather}
\label{oBL_governing_eqs}%
\end{subequations}
which are uncoupled.  At leading order these have the bounded solutions
\begin{subequations}
\begin{gather}
\h{u}_{0}(y) = \bar{u}_{0}\mathrm{e}^{-\eta_{\infty} y}, \quad\quad \h{T}_{0}(y) = \bar{T}_{0}\mathrm{e}^{-\sigma \eta_{\infty} y}, \quad\quad \t{w}_{0}(y) = \bar{w}\mathrm{e}^{-\eta_{\infty} y} + \frac{m p_{0}}{\omega - Mk}, \label{u0t0w0} \\
\t{v}_{0}(y) = \bar{v}_{0} + \i (\omega - M k) \frac{(\gamma - 1)}{\sigma \eta_{\infty}}\bar{T}_{0}\mathrm{e}^{-\sigma \eta_{\infty} y} + \frac{\i k}{\eta_{\infty} }\bar{u}_{0}\mathrm{e}^{-\eta_{\infty} y}, \quad\quad \t{p}_{0} = p_{0}, \label{v0p0}
\end{gather}
\label{oBL_0}%
\end{subequations}
for some constants $\bar{u}_{0}$, $\bar{T}_{0}$, $\bar{w}$, $\bar{v}_{0}$.  At first order the solutions are
\begin{subequations}
\begin{align}
\h{u}_{1}(y) =&\, \bar{u}_{1}\mathrm{e}^{-\eta_{\infty} y} + \frac{1}{2}\bar{u}_{0} y \mathrm{e}^{-\eta_{\infty} y} + \frac{k p_{0}}{\omega - Mk}, \label{ou1} \\ 
\h{T}_{1}(y) =&\, \bar{T}_{1}\mathrm{e}^{-\sigma \eta_{\infty} y} + \frac{1}{2}\bar{T}_{0} y \mathrm{e}^{-\sigma \eta_{\infty} y} + p_{0}, \label{ot1} \\
\t{v}_{1}(y) = &\, \bar{v}_{1} + \!\left(\!\bar{v}_{0} + \frac{(\omega - M k)^{2} - k^{2} - m^{2}}{\omega - M k}\i p_{0}\!\right)\!y + \i (\omega - M k) \frac{(\gamma - 1)}{\sigma \eta_{\infty}}(\h{T}_{1} - p_{0}) \notag \\
&\,+ \frac{\i k}{\eta_{\infty} }\!\left(\!\h{u}_{1} - \frac{k p_{0}}{\omega - Mk}\right)\! + \frac{\i m}{\eta_{\infty} }\!\left(\!\t{w}_{0} - \frac{m p_{0}}{\omega - Mk}\right)\! - \frac{\xi (\gamma - 1)}{2\sigma^{2}}\h{T}_{0} - \frac{\xi k}{2(\omega - M k)}\h{u}_{0}, \label{ov1}\\
\t{p}_{1}(y) =&\, \bar{p}_{1} + \i (\omega - M k) \bar{v}_{0} y + \i \xi (\omega - M k)(\gamma - 1)\big(2 + \beta - 1/\Pr \big)\h{T}_{0}, \label{op1}
\end{align}
\label{oBL_1}%
\end{subequations}
where $\sigma ^{2} = \Pr$ and $\eta_{\infty} ^{2} = \i(\omega - Mk)/\xi$, with $\Re(\eta_{\infty})>0$, and some constants $\bar{u}_{1}$, $\bar{T}_{1}$, $\bar{v}_{1}$ and $\bar{p}_{1}$.  Note that $\eta_{\infty}$ has a branch cut along $k = \omega/M - \i q$ for $q\geq 0$ to ensure that the solutions remain bounded as $y\to\infty$.  In the limit $y\to \infty$, the relations \cref{u0t0w0}, \cref{ou1,ot1} give boundary conditions on $\h{u}$, $\h{T}$ and $\t{w}$,
\begin{equation}
\begin{aligned}
\h{u}_{0}(y) &\to 0,& \,\,
\h{T}_{0}(y) &\to 0,& \,\,
\t{w}_{0}(y) &\to \frac{m p_{0}}{\omega - Mk}, &\qquad \!\!
&\text{as } y \to \infty,
\end{aligned}
\label{bcs0A}
\end{equation}
at $\mathcal{O}(1)$, and
\begin{equation}
\begin{gathered}
\h{u}_{1}(y) \to \frac{k p_{0}}{\omega - Mk}, \qquad\qquad
\h{T}_{1}(y) \to p_{0}, \qquad\qquad
\text{as } y \to \infty,
\end{gathered}
\label{bcs1A}
\end{equation}
at $\mathcal{O}(\delta )$.

To form the effective impedance we match the solutions \cref{v0p0}, \cref{ov1,op1} in the limit $y\to \infty$ to the outer solutions, which are the uniform inviscid acoustics outside the boundary layer, \cref{uniform}. The outer solutions may be expanded near the lining to give
\begin{align}
\tilde{p}_{\mathrm{ui}}(1-\delta y) &= \pinf - \delta y \pinf' + \mathcal{O}(\delta ^{2}),& \tilde{v}_{\mathrm{ui}}(1-\delta y) &= \vinf - \delta y \vinf' + \mathcal{O}(\delta ^{2}), \label{outer1A}
\end{align}
where the derivatives $\pinf'$, $\vinf'$ may be rewritten
\begin{align}
\pinf' &= -\i (\omega -Mk)\vinf,& \vinf' &= \frac{(\omega -Mk)^{2} - k^{2} - m^{2}}{\i (\omega -Mk)}\pinf - \vinf. \label{pvinfdash}
\end{align}
Since we have applied a known normalisation at the lining -- causing constant terms to arise at $\mathcal{O}(\delta)$ in the boundary layer solutions -- we must expand $\pinf = \pinf^{(0)} + \delta \pinf^{(1)}$ and similarly for $\vinf$. Hence the expansions \cref{outer1A} become
\begin{subequations}
\begin{align}
p_{\mathrm{ui}}(1-\delta y) =&\, \pinf^{(0)} + \delta \pinf^{(1)} + \delta y \i (\omega + Mk)\vinf^{(0)} + \mathcal{O}(\delta ^{2}), \\
v_{\mathrm{ui}}(1-\delta y) =&\, \vinf^{(0)} + \delta \vinf^{(1)} + \delta y \!\left(\!\vinf^{(0)} + \frac{(\omega - M k)^{2} - k^{2} - m^{2}}{(\omega - Mk)}\i \pinf^{(0)}\!\right)\! + \mathcal{O}(\delta ^{2}).
\end{align}
\label{outer2A}%
\end{subequations}
These are the outer solutions to which we match our numerical solutions of \cref{BL_governing_eqs} in the limit $y \to \infty$, via the analytical solutions for $y>Y$. The numerical solutions may be found by, for instance, discretising the domain and approximating the $y$ derivatives using finite differences.

If we are solving in a finite numerical domain $y\in[0,Y]$, we may use the relations \cref{v0p0}, \cref{ov1,op1} to extrapolate our solutions out to infinity.  At leading order this is simple due to the exponentially decaying terms; we identify $\bar{v}_{0}$ with $\vinf^{(0)}$ and rearrange \cref{v0p0} for $\vinf^{(0)}$ to find
\begin{equation}
\vinf^{(0)} = \t{v}_{0}(Y) - \frac{\eta_{\infty} \xi}{\sigma }(\gamma - 1)\h{T}_{0}(Y) - \frac{\i k}{\eta_{\infty} }\h{u}_{0}(Y).
\end{equation}
For the pressure we simply find $\pinf^{(0)} = p_{0}$.  At first order consider the following: if
\begin{equation}
\t{v}_{1}(y) = \bar{v}_{1} + (a y + b)\mathrm{e}^{-\eta_{\infty} y} + (c y + d)\mathrm{e}^{-\sigma \eta_{\infty} y} + e y,
\end{equation}
which is the form of \cref{ov1}, then in the limit $y\to\infty$
\begin{equation}
\t{v}_{1}(y) \sim \bar{v}_{1} + e y.\label{eY}
\end{equation}
Evaluating \cref{ov1} at $y=Y$ and rearranging to leave $\bar{v}_{1}$ and the terms linear in $Y$ on the right hand side, as in \cref{eY}, gives
\begin{equation}
\t{v}_{1}(Y) - (a Y + b)\mathrm{e}^{-\eta_{\infty} Y} - (c Y + d)\mathrm{e}^{-\sigma \eta_{\infty} Y} = \bar{v}_{1} + e Y. \label{extrapolate}
\end{equation}
We may identify $\bar{v}_{1}$ with $\vinf^{(1)}$ and $\bar{p}_{1}$ with $\pinf^{(1)}$, and thus use the extrapolated forms of \cref{ov1,op1} -- which are of the form \cref{extrapolate} -- to rearrange for $\pinf^{(1)}$ and $\vinf^{(1)}$:
\begin{subequations}\begin{align}
\pinf^{(1)} =&\, \t{p}_{1}(Y) - \i (\omega - M k) Y \vinf^{(0)} - \i \xi (\omega - M k) (\gamma - 1)\left(2 + \beta - \frac{1}{\Pr}\right)\h{T}_{0}(Y), \label{pinf1}\\
\vinf^{(1)} =&\, \t{v}_{1}(Y) - \left(\!\vinf^{(0)} + \frac{(\omega - M k)^{2} - k^{2} - m^{2}}{(\omega - Mk)}\i \pinf^{(0)}\!\right)Y \notag\\
&\,- \frac{\i (\omega - M k) (\gamma - 1)}{\sigma \eta_{\infty}} \left(\h{T}_{1}(Y) - \pinf^{(0)}\right) - \frac{\i k}{\eta_{\infty}} \left(\h{u}_{1}(Y) - \frac{k \pinf^{(0)}}{\omega - M k}\right)\notag \\ 
&\, - \frac{\i m}{\eta_{\infty}} \left(\t{w}_{0}(Y) - \frac{m \pinf^{(0)}}{\omega - M k}\right) + \frac{\xi (\gamma - 1)}{2 \sigma^{2}} \h{T}_{0}(Y) + \frac{\xi k}{2 (\omega - M k)} \h{u}_{0}(Y). \label{vinf1}
\end{align}\end{subequations}
The effective impedance is then given by
\begin{equation}
Z_{\mathrm{eff}} = \frac{\pinf^{(0)} + \delta \pinf^{(1)}}{\vinf^{(0)} + \delta \vinf^{(1)}};
\end{equation}
this is the function used in the dispersion relation \cref{asym_disp} to find eigenmodes of the $\mathcal{O}(\delta)$ asymptotics.

\section{Solving the high frequency boundary layer equations}\label{sec:B}

Here we solve equations \cref{hf_eqs} to $\mathcal{O}(\ee ^{2})$.  At leading order we find
\begin{gather}
\begin{aligned}
\t{v}_{0}(y, \theta ) &= A_{0}(y),&&\qquad& \hh{u}_{0}(y, \theta ) &= B_{0}(y)\mathrm{e}^{-\theta } - \frac{\i U_{y}}{1-UL}A_{0}(y),  \\
\t{p}_{0}(y, \theta ) &= F_{0}(y),&&\qquad& \hh{T}_{0}(y, \theta ) &= D_{0}(y)\mathrm{e}^{-\sigma \theta } - \frac{\i T_{y}}{1-UL}A_{0}(y),
\end{aligned}\\
\tilde{w}_{0} = G_{0}(y)\mathrm{e}^{-\theta } + \frac{N}{\rho (1-UL)}F_{0}(y),\notag
\end{gather}
where exponentially growing solutions have been excluded.  The explicit $y$ and $\theta $ dependencies will be dropped henceforth. Homogeneous boundary conditions on $\hh{u}_{0}$, $\hh{T}_{0}$ and $\t{w}_{0}$ at $y=0$ give
\begin{align}
B_{0}(0) &= \i U_{y}(0)A_{0}(0),& D_{0}(0) &= \i T_{y}(0)A_{0}(0) = 0,& G_{0}(0) &= -\frac{N}{\rho (0)}F_{0}(0). \label{leading_bcs}
\end{align}
The $D_{0}(0)=0$ relation arises from our isothermal boundary condition $T_{y}(0) = 0$.  Matching $\t{p}$ and $\t{v}$ to the outer solution will fix the values of $A_{0}(0)$ and $F_{0}(0)$ in \cref{sec:B1}, and similarly at subsequent orders.

At first order, we find secularity conditions by disallowing resonant terms. The first order $\t{v}$ equation is
\begin{equation}
\t{v}_{1,\theta} = -\frac{\i (1 - U L)}{\eta T}D_{0}\mathrm{e}^{-\sigma \theta } - \frac{\i L}{\eta }B_{0}\mathrm{e}^{-\theta } - \Big\{\frac{T_{y}}{\eta T}A_{0} - \frac{L U_{y}}{\eta (1 - U L)}A_{0} - \frac{T}{\eta }\Big(\frac{A_{0}}{T}\Big)_{y}\Big\}, \label{v1eq}
\end{equation}
where the curly brackets enclose terms that are functions of $y$ only, and hence are resonant.  To prevent powers of $\theta $ arising, we equate the curly brackets with zero and form the secularity condition for $A_{0}(y)$:
\begin{equation}
A_{0,y} - 2\big(\ln{\eta T}\big)_{y}A_{0} = 0 \quad\quad \Rightarrow \quad\quad A_{0}(y) = \bar{A}_{0}(1 - U L), \label{A0sec}
\end{equation}
where $\bar{A}_{0}$ is a constant. In going from \cref{v1eq} to \cref{A0sec}, the definitions of $\eta (y)$, \cref{eta}, and its derivative are used.  Similarly, the first order $\t{p}$ equation is 
\begin{equation}
\t{p}_{1,\theta} = -\frac{1}{\eta }F_{0,y}, \label{p1eq}
\end{equation}
where the right hand side is a function of $y$ only and hence resonant. As above, we set this to zero to form the secularity condition for $F_{0}(y)$:
\begin{equation}
F_{0}(y) = \bar{F}_{0}, \label{F0sec}
\end{equation}
where $\bar{F}_{0}$ is a constant.  To ascertain $\bar{A}_{0}$ and $\bar{F}_{0}$ we could, for instance, force $\t{v}$ and $\t{p}$ to satisfy some impedance condition at the wall $y=0$; or match to a known solution outside the boundary layer in the limit $y\to \infty$. Solving at first order now gives
\begin{align}
\t{v}_{1} &= A_{1} + \frac{\i L}{\eta } B_{0}\mathrm{e}^{-\theta } + \frac{\i (1 - U L)}{\sigma \eta T}D_{0}\mathrm{e}^{-\sigma \theta },& \t{p}_{1} &= F_{1}.
\end{align}
Expanding the first order equation for $\hh{u}$ we find
\begin{equation}
\begin{multlined}[.85\displaywidth][c]\hh{u}_{1,\theta \theta} - \hh{u}_{1} = \frac{\i U_{y}}{1 - U L}\Big(A_{1} + \frac{\i L}{\eta}B_{0}\mathrm{e}^{-\theta } + \frac{\i (1 - U L)}{\sigma \eta T}D_{0}\mathrm{e}^{-\sigma \theta }\Big) - \frac{\dd L}{\rho (1 - U L)}F_{0} \\
+ \frac{1}{\eta^{2} T}\Big((\eta T B_{0})_{y} + \eta T B_{0,y}\Big)\mathrm{e}^{-\theta } + \frac{\sigma U_{y}}{\eta T}D_{0}\mathrm{e}^{-\sigma \theta }. \end{multlined} \label{u1eq}
\end{equation}
The resonant\footnote{We use the term `resonant' here even though these resonant terms are exponentially decaying.} terms on the right hand side of \cref{u1eq} are those $\propto \exp{(-\theta)}$.  Equating the resonant terms with zero, we find
\begin{equation}
\frac{1}{\eta T}(\eta T B_{0})_{y} - \frac{L U_{y}}{1 - U L}B_{0} + B_{0,y} = 0,
\end{equation}
which may be written
\begin{equation}
B_{0,y} + \frac{3}{2}\big(\ln{\eta T}\big)_{y}B_{0} = 0 \quad\quad \Rightarrow \quad\quad B_{0}(y) = B_{0}(0)(1 - U L)^{-3/4}. \label{B0sec}
\end{equation}
In the same vein, the secularity condition for $G_{0}$ can be found from the first order $\t{w}$ equation
\begin{equation}
\t{w}_{1,\theta \theta} - \t{w}_{1} = -\frac{N}{\rho (1 - U L)}F_{1} + \frac{1}{\eta^{2} T}\Big((\eta T G_{0})_{y} + \eta T G_{0,y}\Big)\mathrm{e}^{-\theta }, \label{w1eq}
\end{equation}
where again the secular terms are $\propto \exp{(-\theta)}$. This leads to
\begin{equation}
G_{0,y} + \frac{1}{2}\big(\ln{\eta T}\big)_{y}G_{0} = 0 \quad\quad \Rightarrow \quad\quad G_{0}(y) = G_{0}(0)(1 - U L)^{-1/4}. \label{G0sec}
\end{equation}
For the first order $\hh{T}$ equation
\begin{equation}
\begin{multlined}[.85\displaywidth][c]\frac{1}{\Pr} \hh{T}_{1,\theta \theta} - \hh{T}_{1} = \frac{\i T_{y}}{1 - U L}\Big(A_{1} + \frac{\i L}{\eta}B_{0}\mathrm{e}^{-\theta } + \frac{\i (1 - U L)}{\sigma \eta T}D_{0}\mathrm{e}^{-\sigma \theta }\Big) - \frac{\dd}{\rho}F_{0} \\
+ \frac{1}{\sigma \eta^{2} T}\Big((\eta T D_{0})_{y} + \eta (T D_{0})_{y}\Big)\mathrm{e}^{-\sigma \theta } + \frac{2 U_{y}}{\eta }B_{0}\mathrm{e}^{-\theta }, \end{multlined}
\end{equation}
the secular terms are $\propto \exp{(-\sigma \theta)}$. Equating these with zero we find
\begin{equation}
\frac{1}{\eta T}(\eta T D_{0})_{y} + \frac{1}{T}(T D_{0})_{y} - \frac{T_{y}}{T}D_{0} = 0,
\end{equation}
which may be written
\begin{equation}
D_{0,y} + \frac{1}{2}\big(\ln{\eta T}\big)_{y}D_{0} = 0 \quad\quad \Rightarrow \quad\quad D_{0}(y) = D_{0}(0)(1 - U L)^{-1/4}. \label{D0sec}
\end{equation}
In fact, the boundary condition $D_{0}(0) = 0$ from \cref{leading_bcs} tells us that $D_{0}(y) \equiv 0$. This is a direct consequence of our isothermal wall condition $T_{y}(0) = 0$.  The first order solutions for $\hh{u}$, $\hh{T}$ and $\t{w}$ are then
\begin{gather}
\hh{u}_{1} = B_{1}\mathrm{e}^{-\theta } - \frac{\i U_{y}}{1-UL}A_{1} + \frac{\dd L}{\rho (1-UL)}F_{0}, \qquad\qquad \t{w}_{1} = G_{1}\mathrm{e}^{-\theta } + \frac{N}{\rho (1-UL)}F_{1}, \notag\\
\hh{T}_{1} = D_{1}\mathrm{e}^{-\sigma \theta } + d_{0}\mathrm{e}^{-\theta } - \frac{\i T_{y}}{1-UL}A_{1} + \frac{\dd}{\rho }F_{0},
\end{gather}
where
\begin{equation}
d_{0}(y) = \frac{\Pr}{1 - \Pr}\Big(2 U_{y} - \frac{L T_{y}}{1-UL}\Big)\frac{B_{0}}{\eta}. \label{d0}
\end{equation}
No slip and isothermal wall boundary conditions at first order lead to
\begin{equation}
\begin{gathered}
B_{1}(0) = \i U_{y}(0)A_{1}(0) - \frac{\dd L}{\rho(0)}F_{0}(0), \quad\quad
D_{1}(0) = -d_{0}(0) - \frac{\dd}{\rho(0)}F_{0}(0), \\
G_{1}(0) = -\frac{N}{\rho (0)}F_{0}(0).
\end{gathered}
\label{first_bcs}%
\end{equation}

At second order we find the secularity conditions for $A_{1}$ and $F_{1}$ by the same method as the preceding order, giving
\begin{subequations}
\begin{gather}
A_{1,y} - 2\big(\ln{\eta T}\big)_{y}A_{1} = \Big[\dd \i (1 - U L) - \frac{\dd \i (L^{2} + N^{2})}{\rho (1 - U L)}\Big]F_{0}, \\
F_{1,y} = \dd \i \rho (1 - U L)A_{0},
\end{gather}
\label{A1F1terms}%
\end{subequations}
which may be solved to find
\begin{subequations}
\begin{align}
A_{1}(y) &= \bar{A}_{1}(1-UL) + \i \dd (1-UL)\int_{0}^{y}\Big(1 - \frac{L^{2} + N^{2}}{\rho (1-UL)^{2}}\Big)F_{0}\mathrm{d}y', \label{A1sec} \\
F_{1}(y) &= \bar{F}_{1} + \i \dd \bar{A}_{0}\int_{0}^{y}\rho(1-UL)^{2}\mathrm{d}y'. \label{F1sec}
\end{align}
\label{A1F1sec}%
\end{subequations}
The solution to the $F_{1}$ secularity condition in \cref{F1sec}, we see, is not in general a constant. The high frequency forces variation in the acoustic pressure over the boundary layer at an order at which it was previously assumed to be constant~\citep{brambley2011b}. (Or, rather, the relationship between frequency and boundary layer thickness assumed here allows the pressure variation to jump to a lower order). Solving at second order with the secular terms removed,
\begin{align}
\t{v}_{2} &= A_{2} + \frac{\i L}{\eta }B_{1}\mathrm{e}^{-\theta } + a_{0}\mathrm{e}^{-\theta } + \frac{\i (1-UL)}{\sigma \eta T}D_{1}\mathrm{e}^{-\sigma \theta },& \t{p}_{2} &= F_{2},
\end{align}
where
\begin{equation}
a_{0} = \frac{\i L T}{\eta }\Big(\frac{B_{0}}{\eta T}\Big)_{y} + \frac{\i(1-UL)}{\eta T}d_{0} + \frac{\i \dd N}{\eta }G_{0}. \label{a0}
\end{equation}

To find the secularity conditions for $A_{2}$ and $F_{2}$ which would close our solutions, we need solutions for $\hh{u}_{2}$ and $\hh{T}_{2}$, and secularity conditions for $B_{1}$ and $D_{1}$.  The latter are found by removing resonant terms, as before, giving
\begin{subequations}
\begin{gather}
B_{1,y} + \frac{3}{2}\big(\ln{\eta T}\big)_{y}B_{1} = -\frac{U_{y}}{2T}d_{0} - \frac{\i U_{y}\eta }{2(1 - U L)}a_{0} + \frac{1}{2\eta T}(T B_{0,y})_{y}, \\
D_{1,y} + \frac{1}{2}\big(\ln{\eta T}\big)_{y}D_{1} = 0.
\end{gather}
\label{B1D1terms}%
\end{subequations}
Solving \cref{B1D1terms} leads to
\begin{subequations}
\begin{gather}
B_{1}(y) = (1-UL)^{-3/4}\Bigg\{\!B_{1}(0) + \!\!\int_{0}^{y}\!(1-UL)^{3/4}\!\bigg[\-\frac{U_{y}}{2T}d_{0} - \frac{\i U_{y}\eta}{1-UL}\frac{a_0}{2} + \frac{1}{2\eta T}(T B_{0,y})_{y}\bigg]\mathrm{d}y\!\Bigg\}\\
D_{1}(y) = D_{1}(0)(1-UL)^{-1/4}.
\end{gather}
\label{B1D1sec}%
\end{subequations}
The solutions for $\hh{u}_{2}$ and $\hh{T}_{2}$ are then found to be 
\begin{subequations}
\begin{gather}
\hh{u}_{2} = B_{2}\mathrm{e}^{-\theta } + b_{0}\mathrm{e}^{-\sigma \theta } - \frac{\i U_{y}}{1-UL}A_{2}  + b_{1}, \\
\hh{T}_{2} = D_{2}\mathrm{e}^{-\sigma \theta } + d_{1}\mathrm{e}^{-\theta } - \frac{\i T_{y}}{1-UL}A_{2} + d_{2},
\end{gather}
\label{u2t2}%
\end{subequations}
where
\begin{subequations}
\begin{align}
b_{0} =&\, \frac{U_{y}}{\sigma \eta T}D_{3}, \qquad\qquad
b_{1} = \, -\frac{\i}{\eta ^{2}T}\Big(\frac{T U_{y} A_{0}}{1-UL}\Big)_{yy} + \frac{\dd L}{\rho (1-UL)}F_{1}, \label{b1} \\
d_{1} =&\, \frac{\Pr}{1-\Pr}\bigg\{\Big(2U_{y} - \frac{LT_{y}}{1-UL}\Big)\frac{B_{1}}{\eta} + \frac{\i T_{y}}{1-UL}a_{0} \notag\\&\qquad\qquad- \frac{2U_{y}}{\eta ^{2}}B_{0,y} + \frac{1}{\Pr \eta^{2} T}\big((\eta T d_{0})_{y} + \eta (T d_{0})_{y}\big)\bigg\}, \label{d1} \\
d_{2} =&\, -\frac{\i}{\eta ^{2}T}\bigg\{\frac{1}{\Pr}\Big(\frac{T T_{y} A_{0}}{1-UL}\Big)_{yy} + \frac{1-UL}{A_{0}}\Big(\frac{TU_{y}^{2}A_{0}^{2}}{(1-UL)^{2}}\Big)_{y}\bigg\} + \frac{\dd}{\rho}F_{1}. \label{d2}
\end{align}
\end{subequations}
Although we are not solving for $\t{v}$ and $\t{p}$ to $\mathcal{O}(\ee^{3})$, we must use the third order equations to form the secularity conditions for $A_{2}$ and $F_{2}$.  As before in \cref{A1F1terms}, we find
\begin{subequations}
\begin{gather}
A_{2,y} - 2\big(\ln{\eta T}\big)_{y}A_{2} = -\frac{\i (1 - U L)}{T}d_{2} - \i L b_{1} + \dd \i \gamma (1 - U L)F_{1} - \frac{\dd \i N^{2}}{\rho (1 - U L)}F_{1}, \\
F_{2,y} = \dd \i \rho (1 - U L)A_{1}.
\end{gather}
\label{A2F2terms}%
\end{subequations}
Solving \cref{A2F2terms} we find
\begin{subequations}
\begin{align}
A_{2}(y) =&\, (1-UL)\bigg\{\!\bar{A}_{2} + \i \dd\!\!\int_{0}^{y}\!\!-\frac{\i}{T}d_{2} - \frac{\i L}{1-UL} b_{1} + \dd \i \gamma F_{1} - \frac{\i \dd N^{2}}{\rho(1-UL)^{2}}F_{1}\,\mathrm{d}y'\bigg\}, \label{A2sec} \\
F_{2}(y) =&\, \bar{F}_{2} + \i \dd \int_{0}^{y}\rho(1-UL)A_{1}\mathrm{d}y', \label{F2sec}
\end{align}
\label{A2F2sec}
\end{subequations}
which are the final conditions needed to close our solutions for $\t{v}$ and $\t{p}$ to $\mathcal{O}(\ee^{2})$.

\subsection{Matching the high frequency solutions to the outer flow}\label{sec:B1}

We must match the inner solutions found above:
\begin{subequations}
\begin{align}
\t{p}(y,\theta) = &\, F_{0}(y) + \ee F_{1}(y) + \ee^{2} F_{2}(y) + \mathcal{O}(\ee ^{3}), \label{hf_pB} \\
\t{v}(y,\theta) = &\, A_{0}(y) + \ee \Big[A_{1}(y) + \frac{\i L}{\eta(y) } B_{0}(y)\mathrm{e}^{-\theta } + \frac{\i (1-U(y)L)}{\sigma \eta(y) T(y)}D_{0}(y)\mathrm{e}^{-\sigma \theta }\Big] \notag \\
&\,+ \ee^{2}\Big[A_{2}(y) + \frac{\i L}{\eta }B_{1}(y)\mathrm{e}^{-\theta } + a_{0}(y)\mathrm{e}^{-\theta } + \frac{\i (1-U(y)L)}{\sigma \eta (y) T(y)}D_{1}(y)\mathrm{e}^{-\sigma \theta }\Big] + \mathcal{O}(\ee ^{3}). \label{hf_vB}
\end{align}
\label{hf_pvB}%
\end{subequations}
with the outer solutions \cref{highfreqouter},
\begin{subequations}
\begin{align}
p_{\mathrm{ui}}(1-\delta y) =&\, \pinf + \i \ee \bar{\delta }(1 - ML) y \vinf + \frac{1}{2}\ee^{2} \bar{\delta }^2 (N^2 - \bar{\alpha }^2) y^2 \pinf + \mathcal{O}(\ee^{3}), \label{phighfreqouterB}\\
v_{\mathrm{ui}}(1-\delta y) =&\, \vinf + \i \ee \bar{\delta } \frac{\bar{\alpha }^{2} - N^{2}}{1 - ML} y \pinf + \frac{1}{2}\ee^{2} \bar{\delta }^2 (N^2 - \bar{\alpha }^{2}) y^2 \vinf + \mathcal{O}(\ee^{3}), \label{vhighfreqouterB}
\end{align}
\label{highfreqouterB}%
\end{subequations}
in the limit $y\to \infty $.

Matching \cref{hf_pvB} with \cref{highfreqouterB} at leading order leads to
\begin{align}
\bar{F}_{0} &= \pinf,& \bar{A}_{0} &= \frac{\vinf}{1-ML}. \label{mleading}
\end{align}
Next, we write the secularity conditions \cref{A1F1sec,A2F2sec} in terms of bounded integrals to aid matching:
\begin{subequations}
\begin{align}
F_{1}(y) =&\, \bar{F}_{1} + \i \dd \bar{A}_{0}(1-ML)^{2}\Big[y - \int^{y}_{0}\cb_{0}\mathrm{d}y'\Big],\\
\frac{A_{1}(y)}{1-UL} =&\, \bar{A}_{1} + \i \dd \bar{F}_{0}\bigg\{\Big(1 - \frac{L^{2}+N^{2}}{(1-ML)^{2}}\Big)y + \frac{L^{2}+N^{2}}{(1-ML)^{2}}\int_{0}^{y}\cb_{1}\mathrm{d}y'\bigg\}, \\
F_{2}(y) =&\, \bar{F}_{2} + \i \dd (1-ML)^{2}\bar{A}_{1}\Big[y - \int^{y}_{0}\cb_{0}\mathrm{d}y'\Big] - \dd ^{2}(L^{2}+N^{2})\bar{F}_{0}\bigg\{I_{1}y \notag \\
&\, - \int_{0}^{y}\cb_{0}(y')\int_{0}^{y'}\cb_{1}(y'')\mathrm{d}y''\mathrm{d}y' + \int_{0}^{y}\Big(\int_{0}^{y'}\cb_{1}(y'')\mathrm{d}y'' - I_{1}\Big)\mathrm{d}y'\bigg\}, \notag \\
&\,- \dd ^{2}(1-ML)^{2}\bar{F}_{0}\Big(1 - \frac{L^{2}+N^{2}}{(1-ML)^{2}}\Big)\Big[\frac{1}{2}y^{2} - \int_{0}^{y}y'\cb_{0}\mathrm{d}y'\Big] \\
\frac{A_{2}(y)}{1-UL} =&\, \i \dd \bar{F}_{1}\bigg\{\!\Big(1 - \frac{L^{2}+N^{2}}{(1-ML)^{2}}\Big)y + \frac{L^{2}+N^{2}}{(1-ML)^{2}}\!\int_{0}^{y}\!\!\cb_{1}\mathrm{d}y'\bigg\}\! + \i \xi (\gamma - 1)^{2}\bar{A}_{0}\!\int_{0}^{y}\!\!\cb_{\mu}\,\mathrm{d}y'\notag \\
&\,+ \bar{A}_{2} - \dd ^{2}(1-ML)^{2}\bar{A}_{0}\bigg\{\Big(1 - \frac{L^{2}+N^{2}}{(1-ML)^{2}}\Big)\frac{y^2}{2} + \frac{L^{2}+N^{2}}{(1-ML)^{2}}\!\int_{0}^{y}\!y'\cb_{1}\mathrm{d}y'\bigg\} \notag \\
&\,+ \dd ^{2}(1-ML)^{2}\bar{A}_{0}\Bigg\{\frac{L^{2}+N^{2}}{(1-ML)^{2}}\!\int_{0}^{y}\!\!\cb_{1}(y')\!\int_{0}^{y'}\!\!\!\cb_{0}(y'')\,\mathrm{d}y''\mathrm{d}y'\notag\\
&\,+ \Big(1 - \frac{L^{2}+N^{2}}{(1-ML)^{2}}\Big)I_{0}y + \!\left(\!1 - \frac{L^{2}+N^{2}}{(1-ML)^{2}}\!\right)\!\!\int_{0}^{y}\!\!\left(\int_{0}^{y'}\!\!\cb_{0}(y'')\mathrm{d}y'' - I_{0}\!\right)\!\mathrm{d}y'\Bigg\},
\end{align}
\label{F1A1F2A2bounded}%
\end{subequations}
where
\begin{align}
I_{0} =&\, \int_{0}^{\infty }\cb_{0}\mathrm{d}y,& I_{1} &= \int_{0}^{\infty }\cb_{1}\mathrm{d}y, \notag\\
\cb_{0} &= 1 - \frac{\rho (1-UL)^{2}}{(1-ML)^{2}},& \cb_{1} &= 1 - \frac{(1-ML)^{2}}{\rho (1-UL)^{2}},
\end{align}\vspace{-0.5\baselineskip}\[
\cb_{\mu}(y) = \frac{1}{1-UL}\Big[\frac{1}{2\Pr}(T^{2})_{yyy} + (TU_{y}^{2})_{y} + \frac{LT}{1-UL}(TU_{y})_{yy}\Big].
\]
Matching at first order provides the relations
\begin{align}
\bar{F}_{1} &= \i \dd I_{0}(1-ML)\vinf,& \bar{A}_{1} &= -\i \dd I_{1}\frac{L^{2}+N^{2}}{(1-ML)^{2}}\pinf, \label{mfirst}
\end{align}
while at second order we find
\begin{subequations}
\begin{align}
\bar{F}_{2} =&\, \dd ^{2}(I_{0}I_{1} + I_{11} - I_{01})(L^{2} + N^{2})\pinf - \dd ^{2}I_{2}\big((1 - ML)^{2} - L^{2} - N^{2}\big)\pinf \\
\bar{A}_{2} =&\, -\i \xi (\gamma - 1)^{2}I_{\mu}\frac{\vinf}{1-ML} + \dd ^{2}(I_{0}I_{1} + I_{3} - I_{10})(L^{2} + N^{2})\frac{\vinf}{1-ML} \notag \\
&\,- \dd ^{2}I_{00}\big((1 - ML)^{2} - L^{2} - N^{2}\big)\frac{\vinf}{1-ML},
\end{align}
\label{msecond}
\end{subequations}
where we have introduced
\begin{align}
I_{2} =&\, \int_{0}^{\infty }y\cb_{0}\mathrm{d}y, \qquad\qquad I_{3} = \int_{0}^{\infty }y\cb_{1}\mathrm{d}y, \qquad\qquad I_{\mu} = \int_{0}^{\infty }\cb_{\mu}\mathrm{d}y, \notag\\
I_{01} =&\, \int_{0}^{\infty }\cb_{0}\int_{0}^{y}\cb_{1}(y')\mathrm{d}y'\mathrm{d}y, \qquad\qquad I_{10} = \int_{0}^{\infty }\cb_{1}\int_{0}^{y}\cb_{0}(y')\mathrm{d}y'\mathrm{d}y, \\
I_{00} =&\, \int_{0}^{\infty }\Big(\int_{0}^{y}\cb_{0}(y')\mathrm{d}y' - I_{0}\Big)\mathrm{d}y, \qquad\qquad I_{11} = \int_{0}^{\infty }\Big(\int_{0}^{y}\cb_{1}(y')\mathrm{d}y' - I_{1}\Big)\mathrm{d}y.\notag
\end{align}

\subsection{The effective impedance} \label{sec:B2}

To form the effective impedance we evaluate the pressure and velocity at the wall, $y=0$:
\begin{align}
\t{p}(0) =&\, F_{0}(0) + \ee F_{1}(0) + \ee ^{2}F_{2}(0),\\
\t{v}(0) =&\, A_{0}(0) + \ee \Big[A_{1}(0) + \frac{\i L}{\eta (0)} B_{0}(0)\Big] \notag\\&\,+ \ee ^{2} \Big[A_{2}(0) + \frac{\i L}{\eta (0)} B_{1}(0) + a_{0}(0) + \frac{\i}{\sigma \eta (0)T(0)}D_{1}(0)\Big],
\end{align}
where the $A_{j}(0)$ and $F_{j}(0)$ are found in the previous section.  The remaining required quantities are
\begin{subequations}\begin{align}
B_{0}(0) =&\, \i U_{y}(0)\frac{\vinf}{1-ML}, \quad \quad \quad B_{1}(0) = \Big[\dd I_{1}U_{y}(0)\frac{L^{2}+N^{2}}{(1-ML)^{2}} - \frac{\dd L}{\rho(0)
}\Big]\pinf,\\
a_{0}(0) =&\, -\frac{\i \dd N^{2}}{\eta(0)\rho(0)}\pinf - \Big[\frac{\Pr}{1-\Pr} \frac{2U_{y}(0)^{2}}{\eta(0)^{2}T(0)} + \frac{5L^{2}U_{y}(0)^{2}}{4\eta(0)^{2}}\Big]\frac{\vinf}{1-ML},\\
D_{1}(0) =&\, -\frac{\dd}{\rho (0)}\pinf - \frac{\Pr}{1-\Pr}\frac{2\i U_{y}(0)^{2}}{\eta(0)}\frac{\vinf}{1-ML},
\end{align}\end{subequations}
The wall impedance $Z = \t{p}(0)/\t{v}(0)$, so we may write
\begin{equation}
Z = \frac{\pinf + \ee \i \dd I_{0}(1-ML)\vinf + \ee ^{2}\bar{\mathcal{A}}\pinf}{\dfrac{\vinf}{1-ML}\Big[1 - \ee \dfrac{LU_{y}(0)}{\eta (0)} + \ee^{2}\bar{\mathcal{B}}\Big] + \pinf \Big[- \ee \i \dd I_{1}\dfrac{L^{2}+N^{2}}{(1-ML)^{2}} + \ee ^{2}\bar{\mathcal{C}}\Big]}
\end{equation}
where
\begin{subequations}\begin{align}
\bar{\mathcal{A}} =&\, \dd ^{2}(I_{0}I_{1} + I_{11} - I_{01})(L^{2} + N^{2}) - \dd ^{2}I_{2}\big((1 - ML)^{2} - L^{2} - N^{2}\big), \\
\bar{\mathcal{B}} =&\, -\i \xi (\gamma - 1)^{2}I_{\mu} + \dd ^{2}(I_{0}I_{1} + I_{3} - I_{10})(L^{2} + N^{2}) - \dd ^{2}I_{00}\big((1 - ML)^{2} - L^{2} - N^{2}\big) \notag\\
&\,+ \frac{\sigma (1-\sigma)}{1-\Pr} \frac{2U_{y}(0)^{2}}{\eta(0)^{2}T(0)} - \frac{5L^{2}U_{y}(0)^{2}}{4\eta(0)^{2}},\\
\bar{\mathcal{C}} =&\, \dd I_{1}\frac{\i LU_{y}(0)}{\eta (0)}\frac{L^{2}+N^{2}}{(1-ML)^{2}} - \frac{\i \dd (L^{2}+N^{2})}{\eta (0)\rho(0)} - \frac{\i \dd (\gamma - 1)}{\sigma \eta (0)}.
\end{align}\end{subequations}
Dividing top and bottom by $\vinf$ allows us to introduce the effective impedance $Z_{\mathrm{eff}} = \pinf/\vinf$; rearranging for $Z_{\mathrm{eff}}$ gives
\begin{equation}
Z_{\mathrm{eff}} = \frac{1}{1-ML}\frac{Z - \ee \dfrac{LU_{y}(0)}{\eta (0)}Z - \ee \i \dd I_{0}(1-ML)^{2} + \ee^{2}\bar{\mathcal{B}}Z}{1 + \ee \i \dd I_{1}\dfrac{L^{2}+N^{2}}{(1-ML)^{2}}Z + \ee ^{2}(\bar{\mathcal{A}} - \bar{\mathcal{C}}Z)}. \label{Zeff_hfB}
\end{equation}
which is equivalent to the result in the main text \cref{Zeff_hf} once the expressions for $\ee$, $\dd$, $\xi$, $\eta(0)$, $L$ and $N$ are substituted, and the $I_{j}$ integrals are written in terms of $r$. If the strict constraint $\delta = \ee^{3}\dd$ is relaxed, and instead the weaker constraint $\delta \sim \ee ^{(2+n)}$, $n>0$, is used, an expansion in powers of the two small parameters $\delta $, $\ee$ can be found to $\mathcal{O}(\delta/\ee^{2})$ and shown to be asymptotically equivalent to $\cref{Zeff_hf}$ at $\mathcal{O}(\ee)$.

\section{Details of numerical method}\label{sec:appendixNUM}

Details of our numerical method are given below.

\subsection{Regularity at $r=0$}

The behaviour of the acoustics near $r=0$ are investigated by assuming for each acoustic quantity a series expansion in $r$, $\t{q} \sim \alpha_{0} + \alpha_{1}r+\dots$, and analysing the $\mathcal{O}(1/r^{n})$ terms of the viscous governing equations \cref{lns}.  By ensuring cancellation at $\mathcal{O}(1/r^{n})$ for each governing equation, consistent regularity conditions are derived. Below, each equation is considered in turn.  In this derivation we assume $m\ge0$, but the same results hold for negative $m$.

Irregular terms appear in the continuity equation \cref{lns_mass} only at $\mathcal{O}(1/r)$:
\begin{equation}
\frac{\t{v}}{r} - \frac{\i m}{r}\t{w} = 0 \qquad \implies \qquad
 \left\{ \begin{array}{rl}
  \t{v}(0) = 0, \quad \quad \,\,\,\, & m = 0, \qquad \t{v} \sim b_{1}r+\dots\\
  \t{v}(0) = \i m \t{w}(0), & m \ne 0.
\end{array} \right.
\label{v0_allm}%
\end{equation}
No other information may be gathered from this equation.

In the axial momentum equation \cref{lns_umom}, the most singular terms at $r=0$ are $\mathcal{O}(1/r^{2})$, which gives us
\begin{equation}
\frac{m^{2}}{r^{2}}\t{u} = 0 \qquad \implies \qquad
 \left\{ \begin{array}{rl}
  \mathrm{Identically}\,\,\mathrm{true}, &  m = 0, \\
  \t{u}(0) = 0, \qquad \quad & m \ne 0, \qquad \t{u} \sim a_{1}r+\dots
\end{array} \right.
\label{u0_allm}%
\end{equation}
where we have set $a_{0}=0$ in the $\t{u}$ expansion for $m\ne0$, but left $a_{1}\ne0$ to provide a contribution at $\mathcal{O}(1/r)$.  The $\mathcal{O}(1/r)$ terms of \cref{lns_umom} give
\begin{equation}
\frac{1}{r}(\H \t{u}_{r} + U_{r}\t{\H}) - \frac{m^{2}}{r^{2}}\H\t{u} - (1+\beta)\frac{1}{r}\H\left(\i k \t{v} + k m \t{w}\right) = 0. \label{r1_xmom}
\end{equation}
In the $m=0$ case, we may use \cref{v0_allm} and the fact that $U_{r}\to 0$ as $r\to 0$ to find $\t{u}_{r}(0) = 0$. If $m\ne0$, \cref{v0_allm} and \cref{u0_allm} imply
\begin{equation}
\frac{(1-m^{2})}{r}\H \t{u}_{r} = 0 \qquad \implies \qquad \left\{ \begin{array}{rl}
  \mathrm{Identically}\,\,\mathrm{true}, & m = 1, \\
  \t{u}_{r}(0) = 0, \qquad\,\,\,\, & m > 1.
\end{array} \right.
\label{u_mg1}
\end{equation}

At $\mathcal{O}(1/r^{2})$ the radial momentum equation \cref{lns_vmom} behaves like
\begin{equation}
-\frac{m^{2}}{r^{2}}\H\t{v} - (2+\beta)\H \frac{\t{v}}{r^{2}} + (3+\beta)\frac{\i m}{r^{2}}\H \t{w} = 0, \label{r2_rmom}
\end{equation}
from which in the $m=0$ case we recover \cref{v0_allm}.  If $m\ne0$, we use the expansions
\begin{align}
\t{v} \sim b_{0} + b_{1}r+\dots, && \t{w} \sim c_{0} + c_{1}r+\dots \label{vw_exp}
\end{align}
in \cref{r2_rmom}, with \cref{v0_allm} implying $b_{0} = \i m c_{0}$, to find
\begin{equation}
\frac{(1-m^{2})}{r^{2}}\t{v} = 0 \qquad \implies \qquad
 \left\{ \begin{array}{rl}
  \mathrm{Identically}\,\,\mathrm{true},& m = 1, \\
  \t{v}(0) = 0, \qquad \quad & m > 1, \qquad \t{v} \sim b_{1}r+\dots
\end{array} \right.
\label{v_mn0}%
\end{equation}
where again $b_{1}$ is left to contribute to the $\mathcal{O}(1/r)$ system.  At $\mathcal{O}(1/r)$ we find, with $\H_{r}\to0$ as $r\to0$,
\begin{equation}
-\frac{m^{2}}{r^{2}}\H\t{v} + (2+\beta)\H \left( \frac{\t{v}_{r}}{r} - \frac{\t{v}}{r^{2}} \right) + (3+\beta)\frac{\i m}{r^{2}}\H \t{w} - (1+\beta)\frac{\i m}{r}\H \t{w}_{r} = 0. \label{r1_rmom}
\end{equation}
If $m=0$, the left hand side is identically zero by the $\t{v}$ expansion in \cref{v0_allm}. If $m\ne0$ we use \cref{vw_exp} and \cref{v0_allm} in \cref{r1_rmom} to find $m b_{1} = 2\i c_{1}$, implying
\begin{equation}
m \t{v}_{r}(0) = 2\i \t{w}_{r}(0). \label{vw2_mn0}
\end{equation}
Now, since for $m=0$ the $\mathcal{O}(1/r)$ system was redundant, we may use the $\mathcal{O}(1)$ system to derive a boundary condition for $\t{p}$.  Using \cref{u0_allm} and setting the derivatives of the base flow to zero at $r=0$, we find for $m=0$
\begin{equation}
\t{p}_{r} = \frac{\H}{\Ren}(2+\beta)\left(\t{v}_{rr} - \frac{\t{v}}{r^{2}} + \frac{\t{v}_{r}}{r}\right). \label{r0_mom}
\end{equation}
Now, using $\t{v}\sim b_{1}r + b_{2}r^{2}+\dots$ from the $m=0$ case of \cref{v0_allm}, the $\mathcal{O}(1)$ contribution of the large bracket is simply $3b_{2}$.  This implies
\begin{equation}
\t{p}_{r}(0) = \frac{3}{2}\frac{\H}{\Ren}(2+\beta)\t{v}_{rr}(0), \qquad m = 0. \label{pr0_m0}
\end{equation}

The azimuthal momentum equation \cref{lns_wmom} at $\mathcal{O}(1/r^{2})$ is
\begin{equation}
-(3+\beta)\frac{\i m}{r^{2}}\H\t{v} - \frac{m^{2}}{r^{2}}\H \t{w} - (1+\beta)\frac{m^{2}}{r^{2}}\H \t{w} - \frac{1}{r^{2}}\H\t{w} = 0. \label{r2_tmom}
\end{equation}
When $m=0$ \cref{r2_tmom} simply reduces to $\t{w}(0) = 0$ (and indeed $\t{w}(r)\equiv 0$).  If $m\ne0$, we may use \cref{v0_allm} to derive the relation
\begin{equation}
\frac{(m^{2}-1)}{r^{2}}\t{w} = 0 \quad \implies \quad \left\{ \begin{array}{rl}
  \mathrm{Identically}\,\,\mathrm{true}, & m = 1, \\
  \t{w}(0) = 0, \qquad\quad & m > 1, \qquad \t{w} \sim c_{1}r+\dots
\end{array} \right.
\label{w_mn0}
\end{equation}
where $c_{1}$ contributes at $\mathcal{O}(1/r)$, and the $m>1$ series expansion is also valid for $m=0$. At $\mathcal{O}(1/r)$, setting gradients of mean flow quantities to zero, \cref{lns_wmom} reduces to
\begin{equation}
\frac{\i m}{r}\t{p} = -\frac{\H}{\Ren}\left\{-(1+\beta)\frac{k m}{r}\t{u} -(3+\beta)\frac{\i m}{r^{2}}\t{v} - (1+\beta)\frac{\i m}{r}\t{v}_{r} - (2+\beta)\frac{m^{2}}{r^{2}}\t{w} + \frac{\t{w}_{r}}{r} - \frac{\t{w}}{r^{2}} \right\}, \label{r1_tmom}
\end{equation}
which is identically zero when $m = 0$ and the series expansion \cref{w_mn0} for $m>1$ is assumed.  When $m\ne0$, we may use \cref{u0_allm} and the expansions \cref{vw_exp}, along with \cref{vw2_mn0} which implies $m b_{1} = 2 \i c_{1}$, to form the boundary condition
\begin{equation}
\t{p}(0) = \frac{\H}{2\Ren}(2+\beta)(4 - m^{2})\t{v}_{r}(0), \qquad m \ne 0.
\end{equation}

The energy equation \cref{lns_energy} is relatively simple. At $\mathcal{O}(1/r^{2})$ we find
\begin{equation}
\frac{m^{2}}{r^{2}}\H\t{T} = 0 \quad \implies \quad \left\{ \begin{array}{rl}
  \mathrm{Identically}\,\,\mathrm{true}, & m = 0, \\
  \t{T}(0) = 0, \qquad\quad & m \ne 0, \qquad \t{T} \sim d_{1}r+\dots
\end{array} \right.
\label{T_allm}
\end{equation}
while at $\mathcal{O}(1/r)$ we find
\begin{equation}
\frac{1}{r}\left(\H \t{T}_{r} + T_{r}\t{H}\right) - \frac{m^{2}}{r^{2}}\H\t{T} = 0, \label{r1_energy}
\end{equation}
where the $\mathcal{O}(1/r)$ contribution of the term $\propto \t{T}/r^{2}$ is considered.  Now, $T_{r}\to0$ as $r\to0$, which leads to
\begin{equation}
\frac{(1-m^{2})}{r}\H \t{T}_{r} = 0 \qquad \implies \qquad \left\{ \begin{array}{rl}
  \mathrm{Identically}\,\,\mathrm{true}, & m = 1, \\
  \t{T}_{r}(0) = 0, \qquad\,\,\,\, & m > 1.
\end{array} \right.
\label{T_mg1}
\end{equation}

Collecting the information above, our regularity conditions at $r=0$ are:
\begin{equation}
    \begin{gathered}
\t{p}_{r} = \frac{3}{2}\frac{\H}{\Ren}(2+\beta)\t{v}_{rr}, \qquad \t{u}_{r} = 0, \qquad \t{v} = 0, \qquad \t{w} = 0, \qquad \t{T}_{r} = 0,\\
    \end{gathered}
\end{equation}
for $m=0$, and
\begin{equation}
    \begin{gathered}
\t{p} = \frac{1}{2}\frac{\H}{\Ren}(2+\beta)(4 - m^{2})\t{v}_{r},\\
\t{u} = 0,\qquad  \t{v} = \i m \t{w}, \qquad  \t{w}_{r} = -\frac{\i m}{2}\t{v}_{r}, \qquad \t{T} = 0,
    \end{gathered}
\end{equation}
for $m\ge1$.

\subsection{Mode finding}

To find acoustic modes, we solve the dispersion relation \cref{dispersion} numerically. This is done by iterating on $k$ (or $\omega$) via a Newton--Raphson procedure, given a fixed $\omega$ (or $k$). For the majority of modes (e.g. the cut-off and cut-on acoustic modes in the $k$-plane) the solutions of the Ingard--Myers dispersion relation
\begin{equation}
Z = \frac{(\omega - M k)^{2}}{\i \omega}\frac{J_{m}(\alpha)}{\alpha J'_{m}(\alpha)} \label{ingardmyersdisp}
\end{equation}
are used as an initial guess.  Then, to find surface modes in the $k$-plane and unstable modes in the $\omega$-plane, a fine two-dimensional mesh of complex-valued initial guesses is fed into the Newton--Raphson solver.  This is done to minimise the chance of missing solutions.

The following test gives an example of the time taken to find modes for each model --- the asymptotic models, using the dispersion relation \cref{asym_disp}, and the numerics, using the dispersion relation \cref{dispersion}. To find $102$ acoustic modes (not surface waves) with Ingard--Myers modes as initial guesses, the time taken was: $40.7\mathrm{s}$ for the high frequency model \cref{Zeff_hf}; $527.4\mathrm{s}$ for the $\mathcal{O}(\delta)$ model \cref{BL_governing_eqs}; and $1416.5\mathrm{s}$ for the LNSE numerics \cref{lns}. This test was performed on a laptop with a $2.5\mathrm{GHz}$ Intel i5 processor.

\subsection{Numerical convergence}

The numerical solver was checked for consistency and convergence in a number of ways: by comparison with analytical solutions in the inviscid uniform flow case; by comparison with asymptotic mode shapes in the boundary layer; by checking convergence of the solver with respect to number of grid points used; and by checking that the asymptotics agree with the numerics to the stated order of accuracy. The first of these is trivial -- the cylindrical solution $\t{p}=J_{m}(\alpha r)$ in the uniform inviscid case is well-known -- and will not be discussed further. The second point may be verified in \cref{fig:asympmodeshapes} --- it is clear that the numerical and asymptotic solutions have the same near-wall behaviour. 

The third point may be assessed by varying the number of grid points and checking the values of $Z$ calculated at the wall in each case, and calculating the relative error with respect to some well-converged case.  
\begin{figure}
\captionsetup[subfigure]{aboveskip=0pt}
    \centering
    \begin{subfigure}[t]{0.49\textwidth}
        \centering        
        \includegraphics*[width=0.99\textwidth]{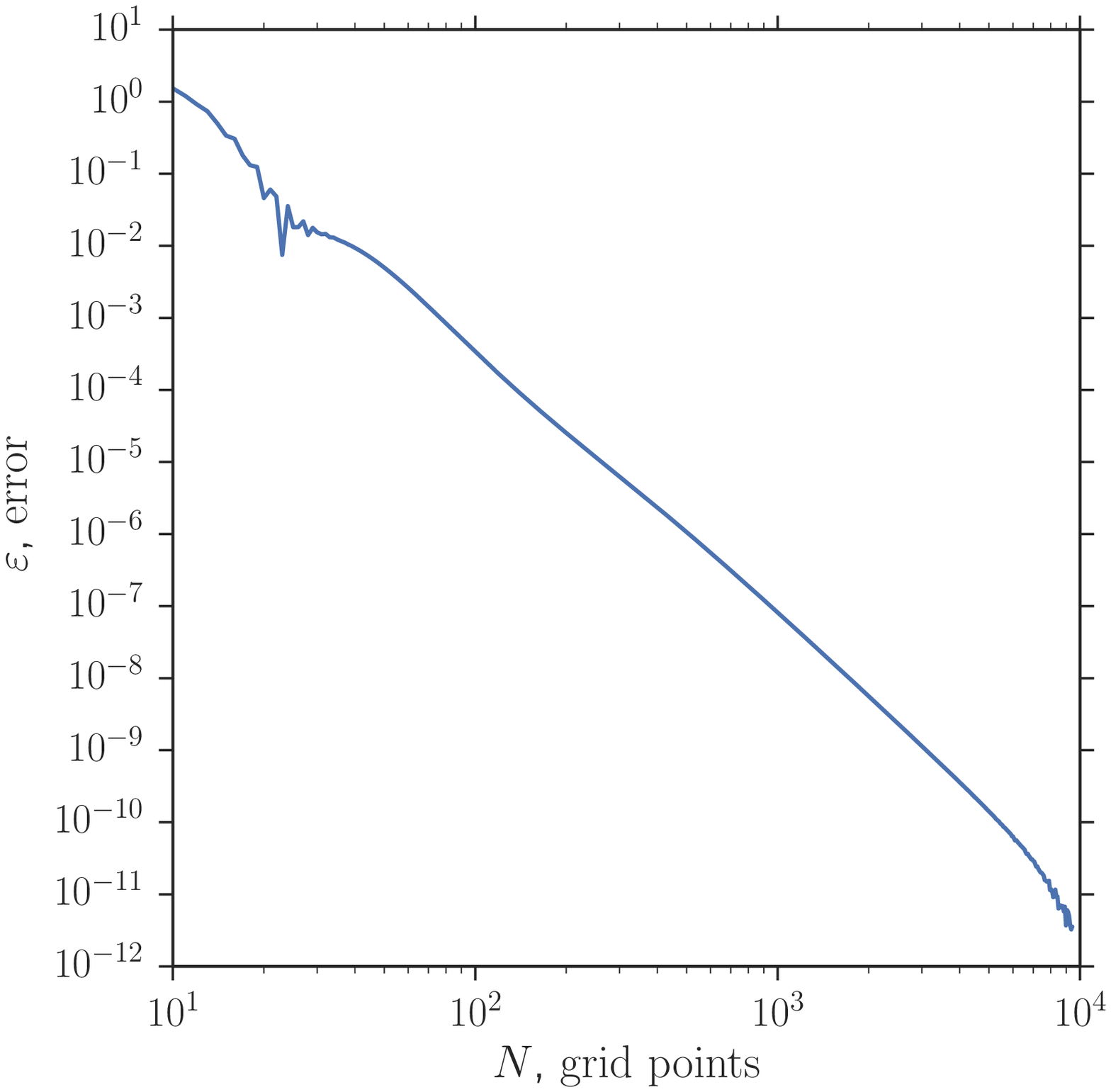}
    \caption{}
    \label{fig:nsconv}
    \end{subfigure}%
    ~
    \begin{subfigure}[t]{0.49\textwidth}
        \centering
        \includegraphics*[width=0.97\textwidth]{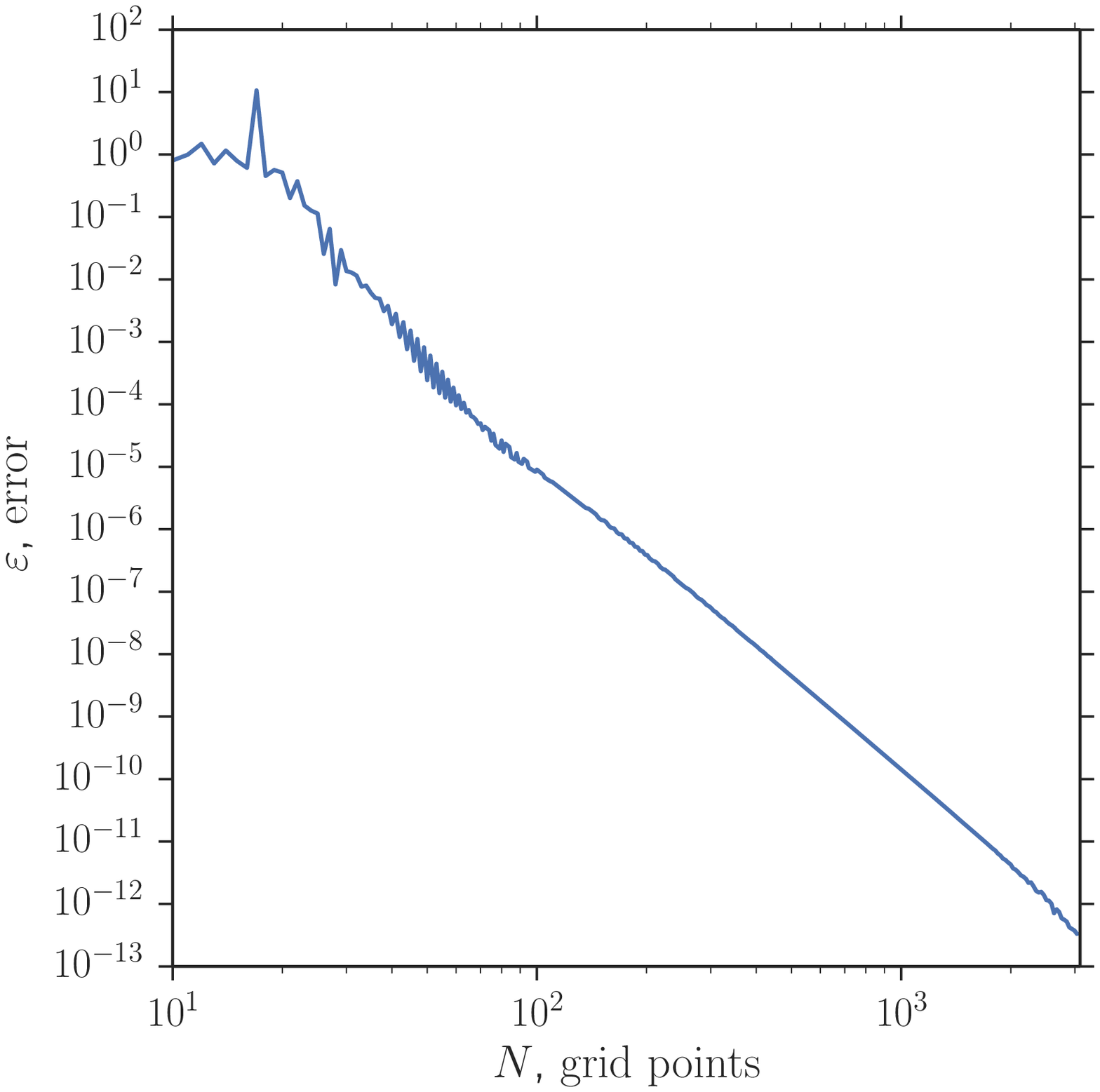}
    \caption{}
    \label{fig:pbconv}
    \end{subfigure}%
\caption{Relative error $\varepsilon = \abs{Z/Z_{c} - 1}$ of calculated impedance value as number of grid points $N$ is increased, with respect to a well-converged case with $N=50000$. (a) Viscous numerics with $\omega = 31$, $k = 15+5\i$, $m = 12$, $M = 0.5$, $\delta = 7\times10^{-3}$, $\Ren = 10^{6}$. (b) Inviscid numerics with $\omega = 5$, $k = 3- 15\i$, $m = 4$, $M = 0.5$ and $\delta = 2\times10^{-3}$. In both (a) and (b) the hyperbolic boundary layer profiles in \cref{baseflow} are used.} \label{fig:numconv}
\end{figure}
The convergence plots \cref{fig:numconv} show that for $N\approx8000$ the numerics are achieving errors of $\lesssim 10^{-8}$ in both the viscous and inviscid cases.  The rate of convergence is set by the treatment of the end-points of the numerical domain, where the 6th-order stencil is reduced to 4th-order to retain the use of central-difference approximations close to the domain edge.  Up-wind and down-wind 4th-order stencils are used at the boundary points.  

The final point is addressed in \cref{fig:asympaccuracy}, in which the accuracy of the asymptotics is measured against the numerical solution. This may be thought of conversely as a consistency check for the numerical solver. The correct gradients of asymptotic error shown in the figure show that the numerical solver are consistent down to very small errors ($\sim 10^{-8}$ or smaller).

\bibliographystyle{jfm}
\bibliography{paper}

\end{document}